\documentstyle[amssymb,epsf,12pt,thmsa,sw20lart]{article}

\input{tcilatex}
\begin{document}
\title{Complex Angular Momentum in General Quantum Field Theory}
\author{{J. Bros $^a$\, and \,G.A. Viano $^b$ }\\[5mm] 
{${}^a\, $ Service de Physique Th\'eorique,\, CEA--Saclay} \\ 
{F--91191 Gif--sur--Yvette Cedex,\,France  \ } \\
{${}^b\,$ Istituto Nazionale di Fisica Nucleare (I.N.F.N.),}\\
{Sezione di Genova,}\\
{ Dipartimento di Fisica dell'Universit\`a,
Genova,Italy}}
\date{November 20, 1998}
\maketitle
\begin{abstract}
\noindent It is proven that for each given two-field channel - called the ``$t-$channel''-   
with (off-shell) ``scattering angle'' $\Theta_t$,   
the four-point Green's function of any scalar Quantum Fields 
satisfying the basic principles of locality, spectral condition together with
temperateness admits a Laplace-type transform in the   
corresponding complex angular momentum variable $\lambda_t$,
dual to $\Theta_t$. This transform enjoys the following properties:\, 
a) it is holomorphic in a half-plane of the 
form $\func{Re} \lambda_t > m$, where $m$ is a certain ``degree of temperateness'' of the fields
considered,  
\,b) it is in one-to-one (invertible) 
correspondence with the (off-shell)  
``absorptive parts'' in the crossed two-field channels, 
\,c) it extrapolates in a canonical way to complex values of the angular momentum the coefficients 
of the (off-shell) $t-$channel partial-wave expansion of the 
Euclidean four-point function of the fields. These properties are established for all
space-time dimensions $d+1$ with $d \ge 2$. 
\end{abstract}

\section{Introduction}

The complex angular momentum analysis was widely used in the sixties, in
particle physics, for describing the high-energy asymptotic behaviour of the
scattering amplitude. With the arrival of QCD much attention was diverted
away from the ``old-fashioned'' approach to the strong interactions.
Interest was reignited (see [1] and references therein) within
the particle physics community with the arrival of colliders capable of
delivering very large centre-of-mass energies (e.g. the HERA collider at
DESY and the Tevatron collider at FNAL); from the theoretical viewpoint, 
this revival was made possible by the much earlier important results of BFKL[2]  
who discovered and characterized the Regge-like asymptotic properties of 
appropriate resummations of perturbative amplitudes 
\footnote{see also [3] for similar results in the case of scalar fields} 
in QCD. 

In the deep inelastic lepton-proton scattering a central role is played by
the so-called ``structure-functions'' which parametrize the structure of the
target as ``seen'' by the virtual photon. They are usually denoted by $F_{%
{\rm i}}\left( x,Q^{2}\right) ,$ where $Q^{2}=-q^{2}$ and $q^{\mu }=k^{\mu
}-k^{\prime ^{\mu }}$ is the momentum transfer $(k^{\mu }$ and $k^{\prime
\mu }$ being respectively the incoming and outgoing lepton four-momenta),
while $x=\frac{Q^{2}}{2\nu },\nu =p\cdot q\  \left( p^{2}={\rm M}^{2},{\rm M}%
\text{ being the proton mass}\right) .$ It is now possible to explore the
structure functions in a region where the momentum transfer is much smaller
than the centre- of - mass energy, i.e. for small values of $x.$ 
\ In the parton model one can show that the 
$x$-dependence, in the limit $x\rightarrow 0,$ is related to the behaviour
of hadronic scattering cross-sections at high energy [4]. This behaviour, 
which appears to exhibit Regge-like asymptotic properties, is reminiscent of   
the concept of ``exchange of families of particles with different
spins''. A detailed analysis of small $x$ structure function measurements, at
fixed target energies [5], show that they are indeed approximately consistent
with the predictions of such a model. We can thus say that, on one
side, the phenomenology calls for an extension of the conventional exchange
process and suggests an exchange mechanism involving families of
particles; on the other side, from a theoretical viewpoint, 
these families could be described by
``moving poles'', namely, poles in a certain ``complex angular momentum plane''.

The complex angular momentum theory originated long ago in connection with some problems of
classical mathematical physics, mainly the diffraction [6]. Then Regge [7]
extended these methods to quantum mechanics and specifically to the
scattering by Yukawian potentials (see also [8]).
In these works, the complex angular momentum analysis 
was produced by a direct analytic interpolation 
to complex values of the angular momentum variable of the 
relevant differential equations for partial waves. 
\ Then several authors (see [9] and references therein) conjectured
that the results proved by Regge, at non-relativistic level, might as
well be applied to the high energy relativistic dynamics where the method could
really display all its power. This relativistic extension, which of course
could no more be justified in a simple framework of differential equations,
was given a tentative formulation [9]   
in the approach of the so-called ``$S$-matrix theory'' 
based on the general, but rather
loose concept of ``maximal analyticity''. 
However, it must be emphasized that since that time 
no genuine relativistic complex angular momentum theory
relying on the general principles of Quantum Field Theory (Q.F.T.) has been given at all.

In view of the considerations developed above, one is then led to set the
following question whose conceptual interest is of primary importance:

Is it possible to find in the framework of general Q.F.T. 
a mathematical structure which leads to poles moving in the complex
angular momentum plane and that are responsible for an exchange mechanism
involving families of particles and giving rise to Regge-like asymptotic properties, 
as suggested by the
analyses of BFKL[2] and Berg\`ere et al.[3] in the philosophy of resummations of perturbative QFT?

We have already announced and briefly sketched a positive answer to this question
in a previous work [10]. Here we shall provide a detailed proof of the first 
basic result of [10], namely the existence of a field-theoretical   
off-shell version of the Froissart-Gribov representation of the
partial-waves [11,12]; the latter had been discovered by these authors in 1961 in the
analytic $S$-matrix approach of particle physics, requiring that the
scattering amplitude should satisfy the Mandelstam representation.\ In order
to prove our field-theoretical result we make use of a basic analyticity property of the
four-point function $ F$ implied by the standard axioms of locality,
spectrum and Lorentz invariance; moreover we use a majorization of $F$
which is a consequence of the ``temperateness axiom'' of quantum field
theory. The result which can be derived from these properties is the following: 
for each given two-field channel called the $t-$channel, with total squared 
energy-momentum $t$ and (off-shell) scattering angle $\Theta_t$, there exists  
an appropriate Fourier--Laplace type transform of $F$
with certain analyticity properties in the complex angular momentum $\lambda_t$
which is the natural conjugate variable of $\Theta_t$.  
One thus obtains a generalization of the relationship 
which exists in the standard Laplace transform
theory between analyticity properties (including possible poles)  
of the transform and the asymptotic behaviour
of the original function.\ From our viewpoint this Laplace-type transform can be
regarded as the mathematical structure which relates the complex angular
momentum poles (moving poles) to the high-energy asymptotic behaviour.\
Moreover this approach presents further advantages:

i) The analysis is completely worked out in the complex momentum space scenario
appropriate to Q.F.T.[13] (see, on this point, our comment below).

ii) It is the joint exploitation of harmonic analysis on orbital manifolds 
of the Lorentz group together with basic analyticity properties of Q.F.T.  
which entails the complex angular momentum structure; 
this method holds in any space-time dimension $d+1$ 
with $d \ge 2$. 

iii) By the use of our Fourier-Laplace-type transformation one can perform a partial
diagonalization (namely a diagonalization with respect to the angular variables) of
the convolution product involved in the Bethe--Salpeter integral
equations. This rigorous mathematical structure, which pertains to the general framework 
of Q.F.T., is thereby directly responsible for the
existence of poles in the 
complex angular momentum variable. This is the content of our second basic result presented in [10],  
whose detailed proof will be given elsewhere[14].

One can specify the advantage mentioned in i) under two respects:   

{\it a) with respect to the S-matrix approach.}
\quad The absorptive parts of $F$ in the crossed two-field channels  
have their supports inside regions of appropriate one-sheeted hyperboloids
determined by the future cone ordering relation (in view of the spectral
conditions). This geometrical property can be properly specified in terms
of energy-momentum configurations, which are of more 
controllable interpretation than the sets of Lorentz
invariants, as they were used in the Mandelstam representation. As a matter of fact, 
the Mandelstam double spectral region (used in [11,12]) corresponds to complex energy-momentum
configurations which have no simple physical interpretation.

{\it b) with respect to the approach of Euclidean Q.F.T..}
\quad The fact that the ``Euclidean partial-waves'' admit an analytic interpolation 
in the complex angular momentum variables is explicitly shown to be 
equivalent to the property of analytic continuation 
of the four-point function from Euclidean momentum-space to Minkowskian momentum-space (through a domain 
which is permitted by the requirements of locality and spectral conditions).  

We now wish to stress that the conceptual interest of the present study can be 
envisaged from two viewpoints, according to whether the fundamental fields considered
are those of the QCD-theory or the ``elementary meson and baryon fields''
used at the age of dispersion theory.   
In the latter case, which is the traditional case of application of the axioms of Q.F.T., 
our results appear as ``off-shell results''; but in order to get rid of this restriction,  
one can use the analytic continuation technique
adopted in the proof of dispersion relations (see [15] and references therein) and/or
positivity constraints (analogous to those used by Martin[16]) to reach the
mass-shell values and possibly a positive interval in the energy variable $t$,
so that a range of possible bound-states might be included in our analysis.
We now conjecture that our results might be applicable with even more interest to 
the former case, in which the phenomenon of confinement is present, so that the 
off-shell character of our study not only remains relevant but is even the only one 
to be relevant! 
\ In fact, it seems admitted that the
general principles used here (locality, spectrum, Lorentz covariance, temperateness) still apply to 
theories of QCD-type in suitable gauges: our results on complex angular momentum analysis then follow  
without requiring the existence of asymptotic elementary particles  
of the fields and are fully consistent with confinement.  
Moreover, the possible production of  
{\it a discrete spectrum of composite particles (namely hadrons and possibly ``glueballs'') appearing as 
``Regge-type particles''} 
via appropriate Bethe-Salpeter-type equations is built-in 
[10,14] in this general field-theoretical framework. 
In the present study, we only considered (for simplicity) the case of 
scalar fields; 
but one can expect that the joint exploitation of harmonic analysis on Lorentz orbital 
manifolds together with axiomatic analyticity still yields similar results for 
more general fields in Lorentz covariant gauges.

The paper is organized as follows. Sec.2 is devoted to an appropriate
analysis of the complex geometry associated with a given two field $t-$channel.\ In
Sec.3 we derive axiomatic analyticity properties and bounds of the
four-point functions with respect to the (off-shell) scattering angle $\Theta_t$ in 
manifolds bordered by the $s-$cut and $u-$cut of the crossed channels. 
It is then shown in Sec.4 that these properties of the four-point function are 
equivalent to the existence of a Laplace-type transform of the latter with respect to 
the corresponding  
complex angular momentum variable $\lambda_t$ .\  
This transform, which is explicitly defined in terms of the (off-shell) absorptive parts
of the crossed $s-$ and $u-$channels,  
is studied in arbitrary space-time dimension $d+1 \ (d\ge 2)$:
analyticity and bounds in a half-plane $\func{ Re } \lambda_t > m$ 
and the property of Carlsonian
interpolation of the Euclidean partial-waves satisfied by this Laplace-type transform 
(Froissart-Gribov-type equalities) are 
established. The inverse of the transformation is also described    
and, as a by-product, 
the connection (mentioned above) between the analytic continuation from Euclidean to  
Minkowskian space and the analytic interpolation  
in the complex angular momentum plane is displayed. 
In Appendix A, we give mathematical tools used for the
analytic completion of Sec.3. Appendix B is devoted to primitives and derivatives of
non-integral order in a complex domain and to their 
Laplace transforms: it provides a complete treatment of the
distribution-like character of the Green functions and absorptive parts 
in Sec.4.

\section{{\bf Complex geometry associated with a two-} -field channel}

Space-time and energy-momentum space are $(d+1)-$dimensional, with $d \ge 2$. 
Vectors in $(d+1)-$dimensional Minkowskian space are represented by
$k = (k^{(0)},\overrightarrow{k}) = (k^{(0)},k^{(1)},\cdots,k^{(d)})$; 
the corresponding scalar product is denoted $ k.k'=k^{(0)}k'^{(0)} 
- k^{(1)}k'^{(1)}\cdots-k^{(d)}k'^{(d)}$ and $ k^2 = k.k = k^{(0)2} - \overrightarrow{k}^2$.

In all the following, a special role is played by a given two-field channel 
(called the $t-$channel)
in which the pairs of incoming and outgoing complex energy-momenta
are denoted respectively $(k_1,\ k_2)$ and $(k_1^{\prime},\ k_2^{\prime})$; we 
choose the corresponding set of independent vector-variables $%
K=k_{1}+k_{2}=k_{1}^{\prime }+k_{2}^{\prime },$ $Z=\dfrac{k_{1}-k_{2}}{2},$ 
$Z^{\prime }=\dfrac{k_{1}^{\prime }-k_{2}^{\prime
}}{2}.$
$K$ is the total energy-momentum vector of this $t-$channel, whose squared energy is 
$t=K^{2}$. In this paper we shall always
assume that $K$ \it is fixed real and space-like \rm, i.e. $t\leqslant 0.$
We shall call $%
\hat M_{K}$ (resp. $\hat M_{K}^{\left( c\right) })$ the space of all real (resp.
complex) momentum configurations $[k] = \left( k_{1},k_{2},k_{1}^{\prime
},k_{2}^{\prime }\right) $ such that $k_{1}+k_{2}=k_{1}^{\prime
}+k_{2}^{\prime }=K. 
\ \ \hat M_{K}$ (resp. $\hat M_{K}^{\left( c\right) }$) is  
isomorphic to 
the real (resp. complex) space 
${\Bbb R}^{2(d+1)}$ (resp. $ {\Bbb C}^{2(d+1)}$) 
of the couple of vectors $(Z, Z'),\ $
$Z$ and $Z'$ being   
respectively the relative incoming and
outgoing (off-shell) $\left( d+1\right)-$momenta of the $t$-channel.

Choosing once for all a time-axis with unit
vector ${\rm e}_0\ \ \left( {\rm e}_{0}^2 = 1\right) $ determines the ``Euclidean subspace'' $%
\hat {\cal E}_{K}$ of $\hat M_{K}^{\left( c\right) }$ in which all 
the energy-momenta are of the form $k_{{\rm i}%
}=\left( {\rm i}q_{{\rm i}}^{\left( 0\right) },\overrightarrow{p_{%
{\rm i}}}\right) ,k_{{\rm i}}^{\prime }=\left( {\rm i}q_{{\rm i}}^{\prime
\left( 0\right) },\overrightarrow{p_{{\rm i}}}^{\prime }\right) $ (with $%
\overrightarrow{p_{{\rm i}}},\overrightarrow{p_{{\rm i}}}^{\prime },q_{{\rm i%
}}^{\left( 0\right) },q_{{\rm i}}^{\prime \left( 0\right) }$ real).

\vskip 0.5cm

\it We shall mainly consider the case  $K\neq 0,$ and   
choose $K$ along the $d$-axis of coordinates: \rm $K=\sqrt{-t}\ {\rm e}%
_{d},$ where ${\rm e}_{d}$ denotes the corresponding unit vector $\left( 
{\rm e}_{d}^{2}=-1\right) .$ 
We also introduce the (off shell) ``scattering angle'' $\Theta
_{t}$ of the $t$-channel as being the angle between the two-planes $\pi $
and $\pi ^{\prime }$ spanned respectively by the pairs of vectors $%
\left( Z,K\right) $ (or $\left( k_{1},k_{2}\right) $) and $\left( Z^{\prime
},K\right) $ (or $\left( k_{1}^{\prime },k_{2}^{\prime }\right) $). It is
convenient to introduce \it (real or complex) unit vectors \rm 
$z,z^{\prime },$ 
(uniquely determined up to a sign) orthogonal to $K$
and belonging respectively to $\pi $ and $\pi ^{\prime }$, such that   
the following orthogonal decompositions hold:

\begin{equation}
Z=\rho z+w\ K,\ \ Z^{\prime }=\rho ^{\prime }z^{\prime }+w^{\prime }K,  \tag{2.1.a}
\end{equation}

\begin{equation}
{\text {\rm with}}\ \ \  \ \  z.K = z'.K = 0 ,\ \ z^2 = z'^2 = -1,   \tag{2.1.b} 
\end{equation}
or equivalently: 


\begin{equation}
k_{1}  =  \rho z  +  \left( w+\frac{1}{2}\right)  K, \ \  
k_{2}  = -\rho z  -  \left( w-\frac{1}{2}\right)  K, \ \ \  
\tag{2.2.a}
\end{equation}

\begin{equation}
k_{1}^{\prime }  =  \rho ^{\prime }z^{\prime}  +  \left( w^{\prime }+\frac{1}{2}%
\right)  K, \ \  
k_{2}^{\prime }  =  -\rho ^{\prime }z^{\prime }  -  \left( w^{\prime }-%
\frac{1}{2}\right)  K. 
\tag{2.2.b}
\end{equation}

\noindent
Then, the ``scattering angle'' $\Theta _{t}$ of the $t-$channel is defined by the equation:
\begin{equation}
\cos \Theta _{t}=-z.z^{\prime }  \tag{2.3}
\end{equation}

\noindent (note that $\Theta _{t}=0$ for $z=z^{\prime }$).

The parameters $\rho ,w$ (resp. $\rho ^{\prime },w^{\prime
} $) introduced in Eqs. (2.1), (2.2) can be computed in terms of the scalar
products $Z^{2},Z.K,K^{2}$ (resp. $Z^{^{\prime }2},Z^{\prime }.K,K^{2}$) or,  
equivalently, in terms of the Lorentz invariants $\zeta _{{\rm i}}=k_{{\rm i}%
}^{2}$ (resp. $\zeta _{{\rm i}}^{\prime }=k_{{\rm i}}^{\prime 2}$), ${\rm i}%
=1,2,$ and $t.$ One readily obtains:

\begin{eqnarray}
w &=&\frac{Z.K}{t}=\frac{\zeta _{1}-\zeta _{2}.}{2t},\quad w^{\prime }=\frac{%
Z^{\prime }.K}{t}=\frac{\zeta _{1}^{\prime }-\zeta _{2}^{\prime }.}{2t} 
\tag{2.4} \\
\rho ^{2} &=&-Z^{2}+w^{2}t=\frac{\Lambda \left( \zeta _{1},\zeta
_{2},t\right) }{4t}  \tag{2.5} \\
\rho ^{\prime 2} &=&-Z^{\prime 2}+w^{\prime 2}t=\frac{\Lambda \left( \zeta
_{1}^{\prime },\zeta _{2}^{\prime },t\right) }{4t},  \tag{2.6}
\end{eqnarray}
where:
\begin{equation}
\Lambda \left( a,b,c,\right) =a^{2}+b^{2}+c^{2}-2\left( ab+bc+ca\right)
=\left( a-b\right) ^{2}-2\left( a+b\right) c+c^{2}  \tag{2.7}
\end{equation}
Finally, the variable $\cos{\Theta}_t$ is also a Lorentz invariant which can
be expressed as follows in terms of $\zeta _{i},\ \zeta _{i}^{\prime} 
\  \left( i=1,2\right) ,t$ and the squared momentum transfer $%
s=\left( k_{1}-k_{1}^{\prime }\right) ^{2}=\left( Z-Z^{\prime }\right) ^{2}:$
\begin{equation}
\cos \Theta _{t}=\frac{s+\rho ^{2}+\rho ^{\prime 2} -\left( w-w^{\prime} 
\right)^2 t}{2\rho \rho ^{\prime }}  \tag{2.8.a}
\end{equation}
\begin{equation}
{\text {\rm or}}\ \ \  
\cos \Theta _{t}=\frac{2st+t^{2}-\left( \zeta _{1}+\zeta _{2}+\zeta
_{1}^{\prime }+\zeta _{2}^{\prime }\right) t+\left( \zeta _{1}-\zeta
_{2}\right) \left( \zeta _{1}^{\prime }-\zeta _{2}^{\prime }\right) }{\left[
\Lambda \left( \zeta _{1},\zeta _{2},t\right) \Lambda \left( \zeta
_{1}^{\prime },\zeta _{2}^{\prime },t\right) \right] ^{1/2}}  \tag{2.8.b}
\end{equation}
The following alternative expression also holds:
\begin{equation}
\cos \Theta _{t}=\frac{-\left( u+\rho ^{2}+\rho ^{\prime 2}\right) +\left(
w+w^{\prime }\right) ^{2}t}{2\rho \rho ^{\prime }},  \tag{2.9}
\end{equation}
where $u$ denotes the squared momentum transfer in the crossed
channel, namely $u = \left( k_{1}-k_{2}^{\prime }\right) ^{2}=\left( Z+Z^{\prime
}\right) ^{2},$
which is such that: 
\[
u = -s-t +{\zeta}_1 +{\zeta}_2 +
{\zeta}_1^{\prime}  +{\zeta}_2^{\prime} .
\]

For $K=0,$ Eqs (2.1) ...(2.9) reduce to the following ones:
\begin{equation}
k_{1} =-k_{2}=Z=\rho z,\quad k_{1}^{\prime }=-k_{2}^{\prime }=Z^{\prime
}=\rho ^{\prime }z^{\prime },  \tag{2.10} \\
\end{equation}
\begin{equation}
\rho ^{2} =-\zeta _{1}=-\zeta _{2},\quad \rho ^{\prime 2}=-\zeta
_{1}^{\prime }=-\zeta _{2}^{\prime }  \tag{2.11} \\
\end{equation}
\begin{equation}
\text{and }\cos \Theta _{t} =-\frac{Z.Z^{\prime }}{\rho \rho ^{\prime }}=%
\frac{s-\zeta _{1}-\zeta _{1}^{\prime }}{2( \zeta _{1}\zeta
_{1}^{\prime })^{1/2}}=-\frac{u -\zeta _{1}-\zeta _{1}^{\prime }%
}{2(\zeta _{1}\zeta _{1}^{\prime }) ^{1/2}}  \tag{2.12}
\end{equation}
\vskip 0.3cm

\noindent
{\sl The space of Lorentz invariants: } 

For any point $[k]=\left( k_{1},k_{2},k_{1}^{\prime },k_{2}^{\prime }\right) $
in $\hat M_{K}^{\left( c\right) },$ we call ${\cal I}\left( [k]\right) $ the 
corresponding set of Lorentz invariants, namely
$\{ \zeta=(\zeta_1, \zeta_2),\ 
\zeta' =(\zeta'_1, \zeta'_2),\ 
\ (s,t,u)$ with $s+t+u=\zeta _{1}+\zeta _{2}+\zeta
_{1}^{\prime }+\zeta _{2}^{\prime } \},$ which vary in a complex space $%
{\Bbb C}_{\left( {\cal I}\right) }^{6}.$ In this space, the choice of
variables adapted to the $t-$channel is specified as follows: ${\cal I}([k]) = 
\left( {\cal I}_{t}\left( [k]\right) ,\cos \Theta _{t}\right) $ with ${\cal I}%
_{t}\left( [k]\right) =\left( \zeta _{1},\zeta _{2},\zeta _{1}^{\prime },\zeta
_{2}^{\prime },t\right) .$

\vskip 0.3cm 

For each $K$ with $t = K^2 \le0$,  
let $\hat{\Omega}_K $ be the subset of all points $[k]$ in $\hat M_{K}^{\left(
c\right) }$ whose parameters $\rho ,w,\rho^{\prime },w^{\prime }$ in the
representation (2.2) are real-valued.
This reality condition is equivalent (in view of
Eqs (2.4), (2.5), (2.6)) to the fact that 
$  \zeta _{{\rm i}},\zeta _{{\rm i}%
}^{\prime },({\rm i}=1,2)$ 
are real and satisfy the 
following inequalities:

\[
\Lambda \left( \zeta _{1},\zeta _{2},t\right) \leqslant 0,\quad \Lambda
\left( \zeta _{1}^{\prime },\zeta _{2}^{\prime },t\right) \leqslant 0, 
\]

\noindent which imply, for $K \neq0$, that the points $\zeta =\left( \zeta _{1},\zeta
_{2}\right) $ and $\zeta ^{\prime }=\left( \zeta _{1}^{\prime },\zeta
_{2}^{\prime }\right) $ belong to the following parabolic region (see Fig 1): 

\begin{equation}
\Delta _{t}=\left\{\zeta= \left( \zeta _{1},\zeta _{2}\right) 
\in {\Bbb R}^2;\ \left( \zeta
_{1}-\zeta _{2}\right) ^{2}-2\left( \zeta _{1}+\zeta _{2}\right)
t+t^{2}\leqslant 0\right\}  \tag{2.13}
\end{equation}
\begin{figure}
\epsfxsize=10truecm
{\centerline{\epsfbox{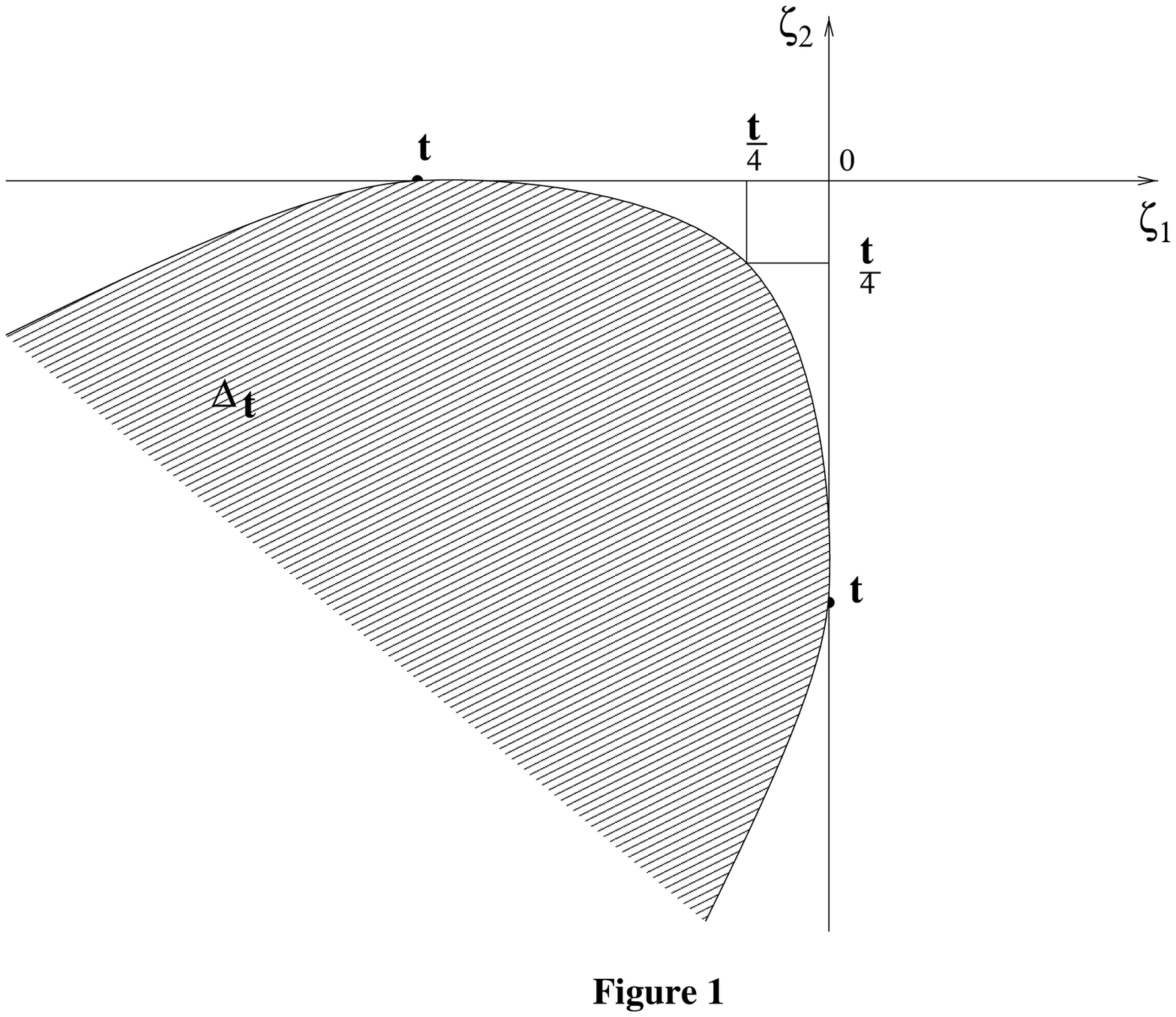}}} 
\end{figure} 

For $K=0,$ the corresponding set $\Delta _{0}$ is (in view of Eqs
(2.11)) the half-line $\zeta _{1}=\zeta _{2} \le 0.$

\subsection{Lorentz 
foliation and the associated complex quadrics; the case $K \neq 0$: }

\noindent 
{\sl Using the $(d-1)-$dimensional unit complex quadric: }

\smallskip 

For $K\neq 0,$ the range of each vector $z, 
z^{\prime }$ in Eqs (2.2) is (in view of Eqs (2.1.b)) a $( d-1)-$dimensional complex
quadric $X_{d-1}^{\left( c\right) }$ in the subspace orthogonal to $K,$
namely:

\begin{equation}
X_{d-1}^{\left( c\right) }= \left\{ z=\left( z^{\left( 0\right) },z^{\left(
1\right) },...z^{\left( d-1\right) }\right) \in {\Bbb C}^{d};\ z^{\left(
0\right) 2}-z^{\left( 1\right) 2}-\cdots-z^{\left( d-1\right)
2}=-1\right\}  \tag{2.14}
\end{equation}

\noindent Two real submanifolds of $X_{d-1}^{\left( c\right) }$ play an
important role:

\noindent a) the one-sheeted hyperboloid $X_{d-1}=X_{d-1}^{\left( c\right)
}\cap {\Bbb R}^{d},$ obtained by restricting $\left( z^{\left( 0\right)
},...,z^{\left( d-1\right) }\right) $ to real values in Eq. (2.14).

\noindent b) the ``euclidean sphere'' ${\Bbb S}_{d-1}=X_{d-1}^{\left(
c\right) }\cap \left( {\rm i}{\Bbb R\times R}^{d-1}\right) ,$ obtained by
putting $z^{\left( 0\right) }={\rm i}y^{\left( 0\right) }$ and $z^{\left(
1\right) }...z^{\left( d-1\right) }$ real in Eq. (2.14).

\smallskip 

\noindent
In view of Eqs (2.2)...(2.6), each point $[k]=\left(
k_{1},k_{2},k_{1}^{\prime },k_{2}^{\prime }\right) $ in $\hat M_{K}^{\left(
c\right) }$ can thus be represented by $\left( {\cal I}_{t}\left( [k]\right) ,\left(
z,z^{\prime }\right) \right) ,$ with ${\cal I}_{t}\left( [k]\right) =\left(
\zeta,\zeta^{\prime },t\right) ,\left( \zeta,\zeta^{\prime }\right) \in {\Bbb C}^{4}$ and the
pair $\left( z,z^{\prime }\right) $ in $X_{d-1}^{\left( c\right) }\times
X_{d-1}^{\left( c\right) }.$

We introduce 
the following {\it Cauchy--Riemann  
submanifold}
$\hat{\Omega}_{K}$ of ${\hat{M}}_K^{(c)}$:  
\[
\hat{\Omega}_{K}=\left\{[k] \equiv ({\cal I}_{t}\left( [k]\right) ,\left( z,z^{\prime
}\right));\ (\zeta,{\zeta}') \in \Delta_t \times \Delta_t,\  
\left( z,z^{\prime }\right) \in X_{d-1}^{\left(
c\right) }\times X_{d-1}^{\left( c\right) }\right\} 
\]

\noindent 
We then distinguish the following
{\it two maximal real submanifolds}
of $\hat{\Omega}_{K}$:  

\smallskip

\noindent a) $\left( z,z^{\prime }\right) $ in $X_{d-1}\times X_{d-1}:$ this  
submanifold is the subset $\hat M_{K}^{\left( sp\right) }$ of $%
\hat M_{K} $ characterized by the condition that the two-planes $\pi $ and $\pi
^{\prime }$ determined respectively by the real vectors 
$\left( k_{1},k_{2}\right) $ (or $(z,K)$) and $\left(
k_{1}^{\prime },k_{2}^{\prime }\right) $ (or $(z^{\prime},K)$) are space-like.

\vskip 0.3cm

\noindent b) $\left( z,z^{\prime }\right) $ in ${\Bbb S}_{d-1}\times {\Bbb S}%
_{d-1}:$ this is the Euclidean subspace $\hat {\cal E}%
_{K} $ of $\hat M_{K}^{\left( c\right) }.$

\vskip 0.3cm

We note that in the representation $\left( {\cal I}_{t}\left( [k]\right)
,\left( z,z^{\prime }\right) \right) $ of $[k] $  the pair $\left( z,z^{\prime
}\right) $ still contains \it one Lorentz invariant,  \rm namely 
$\cos \Theta_t 
=-z.z^{\prime },$ replaced equivalently by $s$ or $u$ according to Eqs
(2.8), (2.9), and that three situations are of special interest: 

\smallskip

\noindent i) $[k]\in \hat {\cal E}_{K}:$ the corresponding condition $\left(
z,z^{\prime }\right) \in {\Bbb S}_{d-1}\times {\Bbb S}_{d-1}$ then implies
that $-1\leqslant 
\cos \Theta_t  
\leqslant 1$

\noindent ii) $[k]\in \hat M_{K}^{(sp)}$ and $s=\left( k_{1}-k_{1}^{\prime }\right)
^{2}>0: $ Eq. (2.8) implies that $\cos \Theta _{t}-1>0;$ the corresponding
pair $\left( z,z^{\prime }\right) $ lies in $X_{d-1}\times X_{d-1}$ in such
a way that the two-plane spanned by $z$ and $z^{\prime }$ is time-like (i.e. 
$\Theta _{t} = iv $ with $v$ real, and  
$z.z^{\prime }=-\cosh\ v $).  

\noindent iii) $[k]\in \hat M_{K}^{(sp)}$ and $u=\left( k_{1}-k_{2}^{\prime }\right) >0:$
Eq. (2.9) implies that $\cos \Theta _{t}+1<0$ and one has: $\left( z,z^{\prime
}\right) \in X_{d-1}\times X_{d-1}$ with $z.z^{\prime }=\cosh v$ (i.e.  
$\Theta _{t}= \pi + iv$).  

\vskip 0.3cm

Let $G$ be the connected Lorentz group acting in the Minkowskian space $%
{\Bbb R}^{d+1},$ namely $G\approx SO_{0}\left( 1,d\right) $ and let $G^{\left(
c\right) }\approx SO_{0}\left( 1,d\right) ^{\left( c\right) }$ be the
complexified of $G,$ acting on ${\Bbb C}^{d+1}.$
Let then $G_{K}$ (resp. $G_{K}^{\left( c\right) }$) be the
stabilizer of $K$ in $G$ (resp. $G^{(c)}$). Since $K$ is real and space-like,
one has $G_{K}\approx SO_{0}\left( 1,d-1 \right) $ and $G_{K}^{\left( c\right)
}\approx SO_{0}\left( 1,d-1\right) ^{\left( c\right) } $ ; $G_{K}$ and $%
G_{K}^{\left( c\right) }$ act transitively respectively on $X_{d-1}$ and $%
X_{d-1}^{\left( c\right) }.$ We also introduce the maximal orthogonal
subgroup $O_{K}\approx SO\left( d\right) $ of $G_{K}^{\left( c\right) }$
which acts transitively on the euclidean sphere ${\Bbb S}_{d-1}$ of $%
X_{d-1}^{\left( c\right) }.$

\vskip 0.3cm

With each $\left( \zeta ,\zeta ^{\prime },K\right) ,\;\zeta \in \Delta_t,\  
\zeta ^{\prime }\in \Delta_t,\ $ 
we associate the 
manifold 
\begin{equation}
\hat \Omega_{\left( \zeta ,\zeta ^{\prime },K\right) }=\left\{
[k] 
\in {\hat M}_{K}^{\left( c\right) }
;\ [k]= [k]_{\left( \zeta ,\zeta ^{\prime },K\right) }\left( z,z^{\prime
}\right) ;\left( z,z^{\prime }\right) \in X_{d-1}^{\left( c\right) }\times
X_{d-1}^{\left( c\right) }\right\},  \tag{2.15}
\end{equation}
where the mapping $[k]= [k]_{\left( \zeta ,\zeta ^{\prime },K\right)
}\left( z,z^{\prime }\right) $ is defined by Eqs (2.2), with the  
parameters $\rho ,w,\rho ^{\prime },w^{\prime }$ reexpressed in terms of $%
\left( \zeta ,\zeta ^{\prime },t\right) $ via Eqs (2.4)...(2.6), namely

\begin{equation}
[k]=[k]_{\left(\zeta,\zeta^{\prime},K\right)}(z,z^{\prime}), \quad
(z,z') \in X_{d-1}^{(c)} \times X_{d-1}^{(c)} :  \tag {2.16.a} 
\end{equation}

\begin{equation}
k_1 = \left[\frac{\Lambda(\zeta_1,\zeta_2,t)}{4t}\right]^{\frac{1}{2}} z 
\ +\ \frac{\zeta_1-\zeta_2+t}{2t} K,\ \  
k_2 = -\left[\frac{\Lambda(\zeta_1,\zeta_2,t)}{4t}\right]^{\frac{1}{2}} z 
\ -\ \frac{\zeta_1-\zeta_2-t}{2t} K,  \tag {2.16.b} 
\end{equation}

\begin{equation}
k_1^{\prime} = \left[\frac{\Lambda(\zeta_1^{\prime},\zeta_2^{\prime},t)}{4t}\right]^{\frac{1}{2}}  
z^{\prime}\ +\ \frac{\zeta_1^{\prime}-\zeta_2^{\prime}+t}{2t} K,\ \  
k_2^{\prime} =- \left[\frac{\Lambda(\zeta_1^{\prime},\zeta_2^{\prime},t)}{4t}\right]^{\frac{1}{2}}  
z^{\prime}\ -\ \frac{\zeta_1^{\prime}-\zeta_2^{\prime}-t}{2t} K.\ \  \tag {2.16.c} 
\end{equation}

\noindent
The set $\{ \hat \Omega_{\left( \zeta ,\zeta ^{\prime },K\right)
};\left( \zeta ,\zeta ^{\prime }\right) \in \Delta_t \times \Delta_t \} $ defines
a \it foliation \rm of $\hat \Omega_K $ whose sheets $\hat
\Omega_{\left( \zeta ,\zeta ^{\prime },K\right) }$ have the following
interpretation: for $\zeta$ and $\zeta' \notin \partial\Delta_t$, 
each submanifold $\hat{\Omega}_{\left( \zeta ,\zeta ^{\prime },K\right) }$
is the product of two $\left( d-1\right) $-dimensional complex quadrics and
can be seen as {\it an orbit of the group }$G_{K}^{\left( c\right) }\times
G_{K}^{\left( c\right) }$via the action $\left( k_{{\rm i}},k_{{\rm i}%
}^{\prime }\right) \rightarrow \left( gk_{{\rm i}},g^{\prime }k_{{\rm i}%
}^{\prime }\right) ,{\rm i}=1,2,\ \left( g,g^{\prime }\right) \in \left(
G_{K}^{\left( c\right) }\times G_{K}^{\left( c\right) }\right) .$
(Note that a similar foliation could be defined for the whole set $\hat M_K^{(c)};$
it is not used in the present paper). 

We also note that the ``Euclidean spheres'' of the manifolds $\hat\Omega_{(\zeta, 
\zeta',K)}$ (obtained by restricting Eq.(2.16.a) to the set $\{ (z,z') \in 
{\Bbb S}_{d-1} \times {\Bbb S}_{d-1} \}$) define correspondingly a foliation of the 
Euclidean subset $\hat{\cal E}_K$ of $\hat\Omega_K$.  

\vskip 0.3cm

\noindent
{\sl Choice of a base point} 
\vskip 0.3cm

Since the group $G_{K}^{\left( c\right) }$ acts transitively on $%
X_{d-1}^{\left( c\right) }$ it is convenient to introduce a ``base-point'' $z_0$ 
on the latter which we choose on the $(d-1)-$axis  
of coordinates, namely $z_{0}=\left(
0,...0,1,0\right) .$
By now assuming that the point $z^{\prime }$ is fixed at $z^{\prime }=z_{0}$
in Eqs (2.2), one obtains a set of definitions which parallel those of the
previous paragraph.

One thus defines   
$M_K^{(c)}$ (resp. $\Omega_K$) as the subset of  
$\hat M_K^{(c)}$ (resp. $\hat \Omega_K$)  
in which the vectors $\left( k_{1}^{\prime
},k_{2}^{\prime }\right) $ are real and 
belong to the $\left( z_{d-1},z_d\right) $-plane of
coordinates.

One  also associates with each
$\left( \zeta ,\zeta ^{\prime },K\right) ,\;\zeta \in \Delta_t,\  
\zeta ^{\prime }\in \Delta_t,\ $ 
the manifold  
\begin{equation}
\Omega _{\left( \zeta ,\zeta ^{\prime },K\right) }=\left\{ [k] 
\in M_{K}^{\left( c\right) }
;[k]=[k]_{\left( \zeta ,\zeta ^{\prime },K\right)
}\left( z,z_{0}\right) ;z\in X_{d-1}^{\left( c\right) }\right\}.  \tag{2.17}
\end{equation}
If $\zeta \notin \partial\Delta_t,\ $  
$\Omega _{\left( \zeta ,\zeta ^{\prime },K\right) }$ is a $%
\left( d-1\right) $ dimensional complex quadric which is an orbit of the
group $G_{K}^{\left( c\right) }$ via the action $\left( k_{i},k_{i%
}^{\prime }\right) \rightarrow \left( gk_{i},k_{i}^{\prime
}\right),\ {i}=1,2,\ g\in G_{K}^{\left( c\right) }.$ The set $\left\{
\Omega _{\left( \zeta ,\zeta ^{\prime },K\right) };\left( \zeta ,\zeta
^{\prime }\right) \in {\Delta_t \times \Delta_t}\right\} $ thus defines a foliation of
$\Omega_K$; 
$\Omega_K$  
is a Cauchy-Riemann submanifold of $M_K^{(c)}$  
whose complex structure is parametrized by the variable $z$ in $%
X_{d-1}^{\left( c\right) }$ and which contains as maximal real submanifolds:

a) the real submanifold $M_{K}^{(sp)}=
M_{K}^{\left( c\right) }\cap \hat M_{K}^{(sp)}$ obtained
for $z$ varying in $X_{d-1}$ and characterized by the property  
that the plane $\pi $ defined by the (real) points $k_{1},k_{2}$ is
space-like.

b) the euclidean subspace ${\cal E}_{K} =
M_{K}^{\left( c\right) } \cap \hat {\cal E}_{K}$ obtained for $z$ varying in  
${\Bbb S}_{d-1}.$
\vskip 0.5cm





Finally, the passage from the vectors to the invariants is summarized in 

\vskip 0.3cm
\noindent {\sl Proposition 1}\  

\it Let ${\cal I}$ be the projection which associates with each  
configuration $[k]\equiv 
({\cal I}_{t}\left(
[k]\right) ,\left( z,z^{\prime }\right)) $ in $\hat M_{K}^{\left( c\right) }$ the
set of invariants ${\cal I}\left( [k]\right) =\left( {\cal I}%
_{t}\left([k]\right) ,\cos \Theta _{t}\right) .$ This projection is
implemented by the mapping $\left( z,z^{\prime }\right) \stackrel{\hat{i}}{%
\rightarrow }\cos \Theta _{t}=-z.z^{\prime }$ which projects  
$X_{d-1}^{\left( c\right) }\times X_{d-1}^{\left( c\right) }$ onto ${\Bbb C}%
. $

Correspondingly, the restriction of ${\cal I}$ to the subspace 
$M_{K}^{\left( c\right) }$ of $\hat M_{K}^{\left( c\right) }$ is
implemented by the mapping $ z \stackrel{i}{%
\rightarrow }\cos \Theta _{t}=-z.z_{0}=z^{\left( d-1\right) }$ which 
projects $X_{d-1}^{\left( c\right) }$ onto the complex $%
z^{\left( d-1\right) }$-plane. \rm  

\subsection{Lorentz 
foliation and the associated complex quadrics; the case $K = 0$: }
\smallskip 

For $K=0,$ the range of the vectors $z$ and $z^{\prime }$ is the complex
quadric:
\begin{equation}
X_{d}^{\left( c\right) }=\left\{ z=\left( z^{\left( 0\right) },z^{\left(
1\right) },...,z^{\left( d\right) }\right) \in {\Bbb C}^{d+1};\ z^{\left(
0\right) ^{2}}-z^{\left( 1\right) ^{2}}-...-z^{\left( d\right)
^{2}}=-1\right\}  \tag{2.18}
\end{equation}
and one similarly introduces the real one-sheeted hyperboloid $%
X_{d}=X_{d}^{_{\left( c\right) }}\cap {\Bbb R}^{d+1}$ and the euclidean
sphere ${\Bbb S}_{d}=X_d^{\left( c\right) }\cap \left( {\rm i}{\Bbb R\times
R}^{d}\right) .$

Each point $[k]=\left( k_{1},\;-k_{1},\;k_{1}^{\prime },\;-k_{1}^{\prime
}\right) $ in $\hat M_{0}^{\left( c\right) }$ is represented by $\left( {\cal I}%
_{0}([k]),\left( z,z^{\prime }\right) \right) $ where ${\cal I}_{0}\left(
[k]\right) =\left( \zeta ,\zeta ^{\prime },0\right) $ with $\zeta =\left(
\zeta _{1},\zeta _{1}\right) ,\;\zeta ^{\prime }=\left( \zeta _{1}^{\prime
},\zeta _{1}^{\prime }\right),\  \zeta _{1}\in {\Bbb C},\;\zeta _{1}^{\prime
}\in {\Bbb C},$ and $\left( z,z^{\prime }\right) \in X_{d}^{_{\left(
c\right) }}\times X_{d}^{_{\left( c\right) }}.$ 
\it We note the degeneracy of the representation $\left( {\cal I}%
_{0}\left([k]\right) ,\left( z,z^{\prime }\right) \right) $ for the space $%
\hat M_{0}^{\left( c\right) },$ namely the fact that the number of Lorentz
invariants in ${\cal I}_{0}\left( [k]\right) $ reduces from four to two,
while the number of ``orbital variables'' $\left( z,z^{\prime }\right) $
increases correspondingly from $2\left( d-1\right) $ to $2d.$ \rm  
\vskip 0.3cm

The set $\hat \Omega_0$ is the following Cauchy-Riemann manifold

\[
\hat{\Omega}_0\ =\left\{[k] \equiv ({\cal I}_0\left( [k]\right) ,\left( z,z^{\prime
}\right));\ (\zeta,{\zeta}') \in \Delta_0 \times \Delta_0,\  
\left( z,z^{\prime }\right) \in X_d^{\left(
c\right) }\times X_d^{\left( c\right) }\right\}, 
\]

\noindent
which contains \it as maximal real submanifolds \rm  
the Minkowskian and Euclidean submanifolds $%
\hat M_{0}^{\left( sp\right) }$ and $\hat {\cal E}_{0}$ of $\hat 
M_{0}^{\left( c\right) },$
obtained respectively for the ranges $\left\{ z,z^{\prime }\in X_{d}\times
X_{d}\right\} $ and $\left\{ \left( z,z^{\prime }\right) \in {\Bbb S}%
_{d}\times {\Bbb S}_{d}\right\} .$
All the previous considerations concerning the variable $%
\cos \Theta _{t}=-z.z^{\prime }$ remain valid in the case $K=0.$

With each point $\left( \zeta,\zeta^{\prime },0\right) ,$ with $\zeta =\left( \zeta
_{1},\zeta _{1}\right) ,\zeta _{1}\le 0,\ \zeta^{\prime }=\left(
\zeta _{1}^{\prime },\zeta _{1}^{\prime }\right) ,\zeta _{1}^{\prime }\le 0  
,$ one now associates the manifold
\[
\hat \Omega_{(\zeta ,\zeta ^{\prime },0)}=\left\{ [k]
\in {\hat M}_{0}^{\left( c\right) }
;\ [k]=[k]_{\left( \zeta ,\zeta ^{\prime },0\right) }\left( z ,z ^{\prime
}\right) ;\ \left( z ,z ^{\prime }\right) \in X_{d}^{\left( c\right)
}\times X_{d}^{\left( c\right) }\right\} 
\]
The set $\left\{ \hat{\Omega}_{(\zeta ,\zeta ^{\prime },0)};\zeta =\left(
\zeta _{1},\zeta _{1}\right) ,\zeta _{1}\le 0,\ \ \zeta ^{\prime }=\left(
\zeta _{1}^{\prime },\zeta _{1}^{\prime }\right) ,\zeta _{1}^{\prime }\le 0 
\right\} $ defines a foliation of $\hat \Omega_0 ;$ in
this foliation, each sheet $\hat{\Omega}_{(\zeta ,\zeta ^{\prime },0)}$ is the
product of two $d$-dimensional complex quadrics and can be seen as {\it an
orbit of the group }$G^{\left( c\right) }\times G^{\left( c\right) }$ via
the action: $\left( k_{{\rm i}},k_{{\rm i}}^{\prime }\right) \rightarrow
\left( gk_{{\rm i}},g^{\prime }k_{{\rm i}}^{\prime }\right) ,{\rm i}%
=1,2,\left( g,g^{\prime }\right) \in G^{\left( c\right) }\times G^{\left(
c\right) }.$

We can make use of the same base point $z_{0}$ as before $\left(
z_{0}=(0,...,0,1,0)\right) $ and introduce the subspace $M_0^{(c)}$ of
$\hat M_0^{(c)}$ in which $k_1^{\prime}= -k_2^{\prime}$ is real and along the 
$z_{d-1}-$axis. 
Then 
for all $(\zeta,\zeta^{\prime}) \in \Delta_0 \times \Delta_0$
the following $d$-dimensional
complex manifolds 
\[
\Omega _{(\zeta ,\zeta ^{\prime },0)}=\left\{ 
[k]
\in M_{0}^{\left( c\right) }
; \  
[k]=[k]_{(\zeta ,\zeta ^{\prime },0)}\left( z,z_{0}\right) ;z\in X_{d}^{\left(
c\right) }\right\} , 
\]
are orbits of the group $G^{\left( c\right) }$ via the
action $\left( k_{i},k_{i}^{\prime }\right) \rightarrow \left(
gk_{i},k_{i}^{\prime }\right) ,{i}=1,2,\ g\in G^{\left(
c\right) }.$

The set $\left\{ \Omega _{\left( \zeta ,\zeta ^{\prime },0\right) };\ \zeta
=\left( \zeta _{1},\zeta _{1}\right) ,\zeta_1  \le 0;\ \zeta ^{\prime
}=\left( \zeta _{1}^{\prime },\zeta _{1}^{\prime }\right) ,\zeta
_{1}^{\prime }\le 0 \right\} $ defines a foliation of the subset $%
\Omega_0$ of $\hat \Omega_0$
in which the vector $k_{1}^{\prime }=-k_{2}^{\prime }$ is real and along the 
$z_{d-1}-$axis.
$\Omega_0$ is a Cauchy-Riemann manifold 
whose complex structure is parametrized by $z\ ( z\in
X_{d}^{\left( c\right) }) ,$ and which contains the real Minkowskian
submanifold $M_{0}^{\left( sp\right)} $ and the Euclidean
subspace ${\cal E}_{0}$ of $M_{0}^{\left( c\right)
}$ (obtained respectively for $z\in X_{d}$ and $z\in {\Bbb S}%
_{d}).$

Proposition 1 remains
true up to obvious changes $%
(X_{d-1}^{\left( c\right) }$ being replaced by $X_{d}^{\left( c\right) }).$
\vskip 1cm

\subsection{\bf The spectral sets $\Sigma_{s}$ and $\Sigma_{u}$}

We define \it the $s$-channel and $u$-channel spectral sets \rm $\Sigma_{s}$ and $%
\Sigma_{u}$ associated with a given field theory as the following analytic
hypersurfaces in complex momentum space ${\Bbb C}_{\left( k\right)
}^{3\left( d+1\right) }:$

\begin{eqnarray}
\Sigma_{s} =\left\{ [k]\equiv \left( K,Z,Z^{\prime }\right) \in {\Bbb C}%
^{3\left( d+1\right) };\  s=\left( Z-Z^{\prime }\right) ^{2} =  
s_{0}+ \tau;\ \tau \ge 0 \right\} , \tag{2.19} \\
\Sigma_{u} =\left\{ [k]\equiv \left( K,Z,Z^{\prime }\right) \in {\Bbb C}%
^{3\left( d+1\right) };\  u=\left( Z+Z^{\prime }\right) ^{2} = 
u_{0}+ \tau;\ \tau \ge 0 \right\} ,  \tag{2.20}
\end{eqnarray}

\noindent where $s_{0}$ and $u_{0}$ are positive numbers interpreted as the
mass thresholds of the corresponding channels.

Since $\Sigma_{s}$ and $\Sigma_{u}$ are Lorentz--invariant sets, their
projections onto the space of Lorentz invariants $\left( {\cal I}_{t}\left(
[k]\right) ;\cos \Theta _{t}\right) $ are analytic hypersurfaces ${\cal I}%
\left( \Sigma_{s}\right) $ and ${\cal I}\left( \Sigma_{u}\right) $ whose
equations result from Eqs. (2.8), (2.9) namely:

\begin{equation}
{\cal I}\left( \Sigma_s\right) \left\{ 
\begin{array}{c}
\cos \Theta _{t}-1=\dfrac{\left[ s_{0}+\left( \rho -\rho ^{\prime }\right)
^{2}-\left( w-w^{\prime }\right) ^{2}t\right] +\tau }{2\rho \rho ^{\prime }};
\\ 
\\ 
\text{with }{\tau}\geqslant 0.\qquad \qquad \qquad \qquad \qquad \qquad \quad
\end{array}
\right.  \tag{2.21}
\end{equation}

\begin{equation}
{\cal I}\left( \Sigma_u\right) \left\{ 
\begin{array}{c}
\cos \Theta _{t}+1=\dfrac{\left[ -u_{0} -\left( \rho -\rho
^{\prime }\right) ^{2}+\left( w+w^{\prime }\right) ^{2}t\right] -\tau }{%
2\rho \rho ^{\prime }} \\ 
\\ 
\text{with }{\tau}\geqslant 0.\qquad \qquad \qquad \qquad \qquad \qquad \quad
\end{array}
\right.  \tag{2.22}
\end{equation}

Let us now consider the intersections of $\Sigma_s$ 
and $\Sigma_u$ with any orbit $\hat{\Omega}_{\left(
\zeta ,\zeta ^{\prime },K\right) }$ in $\hat{\Omega}_{K};$  
it readily follows from Eqs. (2.21) and (2.22)
that these intersections can be parametrized by the variables $z,z^{\prime }$
in the following way:
\[
{\Sigma_s}\cap \hat{\Omega}_{\left( \zeta ,\zeta
^{\prime },K\right) } = 
\]
\begin{equation}
\left\{ [k];[k]=[k]_{\left( \zeta ,\zeta ^{\prime },K\right) }\left( z,z^{\prime
}\right) ;\ \left( z,z^{\prime }\right) \in X_{d-1}^{\left( c\right) }\times
X_{d-1}^{\left( c\right) },\right. \\ 
\left. -z.z^{\prime }=\cosh v,\quad v\geqslant v_{s}\right\}
\tag{2.23} \\
\end{equation}
\[
{\Sigma_u}\cap \hat{\Omega}_{\left( \zeta ,\zeta
^{\prime },K\right) } = 
\]
\begin{equation}
\left\{ [k];[k]=[k]_{\left( \zeta ,\zeta ^{\prime },K\right) }\left( z,z^{\prime
}\right) ;\ \left( z,z^{\prime }\right) \in X_{d-1}^{\left( c\right) }\times
X_{d-1}^{\left( c\right) },\right. \\ 
\left. z.z^{\prime }=\cosh v,\quad v\geqslant v_{u}\right\}
\tag{2.24} \\
\end{equation}
where $v_{s}=v_{s}\left( \zeta ,\zeta^{\prime} ,t\right) $ and $%
v_{u}=v_{u}\left( \zeta,\zeta^{\prime},t\right) $ are defined by the
equations:
\begin{eqnarray}
\cosh v_{s}-1 &=&\frac{s_{0}+\left( \rho -\rho ^{\prime }\right) ^{2}-\left(
w-w^{\prime }\right) ^{2} t}{2\rho \rho ^{\prime }},  \tag{2.25} \\
&&  \nonumber \\
\cosh v_{u}-1 &=&\frac{u_{0}+\left( \rho -\rho ^{\prime }\right) ^{2}-\left(
w+w^{\prime }\right) ^{2}t}{2\rho \rho ^{\prime }},  \tag{2.26}
\end{eqnarray}
with $\rho ,w, \rho ^{\prime },w^{\prime }$ expressed by Eqs. (2.4),
(2.5), (2.6).

We then see that the images of these sets in the $\cos \Theta _{t}$-plane
(by the projection $\hat{i}$ introduced in Proposition 1) are the two real half-lines
\begin{equation}
\underline{\sigma }_{+}\left( v_{s}\right) =\left[ \cosh v_{s}, 
+\infty \left[
\;\ {\rm and}\ \underline{\;\sigma }_{-}\left( v_{u}\right) =\right] -\infty
,-\cosh v_{u}\right]  \tag{2.27}
\end{equation}

In the next section, the previous sets will appear as \it ``cuts'' \rm bordering
analyticity domains, namely the following \it ``cut orbits'' \rm 
(for each $\left( \zeta ,\zeta ^{\prime },K\right)$): 

\begin{equation}
\hat{\Omega}_{\left( \zeta ,\zeta ^{\prime },K\right) }^{\left( {\rm cut}%
\right) }
=\hat{\Omega}_{\left( \zeta ,\zeta ^{\prime },K\right) }\backslash
\left( \Sigma_s \cup \Sigma_u 
\right) ;
\tag{2.28}
\end{equation}
We also introduce correspondingly in $\Omega _{K}$ the \it cut orbits\rm:
\begin{equation}
\Omega _{\left( \zeta ,\zeta ^{\prime },K\right) }^{\left( {\rm cut}\right)} 
=\Omega _{\left( \zeta ,\zeta ^{\prime },K\right) }\backslash \left(
\Sigma_s \cup \Sigma_u \right)
;  \tag{2.29}
\end{equation}
each of them is represented in the parametric variables $z$ by the
complex quadric $X_{d-1}^{\left( c\right) }$ minus the cuts 
\begin{equation}
\Sigma_+^{(c)}(v_s) = \left\{
z\in X_{d-1}^{(c)};\ z^{\left( d-1\right) }\in [\cosh v_{s}, + \infty[ \right\}  
\tag{2.30}
\end{equation}
and
\begin{equation}
\Sigma_-^{(c)}(v_u) = \left\{
z\in X_{d-1}^{(c)};\ z^{\left( d-1\right) }\in\  ]-\infty, -\cosh v_{u}] \right\}. 
\tag{2.31}
\end{equation}
As an immediate consequence of Proposition 1 one then has:
\vskip 0.5cm
\noindent {\sl Lemma 1}

\it  The projection of each set 
$\hat{\Omega}_{\left( \zeta ,\zeta ^{\prime },K\right) }^{\left( {\rm cut}%
\right) }$ 
by $\hat {i}$ and of each set  
$\Omega _{\left( \zeta ,\zeta ^{\prime },K\right) }^{\left( {\rm cut}\right)}$ 
by $i$
onto the $\cos \Theta _{t}$-plane is the corresponding cut-plane 
\begin{equation}
\underline{\Pi}_{(\rho,w,\rho^{\prime},w^{\prime},t)} = 
{\Bbb C} \backslash \left\{\underline{\sigma}_+\left( v_s \right) \cup%
\underline{\sigma }_- \left( v_u \right) \right\} . 
\tag{2.32}
\end{equation}
entirely specified by formulas (2.25), (2.26) and (2.4)...(2.6).  \rm

\vskip 2cm

\section{\bf Perikernel structure of four-point functions in complex
momentum space}

\subsection{\bf Four-point functions of local fields: primitive
analyticity domain and bounds} 

\quad We here recall some basic results of the theory of four-point functions
in the axiomatic framework of quantum field theory (see e.g. [13] and references therein).
In this theory, one deals with the set of ``generalized retarded functions''
which are built from vacuum expectation values of the form:
\[
W
^{\left( {\rm i}\right)} 
{\left( x_{1},x_{2},x_{3},x_{4}\right) }
=\left\langle \Omega ,\phi _{{i}_{1}}\left( x_{{i}_{1}}\right) \phi
_{{ i}_{2}}\left( x_{{i}_{2}}\right) \phi _{{i}_{3}}\left( x_{{i}_{3}}\right)
\phi _{{i}_{4}}\left( x_{{i}_{4}}\right) \Omega \right\rangle , 
\]
$\left( { i}\right) =\left( {i}_{1},{i}_{2},{i}_{3},{i}%
_{4}\right) $ denoting any permutation of $\left( 1,2,3,4\right) ;$ here, the $%
\phi _{j}{\rm \ }^{\prime }s$

\noindent $\left( j=1,2,3,4\right) $ denote local fields which satisfy
mutually the {\it postulate of local commutativity}: $\left[ \phi _{j}\left(
x\right) ,\phi _{\ell }\left( y\right) \right] =0$ if $\left( x-y\right)
^{2}<0\left( j,\ell =1,2,3,4\right) .$ In view of the translation invariance
of the theory, the so-called ``Wightman functions'' 
$W
^{\left( {i}\right)} 
{\left( x_{1},x_{2},x_{3},x_{4}\right) }$ 
are defined in
the space ${\Bbb R}^{4\left( d+1\right) }/{\Bbb R}^{d+1}$ of the vector
variables $\xi =\left( \xi _{{ i}}=x_{{ i}}-x_{{ i}+1},{i}%
=1,2,3\right) ;$ in the standard formulation of Wightman field theory, they
are defined as tempered distributions.

The construction of the generalized retarded functions (g.r.f.) in terms of
the Wightman functions requires the use of the algebra generated by multiple
commutators of the fields together with step functions of the time-coordinates $\theta
\left( x_{i}^{\left( 0\right) }-x_{j}^{\left( 0\right) }\right) $ [17,18,19]. 
The g.r.f. are special elements $r_{\alpha }\left( x\right) $ of this
algebra which have minimal support properties in the configuration space $%
{\Bbb R}^{4\left( d+1\right) }$ in the following sense. 
In $\xi-$space (i.e.
${\Bbb R}^{4\left( d+1\right) }/{\Bbb R}^{d+1}$) 
the support $\Gamma
_{\alpha }$ of each g.r.f. $r_{\alpha }\left( x\right) =\underline{r}%
_{\alpha }\left( \xi \right) $ is a Lorentz-invariant cone whose convex hull $%
\hat{\Gamma}_{\alpha }$ is a salient cone with apex at the origin:  
{\it each cone} $\Gamma _{\alpha }$ {\it is determined explicitly as a consequence of 
the postulate of local commutativity}
.\ It is assumed that the set of g.r.f. 
$ \underline {r}_{\alpha}$ can be defined 
\footnote{%
The use of ``sharp'' time-ordered or retarded products (involving formally the product of
distributions with the ``sharp'' step-function $\theta \left( x^{\left(
0\right) }\right) $) necessitates an extra-postulate with respect to the
Wightman axioms (see e.g. the axiomatic presentations of [20] and [21]).}
{\it as tempered distributions on}
$ {\Bbb R}^{4(d+1)}/ {\Bbb R}^{d+1}$ 
satisfying the previously mentioned support properties.

\vskip 1cm 

\noindent
{\sl Analyticity and bounds in the tubes } ${\cal T}_{\alpha }:$

Analyticity in complex momentum space is readily obtained by introducing the
Fourier-Laplace transforms of the g.r.f. $r_{\alpha }.$

Due to translation invariance, the Fourier transforms of the g.r.f. $%
r_{\alpha }\left( x\right) $ are of the form: $\delta \left(
p_{1}+p_{2}+p_{3}+p_{4}\right) \tilde{r}_{\alpha }\left( p\right) ,$ where
each $\tilde{r}_{\alpha }\left( p\right) $ is a tempered distribution on the
linear space
$ M=\left\{ p=\left( p_{1},...,p_{4}\right) ;p_{1}+p_{2}+p_{3}+p_{4}=0\right\} $. 
Let $M^{\left( c\right) }$ be the complexified of $M,$ whose points are
denoted by $k=p+{\rm i}q=\left( k_{1},k_{2},k_{3},k_{4}\right) ,$ with $%
k_{1}+k_{2}+k_{3}+k_{4}=0.$

The support properties of the distributions $\underline{r}_{\alpha }\left(
\xi \right) =r_{\alpha }\left( x\right) $ 
imply that
one can define the corresponding
Fourier-Laplace transforms (still denoted by) $\tilde{r}_{\alpha }\left(
k\right) ,$ formally given by
\[
\tilde{r}_{\alpha }\left( k\right) =\frac{1}{%
(2\pi)^{2\left( d+1\right) }}\int {\rm e}^{{\rm i}(k.\xi)} 
\underline{r}_{\alpha}\left( \xi \right) d\xi _{1}d\xi _{2}d\xi _{3},
\]
with
\[
\left( k.\xi \right) =k_{1}.\xi _{1}+\left( k_{1}+k_{2}\right). \xi
_{2}+\left( k_{1}+k_{2}+k_{3}\right). \xi _{3}\equiv \sum_{{ i}=1}^{4}k_{%
{i}}x_{{i}}, 
\]
as {\it holomorphic functions} in the respective 
domains $\ \ {\cal T}_{\alpha }=M+{\rm i}\ \ {\cal C}_{\alpha }$ of $M^{(c)}$, called
``the tubes $\ \ {\cal T}_{\alpha }$ with bases $\ \ {\cal C}_{\alpha }$ ''.
For each $\alpha$, ${\cal C}_{\alpha}$ is the (open)
\footnote{$C_{\alpha }$ is open and non-empty, since $\hat
\Gamma_{\alpha }$ is a {\it salient} cone.} 
dual cone of the support $\Gamma
_{\alpha}$ 
of $\underline{r}_{\alpha}$  
(or of the convex hull $\hat{\Gamma}_{\alpha }$ of the latter), 
namely:
\[
{\cal C}_{\alpha }=\left\{ q\in M;\ \left( q.\xi \right) >0{\rm \ for\ all\ }%
\xi \in \hat{\Gamma}_{\alpha }\right\} . 
\]

Moreover, as a consequence of the tempered character of $\underline{r}%
_{\alpha }\left( \xi \right) ,\tilde{r}_{\alpha }\left( k\right) $ 
satisfies a global majorization of the following form
in its domain $\ \ {\cal T}_{\alpha }:$ 
\begin{equation}
\left| \tilde{r}_{\alpha }\left( k\right) \right| \leqslant C\max \left[
\left( 1+\parallel k\parallel \right) ^{m},\left[d\left( q,\partial {\cal C}%
_{\alpha }\right)\right] ^{-n}\right]  \tag{3.1}
\end{equation}
where $\parallel \;\parallel $ denotes a euclidean norm in $%
M^{(c)}\  ,d\left( q,\partial {\cal C}_{\alpha }\right) $ denotes the corresponding
distance from $q$ to the boundary $\partial {\cal C}_{\alpha }$ of ${\cal C}%
_{\alpha },$ and $m \ge 0,\  n \ge 0. $ These numbers characterize the 
{\it ``degrees of temperateness''} of the theory by taking into account respectively
the dominant ultraviolet behaviour and the highest degree of local singularities of the 
four-point function in momentum space.
In view of the role which they will be shown to play in 
complex angular momentum analysis, it is better to
assume that they are general real numbers, i.e. not necessarily integers (as often assumed in 
standard Q.F.T.). 

Under these assumptions, each Fourier transform $\tilde{r}%
_{\alpha }\left( p\right) $ is then rigorously characterized as the
``distribution-boundary value'' of the corresponding holomorphic function $\tilde{r}%
_{\alpha }\left( k\right) $ from the tube$\ {\cal T}_{\alpha }$ namely:
\[
\lim_ 
{q\rightarrow 0, \\ 
q\in{\cal C}_{\alpha }}
\dint \tilde{r}_{\alpha }\left( p+{\rm i}q\right) \varphi \left( p\right)
dp=\left\langle \tilde{r}_{\alpha },\varphi \right\rangle 
\]
for all test-functions $\varphi \left( p\right) $ in the Schwartz
space ${\cal S}\left( M\right) .$

We now recall the definition of the cones$\ {\cal C}%
_{\alpha }$ (see [13] and references therein).
Let $\alpha =\left( \left( \alpha _{i};{i}%
=1,2,3,4\right) ,\left( \alpha _{j\ell };j,\ell =1,2,3,4,j\neq \ell \right)
\right) ,$ where the $\alpha _{i}$ and $\alpha _{j\ell }$ are equal to 
$+1$ or $-1.$
The corresponding cone $\ \ {\cal C}_{\alpha }$ is defined by the 
following conditions:
\begin{equation}
\alpha _{i}q_{i}\in V^{+},\alpha _{j\ell }\left( q_{j}+q_{\ell
}\right) \in V^{+};{i},j,\ell =1,2,3,4,j\neq \ell;  \tag{3.2}
\end{equation}
the condition of non-emptiness of ${\cal C}_{\alpha }$ puts
obvious constraints on the set $\alpha ,$ such as
$\alpha _{j\ell }=-\alpha _{mn}$ if $\left( j,\ell ,m,n\right)
=\left( 1,2,3,4\right),\  
\alpha _{j\ell }=\alpha _{j}$ if $\alpha _{j}=\alpha _{\ell
},\;\alpha _{n}=-\alpha _{j}$ if $\alpha _{j}=\alpha _{\ell }=\alpha _{m}\;$%
etc...

\noindent Each cone$\ {\cal C}_{\alpha }$ 
is represented conveniently by a simplicial
triedron in ${\Bbb R}_{\left( s_{1},s_{2},s_{3}\right) }^{3}$ whose faces
are contained in three of the planes with equations $s_{i}=0,\  
i= 1,2,3,\  s_{4}=-\left( s_{1}+s_{2}+s_{3}\right) =0,\ s_{i}  
+s_{j}=0,\  i,j=1,2,3,$ or equivalently by a triangular cell
determined by these planes on the unit sphere $%
s_{1}^{2}+s_{2}^{2}+s_{3}^{2}=1:$ we thus obtain the so-called \it ``Steinmann
sphere'' representation \rm of the tubes $\ {\cal T}_{\alpha }$ of the four-point
function in complex momentum space. 

\vskip 0.3cm

\noindent
{\sl The coincidence region ${\cal R}:$}

\quad It follows from the spectral conditions of the field theory considered
that all the distributions $\tilde{r}_{\alpha }\left( p\right) $ coincide in
the following region ${\cal R}$ of $M:$
\begin{eqnarray*}
{\cal R} &=&\left\{ p = (p_1,p_2,p_3,p_4) \in M;  
\;p_{j}^{2}<{\rm M}_{j}^{2},\ 1\le j \le 4, \right.\\   
&&\left.  t\equiv \left( p_{1}+p_{2}\right) ^{2}<t_{0,}\ \;s\equiv \left(
p_{1}+p_{3}\right) ^{2}<s_{0},\ \;u\equiv \left( p_{1}+p_{4}\right)
^{2}<u_{0}\right\}, 
\end{eqnarray*}
where the numbers ${\rm M}_{j},\ 1 \le j \le 4$
are mass thresholds associated with
the corresponding fields, and$\ t_0,s_0,u_0$ are the mass
thresholds of the corresponding two-field channels.

The region ${\cal R}$ is a star-shaped region with respect to the origin in $M.$ %
\vskip 0.3cm

\noindent
{\sl The four-point function $H\left( k\right) :$}

Since all the holomorphic functions $\tilde{r}_{\alpha }\left( k\right) $
have boundary values on the reals which coincide on the region ${\cal R},$ they
admit a common analytic continuation denoted by $H\left( k\right) $ whose
existence results from the ``edge-of-the-wedge theorem'' (see [22] and references therein). 
This function $H\left( k\right) ,$ called {\it the analytic
four-point function in complex momentum space} of the set of fields 
considered, is holomorphic in the following complex domain $D$ of $%
M^{\left( c\right) }:$
\[
D=\left( \bigcup\limits_{\alpha }\ {\cal T}_{\alpha }\right) \bigcup {\cal N}%
\left( {\cal R}\right) , 
\]
where ${\cal N}\left( {\cal R}\right) $ is a certain complex neighborhood
of the region ${\cal R}$  
(chosen for example as ${\cal N}\left( {\cal R}\right) =\left\{
k=p+{\rm i}q;\ p\in {\cal R},\parallel q\parallel <\varepsilon _{0}\right\} $).   

The bounds (3.1) on 
$H\left( k\right) $ 
in the tubes ${\cal T}%
_{\alpha }$ imply\footnote{%
This result can be obtained as a direct application of proposition
A.3.} similar majorizations in ${\cal N}\left( {\cal R}\right),$ 
namely:
\begin{equation}
\left| H\left( p+{\rm i}q\right) \right| \leqslant C^{\prime }\max \left[
\left( 1+\parallel p \parallel \right) ^{m},\left[ d\left( p,\partial {\cal R}%
\right) \right] ^{-n}\right],  \tag{3.3}
\end{equation}
where $d\left( p,\partial {\cal R}\right) $ is the distance
from the real point $p$ to the boundary of ${\cal R}.$ 

\vskip 0.3cm
$D$ is called the {\it primitive axiomatic domain }of the four-point
function. $D$ is not a ``natural'' holomorphy domain; this means that it
admits a {\it holomorphy envelope} ${\cal H}\left( D\right) $ in which all
functions which are holomorphic in $D$ can be analytically continued [22]. 

\vskip 0.3cm 
\noindent
{\sl Some general properties of the holomorphy envelope ${\cal H}%
\left( D\right) $}:

The problem of the determination of (parts of) the holomorphy envelope $%
{\cal H}\left( D\right) $ of $D$ by means of various methods (such as the
tube theorem, etc...[22]) is called ``the analytic
completion problem''. 
Although the complete knowledge of ${\cal H}\left( D\right) $ has not been
obtained, the following general properties of this domain have been established [13]  
(the proof of a) and c) requires all the mass thresholds $M_j,\ 1 \le j \le 4,\ s_0,t_0,u_0$
to be strictly positive).

\vskip 0.3cm

\noindent {\sl Theorem}

\it
\noindent a) ${\cal H}\left( D\right) $ is a domain of $M^{\left( c\right) }$
which is star-shaped with respect to the origin, 

\noindent b) ${\cal H}\left( D\right) $ is invariant under the diagonal
action of the {\it complex} Lorentz group $G^{\left( c\right) }=SO_{0}\left(
1,d\right) ^{\left( c\right) },$ namely if $k=\left(
k_{1},k_{2},k_{3},k_{4}\right) \in {\cal H}\left( D\right) ,$ then $%
gk=\left( gk_{1},gk_{2},gk_{3},gk_{4}\right) \in {\cal H}\left( D\right) ,$
for every $g$ in $G^{\left( c\right) },$

\noindent c) For any fixed choice of the time-axis, the corresponding
Euclidean subspace 

\noindent ${\cal E}=\left\{ k=\left( k_{i};1\leqslant {i}%
\leqslant 4\right) \in M^{\left( c\right) };k_{i}=p_{i}+{\rm i}%
q_{i};\ p_{i}=\left( 0,\vec{p}_{i}\right),\ q_{i}%
=( q_{i}^{( 0) },0) \right\} ,$

\noindent is contained in ${\cal H}\left( D\right) .$

\noindent d) The seven ``spectral sets'' \quad  
$ \Sigma_s = \{ k \in M^{(c)}; s= (k_1+k_3)^2 \ge s_0 \}, $ 
$ \Sigma_u = \{ k \in M^{(c)}; u= (k_1+k_4)^2 \ge u_0 \}, $ 
$ \Sigma_t = \{ k \in M^{(c)}; t= (k_1+k_2)^2 \ge t_0 \}, $ 
$ \Sigma^{(j)}= \{ k \in M^{(c)};\  k_j^2 \ge {\rm M}_j^2 \}, \ 1 \le j \le 4, $  
do not intersect $ {\cal H}(D).$

\rm



\vskip 0.3cm
\noindent
{\sl Absorptive parts:}

The axiomatic framework also provides a complete description of the
structure and primitive analyticity domains of the {\it off-shell absorptive
parts,} which are the discontinuity functions $\Delta _{s}H,\Delta
_{u}H,\Delta _{t}H$ of $H$ in the respective $s,u$ and $t$-channels. To be
specific $\Delta _{s}H\left( k\right) $ is a holomorphic function of $k_{1}$
and $k_{2}$ which is defined in the ``face'' $\left\{ k;\ \func{Im}\left(
k_{1}+k_{3}\right) =0\right\}$ of the complex momentum space ``triangulation''
described above; its support is the  
intersection of the face 
$\func{Im}\left(k_1 + k_3 \right) =0$
with the corresponding
``spectral set'' ${\Sigma}_s = \left\{ k \in M^{(c)};\ s \ge s_0 \right\}$ 
(also introduced with the notations of Sec.2) in (2.19)).   
The primitive analyticity domain $D_{s}$ of $\Delta_{s}H$ in $\left( k_{1},k_{2}\right)- $space is
the union of the four tubes $\{ \left( k_{1},k_{2}\right) ;\func{Im}%
k_1\in \varepsilon_1V^{+},$ $\func{Im}k_{2}\in \varepsilon _{2}V^{+};\varepsilon
_{1},\varepsilon _{2}=+{\rm \ or\ }- \} $ connected together by a
complex neighborhood of the region:
\begin{equation}
{\cal R}_{s} =\left\{ k{\rm \ 
\func{real};}\ k\in \Sigma_{s};\  
k_{j}^{2}<{\rm M}_{j}^{2},\ 1 \le j \le 4\right\} . 
\tag{3.4}  
\end{equation}

\vskip 0.4cm 
\noindent
{\sl ``Sections of maximal analyticity'' or ``cut-submanifolds''}

We shall say that a complex submanifold ${\cal L}$ of $M^{\left( c\right) }$
provides \it a section of maximal analyticity or a cut-submanifold of the domain 
$D$ (resp. of its holomorphy envelope ${\cal H}\left( D\right) )$ for the $s$
and $u$-channels \rm if ${\cal L}\cap {\cal D}$ (resp. ${\cal L}\cap {\cal H}%
\left( D\right) )$ is equal to ${\cal L}\backslash 
\left( \Sigma_s 
\cup 
\Sigma_u \right) .$ 
Such sections of ${\cal H}%
\left( D\right) $ will be produced below (see \$3.3); in these sections, the
jumps of $H\left( k\right) $ across 
$ \Sigma_s$ 
and 
$\Sigma_u$ 
are always equal to the analytic continuations of the
corresponding absorptive parts $\Delta _{s}H,\Delta _{u}H.$ In fact, the
existence of the analytic continuation of $H$ in ${\cal L}$ implies that the
jumps $\Delta _{s}H,\Delta _{u}H$ are obtained there as distributions in the
real submanifold of ${\cal L};$ they are defined as differences of boundary
values of holomorphic functions from the complex regions of ${\cal L},$
namely from {\it new} directions of $\func{Im}k$-space which belong to $%
{\cal H}\left( D\right) ,$ although not to 
$D.$ If ${\cal L}$ is one-dimensional, ${\cal L}\backslash 
\left( \Sigma_s 
\cup \Sigma_u\right) $ 
is called
a ``cut-plane section'' of ${\cal H}\left( D\right) .$ 

\vskip 1cm
\noindent
{\sl Complex Lorentz invariance of $H\left( k\right): $}

In the following, we shall restrict ourselves to the case of {\it scalar local
fields}. In this case, the g.r.f. $r_{\alpha }\left( x\right) $ are invariant
under the (diagonal) action of the real connected Lorentz group $G=SO_{0}\left(
1,d\right) ;$ this invariance property is then satisfied by the
corresponding Fourier--Laplace transforms $\tilde{r}\left( k\right) $ and
therefore by $H\left( k\right) $ in its analyticity domain $D$ ($D$ being
itself invariant under this group). By a standard argument (based on the
uniqueness of analytic continuation), it follows that $H\left( k\right) $ is
also invariant under the {\it complex} connected Lorentz group $G^{\left( c\right) },$
i.e. $H\left( k_{{\rm i}},...,k_{4}\right) =H\left( gk_{{\rm i}%
},...,gk_{4}\right) $ for all $g$ in $G^{\left( c\right) },$ this property
being satisfied in the whole holomorphy envelope ${\cal H}\left( D\right) .$ %

\subsection{\bf A simple step in the analytic completion problem}

\quad From now on, we adopt the notations of the $t-$channel kinematics given 
in Sec.2, namely we put $k_{1}^{^{\prime }}=-k_{3},\ k_{2}^{^{\prime
}}=-k_{4}$ so that $k_{1}+k_{2}=k_{1}^{^{\prime }}+k_{2}^{^{\prime }}=K $,  
and we replace the notation $k=(k_1,k_2,k_3,k_4)$ of \S 3.1 by
$[k]=(k_1,k_2,k_1^{\prime},k_2^{\prime})$; accordingly, the four-point function is 
now denoted $H([k])$.

The step of the analytic completion problem which we
shall perform will yield domains in any subspace 
$M_{K}^{\left( c\right)}$ 
such that $K^{2}=t\leqslant 0,$ with the coordinatization of 
Sec.2: $K=\sqrt{-t}\ {\rm e}_{d}\ $ and (as specified in \S 2.1), $k_1^{\prime}$ and 
$k_2^{\prime}$ are real vectors varying in the $({\rm e}_{d-1}, {\rm e}_d)-$plane  
of the completed coordinate system.

This step can be said to be ``simple'' because all the new points obtained 
are boundary points of the primitive domain $D$ described in \S 3.1: in fact,
the simple geometrical property which we use is that each subspace 
$M_{K}^{\left( c\right)}$ is a linear manifold containing a common part
of the boundaries of the following two tubes   
\footnote{these tubes are the analyticity domains of the Laplace transforms 
of the ordinary  
advanced and retarded four-point functions 
$a_{(2)}$ 
and 
$r_{(1)}$} 
${\cal T}_{(1)}^{+}=\left\{ [k];\func{Im}k_{1}^{\prime }\in V^{-},\func{Im}%
k_{2}^{\prime }\in V^{-},\func{Im}k_{1}\in V^{+}\right\} $ 
and ${\cal T}_{(2)}^{-}=\left\{ [k];\func{Im}k_1^{\prime }\in  
V^{+},\func{Im}k_{2}^{\prime }\in V^{+},\func{Im}k_{2}\in  
V^{-}\right\}. $ 
This ``common face'' (carried by the linear manifold $\func{Im}k^{\prime}_1 = 
\func{Im}k^{\prime}_2 = 0$) is the tube 
$$ {\cal T}^{+}_{K} = \{ [k] \in M_K^{(c)};\  
\func{Im}k_{1}=-\func{Im}k_{2} \in  
{V}^{+}_{K}\},$$ 
where 
${V}^{+}_{K}$  
denotes the intersection of the hyperplane orthogonal to $K$  
(namely the space ${\Bbb R}^{d}$ spanned by  
${\rm e}_{0},{\rm e}_{1},..,{\rm e}_{d-1}$)  with
the forward light cone $V^{+}$ of ${\Bbb R}^{d+1}.$ 

Similarly we introduce the opposite tube 
$$ {\cal T}^{-}_{K} = \{ [k] \in M_K^{(c)};\  
\func{Im}k_{1}=-\func{Im}k_{2} \in  
{V}^{-}_{K}\} ,$$ 
where 
${V}^{-}_{K}= - V^{+}_{K}.$  
$ {\cal T}^{-}_{K}$ is the common face (in $ M_K^{(c)}$) of   
the tubes ${\cal T}^-_{(1)} = - {\cal T}^+_{(1)}$ and  
${\cal T}^+_{(2)} = - {\cal T}^-_{(2)}$ of the primitive domain $D$.   

The following statement is then contained in Theorem 4 of [23], but for 
simplicity and self-consistency of the present paper, we prefer to 
give here a direct \footnote{the proof given in [23] makes use of
a theorem by Bremermann and relies on a condition of coincidence for adjacent tubes
which, for simplicity, we have omitted in \S 3.1.}
proof of this result (with the help of Appendix A). 
\vskip 0.5cm

\noindent 
{\sl Proposition 2:}

\it
a) \ ${\cal H}\left( D\right) $ contains the
set of all points $[k]$ in 
$ {\cal T}^{+}_{K} \cup   
{\cal T}^{-}_{K};$
moreover,at all the points in 
$ {\cal T}^{+}_{K}$ (resp.   
${\cal T}^{-}_{K}$), 
$H([k])$ admits a common analytic continuation from both tubes  
${\cal T}^+_{(1)}$ and  
${\cal T}^-_{(2)} $ (resp.   
${\cal T}^-_{(1)} $ and  
${\cal T}^+_{(2)} $) of the primitive domain $D$.   

b)\   
The two sets 
$ {\cal T}^{+}_{K}$ and     
${\cal T}^{-}_{K}$
are connected in 
${\cal H}\left( D\right) $ 
by 
${\cal N}({\cal R}) \cap M_K^{(c)}.$
\rm 

\vskip 0.5cm

\noindent {\sl Proof:} 

Let $[\hat{k}]=( \hat{k}_{1},\hat{k}_{2},\hat{k}_{1}^{^{\prime
}},\hat{k}_{2}^{^{\prime }}) $ be any real momentum configuration in $M_K^{(c)}$  
contained in the $\left( {\rm e}_{1},{\rm e}_{2},...,{\rm e}_{d}\right)-$hyperplane  
of coordinates ; $[\hat{k}]$ belongs to the region ${\cal R},$ since all
quantities  $\hat{k}_{i}^{2},\ \hat{k}_{ i}^{^{\prime }2},\ (\hat k_{%
i}-\hat k_{j}^{\prime }) ^{2},\ i,j=1,2$ and  
$(\hat k_{1}+\hat k_{2})^{2} = t$ are $\le 0.$ 
We shall now exhibit a two-dimensional (complex) section of the tubes   
${\cal T}_{(1)}^{+}$  
and
${\cal T}_{(1)}^{-}$
of the domain $D$  
by putting
\begin{equation}
k_{1}=\hat{k}_{1} +\eta {\rm e},\ \ 
k_{1}^{^{\prime }}=\hat{k}_{1}^{^{\prime
}}-\eta ^{^{\prime }}{\rm e},\ \ 
k_{2}^{^{\prime }}=\hat{k}_{2}^{^{\prime }}-\eta
^{^{\prime }}{\rm e},  \tag{3.5}
\end{equation}
with ${\rm e}$ fixed in ${V}_{K}^{+}.$ 

This two-dimensional section is represented by the union of the tubes ${\cal T}_{+},{\cal T}_{-}$
of $\left( \eta ,\eta ^{^{\prime }}\right) $ space described in Proposition
A-1. Now, it is clear that there exists a square of the form $\left| \eta
\right| <a,\left| \eta ^{^{\prime }}\right| <a$ in ${\Bbb {R}}^{2}$ whose
image by the mapping (3.5) belongs to the region ${\cal R}$ (since the point $\eta =\eta
^{^{\prime }}=0$ represents the configuration $[\hat{k}]$ which belongs to $%
{\cal R}).$

\noindent Corollary A-2 then implies that all the points $%
[k]=( k_{1}=\eta {\rm e}+\hat{k}_{1},\hat k_{1}^{^{\prime }},\hat k_{2}^{^{\prime
}}) $ such that either $\func{Im}\eta >0$ or $\func{Im}\eta <0$ or   
$\eta \in \left] -a,+a \right[ $ lie in ${\cal H}\left( D\right) $ ;
since this holds for every choice of ${\rm e}$ in $V_{K}^{+}$ and $%
\hat{k}_{1}$ in the $\left( {\rm e}_{1},{\rm e}_{2},...{\rm e}_{d}\right)-$ 
hyperplane, it is thus proved that all points in  
$ {\cal T}^{+}_{K}$ (resp.   
${\cal T}^{-}_{K}$) appear as points of analyticity for $H([k])$ obtained  
from the tube  
${\cal T}^+_{(1)}$   
(resp.   
${\cal T}^-_{(1)}$).   

A similar argument based on a two-dimensional section of 
${\cal T}^-_{(2)}    
\cup {\cal T}^+_{(2)} $ would exhibit all the points in     
$ {\cal T}^{+}_{K}$ (resp.   
${\cal T}^{-}_{K}$) as points of analyticity obtained  
from the tube  
${\cal T}^-_{(2)}$   
(resp.   
${\cal T}^+_{(2)} $).   

The fact that the analytic continuations of $H([k])$ obtained by these two procedures 
coincide results from the principle of uniqueness of analytic continuation, 
since both of them coincide in the intersection of
${\cal T}^{\pm}_{K}$ with the  
edge-of-the-wedge neighborhood ${\cal N}({\cal R})$ (contained in $D$).   

Point a) of Proposition 2 is thus proved and point b) is then trivial.

\vskip 0.5cm

We shall now restate the result of Proposition 2 in terms of the variables introduced in Sec.2.   
A parametrization of $M_{K}^{(c)}$ is given by  
Eqs (2-2) 
in which $\rho z =  \underline{k}= (\underline{k}^{(0)},...,\underline{k}^{(d-1)},0) $ 
varies in ${\Bbb {C}}^{d}$,  $z' =z_0 = (0,...,0,1,0)$ and $w, \rho', w'$ are real ($\rho' \ge 0$), namely 
\begin{eqnarray}
&[k] =[k]( \underline{k};\ w,\rho ^{\prime },w^{\prime},K) \equiv
\nonumber \\
&k_{1}=\underline{k}+( w+\frac{1}{2}) K,\ k_{2}=-\underline{%
k}-( w-\frac{1}{2}) K,  \tag{3.6} \\
&k_{1}^{^{\prime }}=\rho ^{\prime }z_0+( w^{\prime }+\frac{%
1}{2}) K,\ k_{2}^{\prime }=-\rho ^{\prime}z_0-( w^{\prime}-%
\frac{1}{2}) K ,  \nonumber
\end{eqnarray}

\noindent {\sl Proposition 3:}

\it Let $\underline{D}_{(w, w^{\prime },\rho^{\prime })
} $ be the following domain in the space ${\Bbb {C}}^d$ of the complex
vector $ \underline{k}$:\  
$\underline{D}_{( w,w^{\prime },\rho^{\prime} ) }=\underline{%
{\cal T}}^{+}\cup \underline{{\cal T}}^{-}\cup \underline{{\cal N}}\left( 
\underline{{\cal R}}_{(w,w^{\prime },\rho ^{\prime })}\right) ,$ where: 

a)\  $\underline{{\cal T}}^{\pm }={\Bbb R}^{d}+{\rm i}\underline{V}^{\pm },$ 
\ with 
$$\underline{V}^+ = - \underline{V}^- = \{ \underline{q}\in {\Bbb R}^d;\ \underline{q}^{(0)} > 
\left[ {\underline{q}^{(1)}}^2 + \cdots + {\underline{q}^{(d-1)}}^2 \right]^{1\over 2} \}$$ 

b)\  ${\cal %
N}\left( \underline{{\cal R}}_{(w,w^{\prime },\rho ^{\prime })}\right) $
is a suitable complex neighborhood of the following region

\noindent $\underline{{\cal R}}_{( w,w^{\prime},\rho ^{\prime })} 
=\left\{ \underline{k}\in {\Bbb R}^{d};\ \underline{k}^{2}<\mu
^{2},( \underline{k}-\rho^{\prime }z_o) ^{2}<\mu
_{s}^{2},( \underline{k}+\rho^{\prime}z_o) ^{2}<\mu_{u}^{2}  
\right\} ;$ in the latter, the constants, $\mu ^{2},\mu _{s}^{2},\mu _{u}^{2}$
are defined in terms of the mass thresholds ${\rm M}_{1}^{2},{\rm M}_{2}^{2},s_{0},u_{0}$
by the following expressions
$$\mu ^{2}= \min \left({\rm M}_{1}^{2}-t\left( w+\frac{1}{2}\right)
^{2},{\rm M}_{2}^{2}-t\left( w-\frac{1}{2}\right) ^{2}\right),$$
$$\mu _{s}^{2}=s_{0}-t\left( w-w^{^{\prime }}\right) ^{2},
\ \mu _{u}^{2}=u_{0}-t\left( w+w^{^{\prime }}\right) ^{2}.$$
\vskip 0.3cm
Then ${\cal H}\left( D\right) $ contains the union of the
following sets:

\noindent $\underline{\hat{D}}_{( w,w^{\prime},\rho ^{\prime})}  
=\left\{ [k]\in M_{K}^{( c)} 
;[k]=[k]\left( \underline{k};\ w,\rho ^{\prime },w^{\prime }, K \right) ;%
\underline{k}\in \underline{D}_{( w,w^{\prime},\rho^{\prime} ) }\right\} ,$ for all
real values of $w,w^{^{\prime }}$ and $\rho ^{^{\prime }}\left( \rho
^{^{\prime }}\geqslant 0\right) .$
\rm 

\vskip 0.3cm
\noindent {\sl Proof:}

This
statement is a direct consequence of Proposition 2  
since (in view of the parametrization (3.6)) 
$\underline{{\cal R}}_{( w,w^{\prime },\rho ^{\prime }) }$ is the
trace of ${\cal R}$ in complex $\underline{k}-$space, for fixed values of $%
w,w^{^{\prime }},\rho ^{^{\prime }}.$

\vskip 0.3cm

We shall now prove that bounds of the type (3-1) are satisfied by the
analytic continuation of the four-point function $H\left( [k]\right) $ in the
regions described in Proposition 3\ ; this is a simple example of
the extension to points of ${\cal H}\left( D\right) $
of bounds which are prescribed in the primitive domain $D$. 

\vskip 0.3cm

\noindent {\sl Proposition 4:}

\it Bounds of the following form are satisfied by $H\left( [k]
( \underline{k};\ w,\rho^{\prime} ,w^{\prime },K)\right) $ for $\underline{k}$
varying in the domains \underline{$D$}$_{( w,w^{\prime},\rho^{\prime} ) }$
of Proposition 3. 
\begin{equation}
\left| H\left( [k]\left( \underline{k};w,\rho^{\prime} ,w^{\prime }, K \right) \right)
\right| \leqslant C_{w,\rho^{\prime} ,w^{\prime }}\max \left[ \left( 1+\left\| 
\underline{k}\right\| \right) ^{m},\;d\left( \underline{k},\partial \underline{%
D}_{(w,w',\rho')}\right) ^{-n}\right]  \tag{3.7}
\end{equation}
where 
$${\parallel \underline{k}\parallel}^2  =\sum_{\ 0\leqslant 
{i}\leqslant d-1}{\left|\underline{k}^{\left( {i}\right) }\right|}^2 $$ 
and $m,n$
are the same numbers as in formula (3.1) $ (m \ge 0, n \ge 0)$ .  \rm  

\vskip 0.5cm

\noindent {\sl Proof:} 

For $w,w^{\prime },\rho ^{\prime }$ fixed, we
consider the section of the primitive domain $D$ by
the following complex submanifold parametrized by $\underline{k}$ and $\eta,\  \underline{k}%
\in {\Bbb C}^{d},\ \eta \in {\Bbb C}:$  

\begin{equation}
[k]=[k]\left( \underline{k},\eta \right) \left| 
\begin{array}{c}
k_{1}=\underline{k}+\left( w+\frac{1}{2}\right) K+\eta {\rm e}_{0} \\ 
k_{2}=-\underline{k}-\left( w-\frac{1}{2}\right) K+\eta {\rm e}_{0}
\end{array}
\right| 
\begin{array}{c}
k_{1}^{\prime }=\underline{\rho }^{\prime }z_0+\left( w^{\prime }+\frac{1}{2}%
\right) K+\eta {\rm e}_{0} \\ 
k_{2}^{\prime }=-\underline{\rho }^{\prime}z_0 -\left( w^{\prime }-\frac{1}{2}%
\right) K+\eta {\rm e}_{0} 
\end{array}
\tag{3.8}
\end{equation}

Let us first consider the case when $\underline{k}$ varies in the tube $%
\underline{{\cal T}}^{+}.$ One then checks that if $\eta $ varies in a strip 
$0<\func{Im}\eta <h\left( \underline{k}\right) $ such that $\func{Im}%
\underline{k}-h\left( \underline{k}\right) {\rm e}_{0}\in \partial V_{+},$ the
corresponding point $[k]=[k]\left( \underline{k},\eta \right) $ defined by (3.8)
varies in 
the tube 
${\cal T}_{(2)}^{-}=\left\{ [k];\func{Im}k_1^{\prime }\in  
V^{+},\func{Im}k_{2}^{\prime }\in V^{+},\func{Im}k_{2}\in  
V^{-}\right\} $ 
of the domain $D.$ Similarly, for $\eta$ varying in the strip $%
-h\left( \underline{k}\right) <\func{Im}{\eta }<0$,
the corresponding point $[k] =[k] (\underline {k}, \eta) $ varies in the tube 
${\cal T}_{(1)}^{+}=\left\{ [k];\func{Im}k_{1}^{\prime }\in V^{-},\func{Im}%
k_{2}^{\prime }\in V^{-},\func{Im}k_{1}\in V^{+}\right\}. $ 

Moreover, when $\eta $ varies in a real interval $\left[
-a,+a\right] $ such that $k_{1}^{\prime 2},k_{2}^{\prime 2}$ and $t=\left(
k_{1}^{\prime }+k_{2}^{\prime }\right) ^{2}$ remain negative,   
the boundary values of $H\left( [k]\right) $ from the two
previous tubes (i.e.\ from the sides $\func{Im}\eta >0,$ and $\func{Im}\eta
<0)$ define a common analytic continuation of the corresponding two
branches of $H\left( [k](\underline{k}, \eta) \right) $ across the interval $\left[ -a,+a\right] :$
this follows from the application of Proposition 2
to all situations such that $k'_1 +k'_2 = K +2 {\eta}{\rm e}_0$ (which is legitimate 
for $\eta \in [-a,+a]$).  

We shall now consider the majorizations (3.1) of $H\left( [k]\right) $ in the
tubes ${\cal T}_{(2)}^{-}$ and ${\cal T}_{(1)}^{+}$ and give their expressions
in terms of the complex variables $\underline{k}$ and $\eta$ when $[k]$
belongs to the submanifold (3.8). 

For $\underline{k}$ in $\underline{\cal T}^{+},$ the following majorizations follow
from (3.1), if $\eta $ varies in the set
$\left\{ \eta \in {\Bbb C};\left| \func{Re}{\eta}\right| <a,\ 0<\left| \func{%
Im}{\eta}\right| <h\left(\underline{k}\right) \right\}:$

\begin{equation}
\left| H\left( [k] \left( \underline{k},{\eta}\right)\right) \right|
\leqslant \underline{C}\max \left[ \left( 1+\parallel \underline{k}\parallel 
\right) ^{m},\left| 
\func{Im}{\eta}\right| ^{-n},\left( h\left(\underline{k}\right) -\left| \func{Im}%
{\eta}\right| \right) ^{-n}\right]
\tag{3.9}
\end{equation}

\noindent where $\underline{C}$ is a suitable constant.   

In order to obtain a bound for $H\left( [k]( \underline{k},{\eta}) %
\right) $at ${\eta}=0,$ i.e. at the corresponding point $[k]=[k]\left( 
\underline{k};w,\rho ^{\prime },w^{\prime },K\right) $ (see (3.6)) of the set $%
\underline{\hat{D}}_{\left( w,w^{\prime },\rho ^{\prime }\right) },$ it is
appropriate to apply Proposition A-3 to the function $f\left( {\eta}%
\right) =H\left( [k]\left( \underline{k},{\eta}\right) \right) $ in a square 
$\Delta _{b}$ such that $b = \min \left( \frac{h\left( \underline{k}%
\right) }{\sqrt 2},\left( 1+\parallel \underline{k}\parallel \right) ^{-\frac{m}{n}%
},a\right) .$
In fact, one checks that in this domain $\Delta _{b}$ the majorization (A-1) is
implied by (3.9) (with $M=\underline{C}(\sqrt 2 -1)^{-n} ).$
We can therefore apply the majorization (A-2) which yields (for the chosen value of $b$): 
\begin{equation}
\left| H\left( [k]\left( \underline{k},0\right) \right) \right| \leqslant c_{n}%
\underline{C}\max \left[ \left( 1+\parallel \underline{k}\parallel \right)
^{m},\left( \frac{h\left( \underline{k}\right) }{\sqrt 2}\right)
^{-n},a^{-n}\right] .  \tag{3.10}
\end{equation}
Since $d\left( \underline{k},\partial\underline {\cal T}^{+}%
\right) =\dfrac{h\left( \underline{k}\right) }{\sqrt{2}}$ and the constant $a$  
is independent of $\underline{k},$ the inequality (3.10) can be replaced by
\begin{equation}
\left| H\left( [k]\left( \underline{k},0\right) \right) \right| \leqslant 
\underline{C}^{\prime}\max \left[ \left( 1+\parallel \underline{k}\parallel \right)
^{m},\left( d\left( \underline{k},\partial \underline{\cal T}^{+}\right)
\right) ^{-n}\right] ,  \tag{3.11}
\end{equation}
($\underline{C}^{\prime}$ being a new constant).

The previous argument holds similarly, when $\underline{k}$ varies
in the tube $\underline{\cal T}^{-}$ (the tubes ${\cal T}_{(1)}^{+}$ and $%
{\cal T}_{(2)}^{-}$ being now replaced by their
opposites) and yields:
\begin{equation}
\left| H\left( [k]\left( \underline{k},0\right) \right) \right| \leqslant 
\underline{C}^{\prime} \max \left[ \left( 1+\parallel \underline{k}\parallel \right)
^{m},\left( d\left( \underline{k},\partial \underline{\cal T}^{-}\right)
\right) ^{-n}\right]  \tag{3.11'}
\end{equation}

Finally, when $\underline{k}$ belongs to $\underline{{\cal N}}\left( 
\underline{{\cal R}}_{\left( w,w^{\prime},\rho^{\prime} \right) }\right) ,$ the
majorization (3.3) yields immediately
\footnote{The majorization (3.12) can also be obtained directly from (3.11), (3.11')
and the analyticity of $H$ in ${\cal R}_{(w,w',\rho')}$ by applying again Proposition A.3.}
(by restriction to the submanifold 
(3.8)):
\begin{equation}
\left| H\left( [k]\left( \underline{p}+{\rm i}\underline{q},0\right) \right)
\right| \leqslant \underline{C}'' \max \left[ \left( 1+\parallel 
\underline{p}\parallel \right) ^{m},\left( d\left( p,\partial \underline{{\cal R}%
}_{\left( w,w^{\prime},\rho^{\prime}\right) }\right) \right) ^{-n}\right] 
\tag{3.12}
\end{equation}

The set of majorizations (3.11), (3.11'), (3.12) is equivalent to the global 
majorization (3.7) 
in the domain $\underline{D}_{\left(
w,w^{\prime},\rho^{\prime}\right) }.$ 

\vskip 0.8cm
\noindent
\subsection{{\bf The perikernel structure in the space $\hat M_{K}^{\left(
c\right) }$}}

In this subsection, we shall establish the analyticity properties and bounds
of the four-point function $H\left( [k]\right) $ which are necessary for
introducing (in the next section) an appropriate Laplace-type transform of $%
H $ in a complex angular-momentum variable associated with the $t$-channel.
These analyticity properties and bounds are direct applications of the
results of Propositions 3 and 4, which will be completed in a second step by
making use of the property of complex Lorentz invariance of $H\left(
[k]\right) .$

We first consider the following family of one-dimensional complex
submanifolds $\omega_{\left( \zeta ,\zeta ^{\prime },K\right) }$ in $M_K^{(c)}.$ 
With each
$\left( \zeta ,\zeta ^{\prime },K\right) $
such that $t<0,\zeta \in \Delta _{t}\backslash \partial \Delta _{t},$\ $\zeta 
^{\prime}\in \Delta _{t}$ (see Eq. (2.13)),   
we associate the complex hyperbola 
\begin{equation}
\omega _{\left( \zeta ,\zeta ^{\prime },K\right) }=\left\{ [k];[k]=[k]_{\left(
\zeta,\zeta^{\prime},K \right) }\left( z,z_{0}\right) ;\ z=\left( -%
{\rm i}\sin \theta ,0,...,0,\cos \theta \right) ,\theta \in {\Bbb C}\right\} 
\tag{3.13}
\end{equation}
where $[k]=
[k]_{\left(\zeta,\zeta^{\prime},K \right) }\left( z,z^{\prime}\right)$ 
is the mapping defined by Eqs (2.16). 

Each hyperbola $\omega _{\left( \zeta ,\zeta ^{\prime },K\right) }$ appears
to be the \it  meridian hyperbola in the $\left( {\rm e}_{o},{\rm e}%
_{d-1}\right) $-plane \rm  of the corresponding hyperboloid $\Omega
_{\left( \zeta ,\zeta ^{\prime },K\right) }$ (see Eq.
(2.17)).

We shall then prove:

\vskip 0.8cm
\noindent
{\sl Proposition 5:}

\it
a) For each {\it \ }$\left( \zeta ,\zeta ^{\prime },K\right) $ 
with $\zeta 
\in \Delta_t \backslash \partial\Delta_t,\ \zeta^{\prime} \in \Delta_t, $ 
the submanifold $\omega _{\left( \zeta ,\zeta
^{\prime },K\right) }$ provides a section of maximal analyticity of 
${\cal H}\left( D\right) $ which is the cut-domain $%
\omega _{\left( \zeta ,\zeta ^{\prime },K\right) }^{\left( {\rm cut}\right)
}=\omega _{\left( \zeta ,\zeta ^{\prime },K\right) }\backslash \left(
\Sigma_{s}\cup \Sigma_{u}\right);$  
$\Sigma_{s},\Sigma_{u} $
are the spectral sets defined by Eqs (2.19) (2.20).

b) The domain $\omega _{\left( \zeta ,\zeta ^{\prime },K\right) }^{\left( 
{\rm cut}\right) }$ is represented in the $2\pi $-periodic $\theta $-plane
as the following cut-plane:
\begin{equation}
\Pi _{\left( \rho ,w,\rho ^{\prime },w^{\prime },t \right) }={\Bbb C}\backslash
\left\{ \sigma _{+}\left( v_{s}\right) \cup \sigma_{-}\left( v_{u}\right)
\right\} ,  \tag{3.15}
\end{equation}
where:
\begin{eqnarray}
\sigma _{+}\left( v_{s}\right) &=&\left\{ \theta \in {\Bbb C};\ \theta ={\rm i}%
v+2\ell \pi ,\;\left| v\right| \geqslant v_{s},\;\ell \in {\Bbb Z}\right\} ,
\tag{3.16} \\
\sigma _{-}\left( v_{u}\right) &=&\left\{ \theta \in {\Bbb C};\ \theta ={\rm i}%
v+\left( 2\ell +1\right) \pi ,\;\left| v\right| \geqslant v_{u},\;\ell \in 
{\Bbb Z}\right\} .  \tag{3.17}
\end{eqnarray}
(with $v_{s},v_{u}$ defined by Eqs (2.25), (2.26)).

c) The restriction of the function $H\left( [k]\right) $ to each submanifold $%
\omega _{\left( \zeta ,\zeta ^{\prime },K\right) }$ is well defined as a $%
2\pi $-periodic function:
\begin{equation}
H_{\omega _{\left( \zeta ,\zeta ^{\prime },K\right) }}\left( \theta \right)
=H \left( [k]_{\left( \zeta,\zeta^{\prime},K \right) }\left(
z,z_{0}\right) \right) _{\left| _{z=\left(-{\rm i}\sin \theta ,0,...,0,\cos
\theta \right) }\right. },  \tag{3.18}
\end{equation}
which is holomorphic in the domain $\Pi _{\left( \rho ,w,\rho
^{\prime },w^{\prime },t\right) }$ and satisfies bounds of the following form:
\begin{equation}
\left| H_{\omega _{\left( \zeta ,\zeta ^{\prime },K\right) }}\left( \theta
\right) \right| \leqslant C_{\left( \zeta ,\zeta ^{\prime },K\right) }{\rm e}%
^{m_{*}\left| \func{Im}\theta \right| }\left[ d\left( \theta ,\sigma
_{+}\left( v_{s}\right) \cup \sigma _{-}\left( v_{u}\right) \right) \right]^{-n} .
\tag{3.19}
\end{equation}
in the latter $m_{*}=\max \left( m,n\right) $ and $C_{\left(
\zeta ,\zeta ^{\prime },K\right) }$ is a suitable constant.
\rm 

\vskip 0.5cm

\noindent
{\sl Proof:} 

a) we shall prove that for every $\left( \zeta ,\zeta ^{\prime },K\right) $
with $\zeta 
\in \Delta_t \backslash \partial\Delta_t,\ \zeta^{\prime} \in \Delta_t, $ 
the cut-domain $\omega _{\left( \zeta
,\zeta ^{\prime },K\right) }^{\left( {\rm cut}\right) }$ is contained in the
corresponding subset $\underline{\hat{D}}_{\left( w,w^{\prime},\rho ^{\prime} 
\right) }$ of ${\cal H}\left( D\right) $ obtained in
Proposition 3. In fact, each point $[k]$ in $\omega _{\left( \zeta ,\zeta
^{\prime },K\right) }$ is such that $[k]=[k]\left( \underline{k};w,\rho ^{\prime
},w^{\prime },K \right) $ (see Eqs (3.6)), with $\underline{k}=\rho z,z=\left( -%
{\rm i}\sin \theta ,0,...,0,\cos \theta \right) .$

\noindent By putting $\theta =u+{\rm i}v,$ we check that:

\begin{equation}
\left( \func{Im}\underline{k}\right) ^{2}=\rho ^{2}\left( \func{Im}z\right)
^{2}=\rho ^{2}\sin ^{2}u\geqslant 0  \tag{3.20}
\end{equation}

Since we have assumed that 
$\zeta \notin \partial \Delta _{t},$ 
i.e. $\rho \neq 0,$  
we see from (3.20) that all the complex
points $[k]$ of $\omega _{\left( \zeta ,\zeta ^{\prime },K\right) }$ are
represented by vectors $\underline{k}$ such that $ (\func{Im} \underline{k})^2 > 0,$
which means that $\underline{k}$
belongs either to $\underline{{\cal T}%
}^{+}$ or to $\underline{{\cal T}}^{-}$ and therefore to the domain $%
\underline{D}_{\left( w,w^{\prime},\rho ^{\prime }\right) }$ of Proposition
3.

Moreover, the real points of $\omega _{\left( \zeta ,\zeta ^{\prime
},K\right) }$ are represented by real vectors $\underline{k},$ such that $%
\underline{k}^{2}=\rho ^{2}z^{2}=-\rho ^{2}<0;$ therefore, they belong to $%
\underline{D}_{\left( w,w^{\prime},\rho ^{\prime }\right) },$ i.e. to the
region $\underline{{\cal R}}_{\left( w,w^{\prime},\rho ^{\prime }\right) },$
if and only if they do not belong to the union of the spectral sets $%
\Sigma_{s}$ and $\Sigma_{u}.$ This proves that the domain $%
\omega _{\left( \zeta ,\zeta ^{\prime },K\right) }^{\left( {\rm cut}\right)
}=\omega _{\left( \zeta ,\zeta ^{\prime },K\right) }\backslash \left(
\Sigma_{s}\cup \Sigma_{u}\right) $ is contained in ${\cal H}%
\left( D\right) .$ Since all points in $%
\Sigma_{s}\cup \Sigma_{u}$ are outside ${\cal H}%
\left( D\right) $ 
(see d) of Theorem in \S 3.1), 
$\omega _{\left( \zeta ,\zeta ^{\prime },K \right)%
}^{\left( {\rm cut}\right) }$ is actually the intersection of ${\cal H}%
\left( D\right) $ with $\omega _{\left( \zeta ,\zeta ^{\prime },K \right) }.$

b) In view of (3.20), all the complex points of $\omega _{\left( \zeta
,\zeta ^{\prime },K\right) }^{\left( {\rm cut}\right) }$ are represented in
the $\theta $-plane by the set $\left\{ \theta =u+{\rm i}v;u\neq \ell \pi 
,\;\ell \in {\Bbb Z}\right\} ;$ the real points form two disjoint
sets, represented respectively by
$\left\{ \theta ={\rm i}v+2\ell \pi ;\left| v\right| <v_{s},\;\ell
\in {\Bbb Z}\right\} $ and
$\left\{ \theta ={\rm i}v+\left( 2\ell +1\right) \pi ;\left|
v\right| <v_{u},\;\ell \in {\Bbb Z}\right\} .$ This shows that $\omega
_{\left( \zeta ,\zeta ^{\prime },K\right) }^{\left( {\rm cut}\right) }$ is
represented by the periodic cut-plane $\Pi _{\left( \rho,w,\rho ^{\prime
},w^{\prime },t \right) }.$

c) We shall apply the majorizations of proposition 4 to the present
situation, in which $\underline{k}=\rho z,z=\left( -{\rm i}\sin \theta
,0,...,0,\cos \theta \right) ,\;\theta =u+{\rm i}v.$ Since  

$\func{Re}\underline{k}=\left( \rho \cos u\ \sinh v, 0,...,0,\rho \cos u
\cosh v \right) $

$\func{Im}\underline{k}=\left( -\rho \sin u\ \cosh v
,0,...,0,-\rho \sin u \ \sinh v \right) $

\noindent we get:
\begin{equation}
\parallel \underline{k}\parallel ^{2}=\parallel \func{Re}\underline{k}%
\parallel ^{2}+ \parallel \func{Im}\underline{k}\parallel ^{2}=\rho ^{2}\left( 2\cosh
^{2}v-1\right)  \tag{3.21}
\end{equation}
On the other hand, for $\underline{k}\in \underline{{\cal T}}^{\pm },$ we
have:
\begin{equation}
d \left( \underline{k},\partial \underline{{\cal T}}^{\pm }\right)
=\inf \left( \left| \func{Im}\left( \underline{k}^{\left( 0\right) }+%
\underline{k}^{\left( d-1\right) }\right) \right| ,\;\left| \func{Im}\left( 
\underline{k}^{\left( 0\right) }-\underline{k}^{\left( d-1\right) }\right)
\right| \right) =\rho \left| \sin u \right| {\rm e}^{-\left| v\right| }.
\tag{3.22}
\end{equation}
In this situation, which corresponds to $\sin u \neq 0,$ the
majorizations (3.11), (3.11') therefore yield (in view of Eqs (3.21), (3.22)):
\begin{equation}
\left| H_{\omega _{\left( \zeta ,\zeta ^{\prime },K\right) }}\left( u+{\rm i}%
v\right) \right| \leqslant C\max \left[ \left( 1+\rho \sqrt{2}\cosh v  
\right) ^{m},\frac{{\rm e}^{n\left| v\right| }}{\rho ^{n} {\left| \sin u 
\right|}^n }\right],  \tag{3.23}
\end{equation}
which implies a bound of the form (3.19) when $\theta $ varies in $%
\Pi _{\left(\rho,w,\rho ^{\prime },w^{\prime },t \right) }$ {\it by staying
outside neighborhoods of the intervals }$\left\{ \theta ={\rm i}v+2\ell \pi
;\left| v\right| <v_{s},\;\ell \in {\Bbb Z}\right\} $ and $\left\{ \theta =%
{\rm i}v+\left( 2\ell +1\right) \pi ,\;\left| v\right| <v_{u};\;\ell \in 
{\Bbb Z}\right\} .$ When $\theta $ varies in these neighborhoods, one makes use of  
the majorization (3.12) (since in this case $\underline{%
k}\in \underline{\cal N}\left(\underline{{\cal R}}_{(w,w^{\prime},\rho ^{\prime})}\right)$),  
which completes the proof of the bound (3.19). \vskip 0.5cm

\noindent
{\sl Remark:}   
\it In the limiting case where $\rho =0,$ i.e. $\zeta \in \partial \Delta _{t},$
the set $\omega _{\left( \zeta ,\zeta ^{\prime },K\right) }$ reduces to a
single point $[k],$ which belongs to $\underline{{\cal R}},$ and therefore to $%
{\cal H}\left( D\right) ,$ but the statement of proposition 5 has a trivial
content; we notice that (according to the expressions (2.25), (2.26) of 
$v_{s},v_{u})$ the cuts $\sigma _{+}\left( v_{s}\right) ,\;\sigma _{-}\left(
v_{u}\right) $ of $\Pi _{\left( \rho ,w,\rho ^{\prime },w^{\prime }\right) }$
are shifted up to infinity when $\rho $ tends to zero. \rm  

\vskip 0.5cm

We shall now extend the previous analyticity properties of $H\left( [k]\right) 
$ to the manifolds $\Omega _{\left( \zeta ,\zeta ^{\prime },K\right) }$ and $%
\hat \Omega_{\left( \zeta ,\zeta ^{\prime },K\right) }$ (see Eqs (2.17), (2.15))
by exploiting the
Lorentz invariance of $H.$
We shall first use the invariance of $H\left( [k]\right) $ under the subgroup
of complex Lorentz transformations which leave the vectors $k^{\prime}_1,\ 
k^{\prime}_2$ unchanged. When $K \neq 0,$ 
this is the subgroup $\underline{G}%
^{\left( c\right) }=SO_{0}^{\left( c\right) }\left( 1,d-2\right) $ which
leaves the $\left( {\rm e}_{d-1},{\rm e}_{d}\right) $-plane of coordinates
unchanged.
In this case, $H\left( [k]\right) $ is then holomorphic and constant at all
points $[k]=[k]\left( g\underline{k};w,\rho ^{\prime },w^{\prime }\right) $
deduced from the points $[k]\left( \underline{k};w,\rho ^{\prime
},w^{\prime }\right) $ in $\omega _{\left( \zeta ,\zeta ^{\prime },K\right)
}^{\left( {\rm cut}\right) }$ by the action of any element $g$ in $%
\underline{G}^{\left( c\right) }.$ For $K=0,$ the analysis is
similar, except that the group $\underline{G}^{\left( c\right) }$ is now the
subgroup $SO_{0}^{\left( c\right) }\left( 1,d-1\right) $ which leaves the
point $z_{0}$ (i.e. ${\rm e}_{d-1}$) unchanged.

In particular, for each point $[\hat{k}]$ in $\omega _{\left( \zeta
,\zeta ^{\prime },K\right) },$ represented by a vector $\underline{\hat{k}}%
=\rho \hat{z}$ with $\hat{z}=\left( -{\rm i}\sin \theta ,0,...,0,\cos \theta
\right) ,$ the corresponding point $[\check{k}]=[k]\left( \underline{\check{k}}%
;w,\rho ^{\prime },w^{\prime }\right) ,$ obtained by the symmetry $\theta
\rightarrow -\theta $ (namely such that $\underline{\check{k}}=\rho \check{z},$
with $\check{z}=\left( {\rm i}\sin \theta ,0,...,0,\cos \theta \right) )$
belongs to the orbit $\{ [k]=[k]( g\underline{\hat{k}};w;\rho ^{\prime
},w^{\prime }) \} $ of $\underline{G}^{\left( c\right) }$ 
(in fact, one can find an element $g_{\hat k \rightarrow \check k}$ of $\underline G^{(c)}$ 
such that: $\check{k}=g_{\hat {k}\rightarrow \check{k}}(\hat{k})).$  
It follows that $H([\hat k]) =H([\check k]); $ 
correspondingly $H_{\omega _{\left( \zeta ,\zeta ,K\right) }}\left(
\theta \right) $ is an {\it even} function of $\theta $ and therefore a
holomorphic function of $\cos \theta $ which we denote by 
$\underline{H}%
_{\left( \zeta ,\zeta ^{\prime },K\right) }\left( \cos \theta \right) .$ 
The domain of the latter, which is the image of 
$\Pi _{\left(\rho,w,\rho',w',t\right)}$ 
onto the $\cos \theta-$plane, is the cut-plane
$\underline{\Pi}_{\left(\rho,w,\rho',w',t\right)}$ 
introduced in (2.32). 

In view of Lemma 1 one can now  
define $H([k])$ in each cut-domain 
$ \Omega _{\left( \zeta ,\zeta ^{\prime },K\right) }^{({\rm cut})}$ of 
$ \Omega _{\left( \zeta ,\zeta ^{\prime },K\right) }$ 
(see Eqs (2.17),(2.29)) as the
$\underline{G}^{(c)}-$invariant function 
\begin{equation}
H([k]) = 
\underline{H}%
_{\left( \zeta ,\zeta ^{\prime },K\right) }\left( \cos \Theta_t \right),  
\tag{3.24} 
\end{equation}
where $[k]\equiv \left((\zeta, \zeta^{\prime}, K), (z,z_0) \right)$ and
$\cos \Theta_t = -z.z_0 = z^{(d-1)};$
$\Theta_t$ is the off-shell scattering angle introduced in (2.3) (with here $z' = z_0$) 
and $\cos \Theta_t$ therefore coincides with the variable $\cos \theta$ of 
the parametrization (3.13) when $z$ belongs to the meridian hyperbola $\omega_{(\zeta,\zeta^{\prime},K)}$
of $ \Omega _{\left( \zeta ,\zeta ^{\prime },K\right) }.$ 
We can thus state

\vskip 1cm

\noindent
{\sl Proposition 6}

\it
For every submanifold $\Omega _{\left( \zeta ,\zeta ^{\prime },K\right) }$
in $\Omega _{K}$ (with $\zeta \in \Delta_t \backslash \partial \Delta_t,\ 
\zeta^{\prime} \in \Delta_t$ ), 
the corresponding ``cut-submanifold'' $\Omega _{\left( \zeta ,\zeta ^{\prime
},K\right) }^{\left( {\rm cut}\right) }$ belongs to ${\cal H}\left( D\right).$ 
In each of these submanifolds, the restriction of the 
function $H\left( [k]\right) $ is invariant
under the group $\underline{G}^{\left( c\right) }$ and can be identified  
with a holomorphic function $\underline{H}_{\left( \zeta ,\zeta ^{\prime
},K\right) }\left( \cos \theta \right) $ whose domain is the cut-plane $%
\underline{\Pi }_{\left(\rho,w,\rho^{\prime },w^{\prime}, t \right) }.$
Moreover, the jumps of $H\left( [k]\right) $ across the two cuts $%
\Sigma_{s}$ and $\Sigma_{u}$ in $\Omega _{\left( \zeta ,\zeta
^{\prime },K\right) }$ (or equivalently the jumps of $\underline{H}_{\left(
\zeta ,\zeta ^{\prime },K\right) }\left( \cos \theta \right) $ across the
cuts $\underline{\sigma }_{+}\left( v_{s}\right) $ and $\underline{\sigma }%
_{-}\left( v_{u}\right) $) are the corresponding restrictions of the
absorptive parts $\Delta _{s}H$ and $\Delta _{u}H$ of $H.$
\rm

\vskip 0.3cm

\noindent
(For a complete justification of the last
statement in Proposition 6, we refer the reader to the paragraph {\it ``Absorptive parts''} in {\S 3.1}.)  

\vskip 0.5cm
One similarly extends $H([k])$ to the cut-domains  
$ \hat\Omega _{\left( \zeta ,\zeta ^{\prime },K\right) }^{({\rm cut})}$ of the submanifolds  
$ \hat \Omega _{\left( \zeta ,\zeta ^{\prime },K\right) }$ (see Eqs (2.15), (2.16), (2.28)) 
by now using formula (3.24) with  
$[k]\equiv \left((\zeta, \zeta^{\prime}, K), (z,z^{\prime}) \right)$ and
$\cos \Theta_t = -z.z^{\prime}.$
By also taking into account the bounds (3.19)
on $H_{\omega_{(\zeta,\zeta^{\prime},K)}}(\Theta_t) =  
\underline{H}%
_{\left( \zeta ,\zeta ^{\prime },K\right) }\left( \cos \Theta_t
\right),$
one can then state:

\vskip 0.5cm
\noindent
\begin{theorem}
For every submanifold $\hat \Omega _{\left(
\zeta ,\zeta ^{\prime },K\right) }$ in $\hat{\Omega}_{K}$ (with $\zeta,\zeta' 
\notin \partial \Delta _{t}),$ the corresponding ``cut-submanifold'' $\hat{%
\Omega}_{\left( \zeta ,\zeta ^{\prime },K\right) }^{\left( {\rm cut}\right)
} $ belongs to ${\cal H}\left( D\right) .$ The restriction of $H\left(
[k]\right) $ to each of these submanifolds defines an ``{\it invariant
perikernel of moderate growth with distribution boundary values''} on the
corresponding complexified hyperboloid $X_{d-1}^{\left( c\right) }\left( \text{%
if\ }K\neq 0\right) $ or $X_{d}^{\left( c\right) }\left( \text{if\ }%
K=0\right) .$

This invariant perikernel $H\left( [k]_{\left(\zeta,\zeta^{\prime},K  
\right) }\left( z,z^{\prime }\right) \right) $ is
holomorphic on the domain of $X_{d-1}^{\left( c\right) }\times
X_{d-1}^{\left( c\right) }$ (resp. $X_{d}^{\left( c\right) }\times
X_{d}^{\left( c\right) })$ which is defined as the complement of the union of
the cuts $\left\{ \left( z,z^{\prime }\right) ;\ z.z^{\prime }\leqslant
-\cosh v_{s}\right\} $ and $\left\{ \left( z,z^{\prime }\right) ;\    
z.z^{\prime }\geqslant \cosh v_{u}\right\} .$ It can be identified with the
holomorphic function $\underline{H}_{\left( \zeta,\zeta^{\prime }, K \right) }\left(
-z.z^{\prime }\right) $ of the single variable $\cos \Theta_t = -z.z',\ \Theta_t$ being 
the off-shell scattering angle of the $t-$channel. The domain of this function is the cut-plane 
$\underline{\Pi }_{\left(\rho,w,\rho^{\prime},w',t 
\right) }$ and its growth 
is controlled by the following bounds in terms of $u= \func{Re}\Theta_t$ and 
$v= \func{Im}\Theta_t$: 
\begin{equation}
\left| \underline{H}_{\left( \zeta,\zeta^{\prime },K\right) }\left( \cos \left( u+{\rm %
i}v\right) \right) \right| \leqslant \underline{C}_{\left( \zeta,\zeta^{\prime
},K\right) }{\rm e}^{m_{*}\left| v\right| }\left| \sin u\right| ^{-n} 
\tag{3.25}
\end{equation}
\noindent if $\cos \Theta_t \notin {\Bbb R},$ and:

\begin{equation}
\left| \underline{H}_{\left( \zeta,\zeta^{\prime },K\right) }\left( \cos \Theta_t 
\right) \right| \leqslant \underline{C}_{\left( \zeta,\zeta^{\prime },K\right) }\left|
\cos \Theta_t -\cosh v_{s}\right| ^{-n}\left| \cos \Theta_t +\cosh v_{u}\right|^{-n} ,
\tag{3.26}
\end{equation}

\noindent if $\cos \Theta_t $ belongs to a neighborhood of the real interval $%
\left] -\cosh v_{u},\cosh v_{s}\right[ .$

\noindent In these bounds, $m^* = {\rm max}(m,n),$  $m$ and $n$ being the ``degrees of 
temperateness ''of the theory (introduced in (3.1)).

\end{theorem}

\vskip 0.5cm

The notion of ``invariant perikernel
of moderate growth on a complexified hyperboloid''   
has been introduced in [24,25] as an appropriate notion for studying    
the Laplace transformation associated with the complexified Lorentz group.   
While the term ``perikernel'' refers to the analyticity property of 
$H\left( [k]_{\left( \zeta ,\zeta ^{\prime }K\right) }\left(
z,z^{\prime }\right) \right)$ 
in the cut-domain 
described above, its ``invariant'' character  means that it satisfies the condition 
$H\left( [k]_{\left( \zeta ,\zeta ^{\prime }K\right) }\left( gz,gz^{\prime
}\right) \right) =
H\left( [k]_{\left( \zeta ,\zeta ^{\prime }K\right) }\left(
z,z^{\prime }\right) \right)$ 
for all $g$ in $G_{K}^{\left( c\right) }.$
Finally the property of ``moderate growth'', 
characterized by the bounds (3.25), (3.26),  
fits with the definition given in [25] 
as far as the behaviour at infinity is concerned. However, the present perikernels have
distribution-like (instead of continuous) boundary values on the reals.

\vskip 0.3cm
\noindent {\sl Remark.}
\ {\it The condition $\zeta$ (and $\zeta'$)$\notin \partial \Delta_t$ in Proposition 6
and Theorem 1 simply expresses the non-degeneracy of the corresponding submanifolds 
$\Omega _{\left(
\zeta ,\zeta ^{\prime },K\right) }$  
and $\hat \Omega _{\left(
\zeta ,\zeta ^{\prime },K\right) }$  
In the degenerate cases these sets are trivially contained in ${\cal H}(D)$ (see our previous remark
after Proposition 5).} 

\vskip 5cm

\section{Harmonic analysis of the four-point functions of scalar fields}

\quad Having established in Theorem 1 the perikernel structure of a
four-point function $H\left( [k]\right) $ relatively to a given $t$-channel we
are now in a position to apply the results of [25] which concern the
harmonic analysis of invariant perikernels of moderate growth on the
complexified (unit) hyperboloid $X_{d-1}^{\left( c\right) }.$ We shall give
a self-contained account of these results in \S 4-1 for the case $%
d=2$ and in \S 4-2 for the general case $ d>2 .$
As a matter of fact, we need to present an extended version of the
results of [25] which includes:

a) the presence of two cuts (instead of one, as in [25]) in the definition
of the analyticity domain of the perikernels (the corresponding results have
already been announced and their derivation outlined in [10])

b) the occurrence of perikernels with distribution-like boundary values on
the reals (as previously noticed).

Concerning the rigourous treatment of b), our arguments will make use of
results proved in Appendix B.

In \S 4-3, we come back to our analysis of the four-point function $H([k])$ 
of scalar local fields, also written in terms of the $t$-channel
variables as follows:

\begin{equation}
H([k]) =F\left( \zeta ,\zeta ^{\prime };t,\cos \Theta _{t}\right) .
\tag{4.1}
\end{equation}

The restrictions of $H$ to the manifolds $\hat{\Omega}_{\left( \zeta ,\zeta
^{\prime },K\right) }$ of the ``Lorentz-foliation'' of $\hat\Omega_K$ defined in Eqs (2.15), (2-16) can
then be identified with the perikernels of Theorem 1 (in their reduced form $%
\underline{H}),$ namely:

\[
{\rm for\ all\ }\left( \zeta ,\zeta ^{\prime },t\right) {\rm \ with\ }\left(
\zeta ,\zeta ^{\prime }\right) {\rm \ in\ }\Delta _{t}\times \Delta _{t},  
\]
\begin{equation}
F\left( \zeta ,\zeta ^{\prime };t,\cos \Theta _{t}\right) =\underline{H}%
_{\left( \zeta ,\zeta ^{\prime },K\right) }\left( \cos \Theta _{t}\right) 
\tag{4.1'}
\end{equation}

Applying the results of \S 4.1 and \S 4.2 will then directly lead us to introduce
and describe the properties of a Fourier--Laplace-type integral transform $%
\tilde{F}\left( \zeta ,\zeta ^{\prime };t,\lambda_t \right) $ of $F\left( \zeta
,\zeta ^{\prime };t,\cos \Theta _{t}\right) $ which interpolates
analytically in an appropriate way the set of (off-shell) $t$-channel
partial-waves:
\begin{equation}
f_{\ell }\left( \zeta ,\zeta ^{\prime },t\right) ={\omega _{d-1}}
\int_{-1}^{+1}P_{\ell }^{\left( d\right) }\left( \cos \Theta
_{t}\right) F\left( \zeta ,\zeta ^{\prime };t,\cos \Theta _{t}\right) \left[
\sin \Theta _{t}\right] ^{d-3}d\cos \Theta _{t}  \tag{4.2}
\end{equation}
in a half-plane of the complex variable $\lambda_t $. In Eq.(4.2), the functions $%
P_{\ell }^{\left( d\right) }$ are the ``ultraspherical Legendre
polynomials'' considered in [26] 
\footnote{These polynomials are proportional to the Gegenbauer polynomials $C_{\ell}^p$  
considered in chapter IX of [27] (see in the latter Eq.(6) of \S 4.7, 
which coincides with (4.3) for $p= \frac{d-2}{2}$ up to the normalization constant).}  
(chapter 1 \S 2), 
which reduce  
to $\cos \ell \Theta _{t}$ for $%
d=2$ and to the Legendre polynomials $P_{\ell }$ for $d=3$; they are given 
(for $d \ge 3$) by the following integral
representation (see Eq.(III.18) of [25c)]): 
\begin{equation}
P_{\ell}^{(d)}(\cos t)= \frac{\omega_{d-2}}{\omega_{d-1}} \int_0^{\pi}
(\cos t + {\rm i} \sin t \cos \phi)^{\ell} (\sin \phi)^{d-3} d\phi. 
\tag{4.3}
\end{equation}
In (4.2) and (4.3), $\omega _{d-1}$ denotes the area of the
sphere $S_{d-2}.$

In \S 4.4, it is shown that the previous property of analytic interpolation of the 
$f_{\ell}$ in the variable $\lambda_t$ is indeed equivalent to 
the property of analytic continuation of the Euclidean four-point function 
into the Lorentz-foliation of $\hat\Omega_K$: more precisely, this structure is characterized by   
{\it kernels of the Euclidean sphere-foliation of $\hat{\cal E}_K$ which are 
analytically continued into perikernels.}  

\subsection{Fourier-Laplace transformation on cut-domains of the
complexified hyperbola $X_{1}^{\left( c\right) }$}

On the complex hyperbola $X_{1}^{\left( c\right) }=\left( z^{\left( 0\right)} =  
-{\rm i}\sin \theta ,z^{\left( 1\right) }=\cos \theta ,\theta =u+{\rm i}%
v\right) ,$ one considers the domain $D=X_{1}^{\left( c\right) }\backslash
\left( \Sigma_{+}^{\left( c\right) }\cup \Sigma_{-}^{\left( c\right) }\right) $
(Fig.2), whose representation in the $\theta $-plane is the periodic
cut-plane $\Pi ={\Bbb C}\backslash \left( \sigma _{+}\cup \sigma _{-}\right).$ 
\begin{figure}
\epsfxsize=8
truecm
{\centerline{\epsfbox{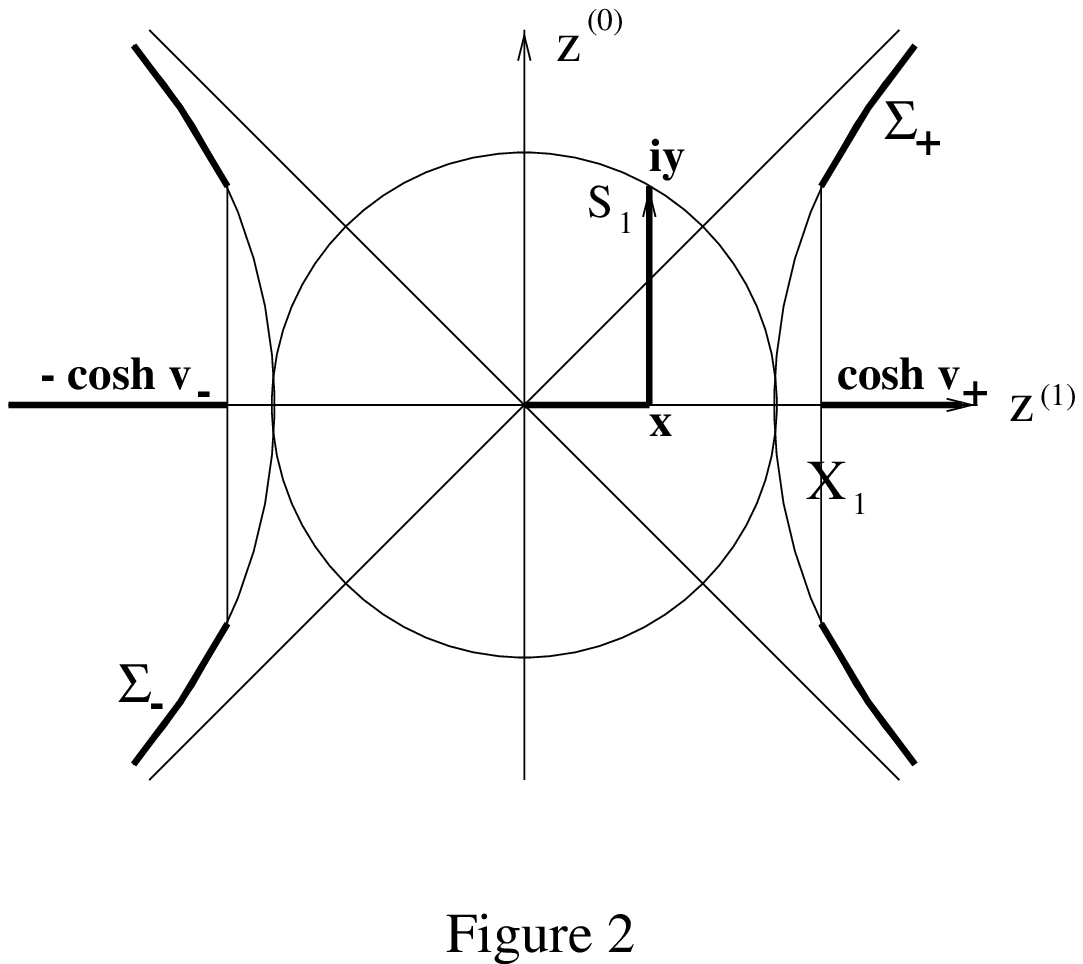}}} 
\end{figure} 
\begin{figure}
\epsfxsize=8truecm
{\centerline{\epsfbox{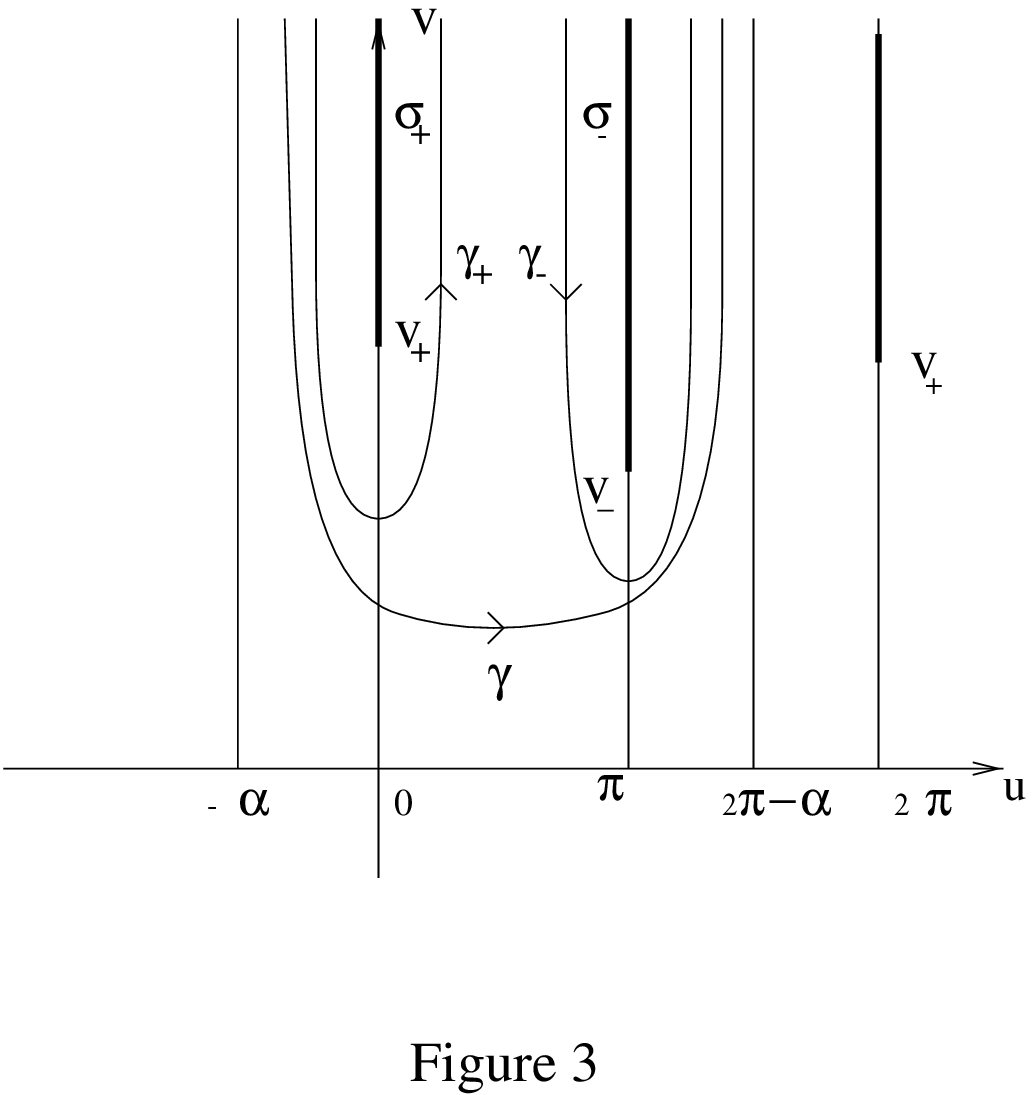}}} 
\end{figure} 
The cuts $\sigma _{+},\sigma _{-}$ are of the form (3.16), 
(3.17) (with $v_s =v_{+},\ v_u =v_{-})$ and the
corresponding subsets $\Sigma_{+}^{\left( c\right) },\Sigma_{-}^{\left( c\right)
}$ of $X_{1}^{\left( c\right) }$ are given (as in (2.30) and (2.31)) by:
\begin{equation}
\Sigma_{\pm }^{\left( c\right) }=\left\{ z=\left( z^{\left( 0\right)
},z^{\left( 1\right) }\right) \in X_{1}^{\left( c\right) };\pm z^{\left(
1\right) }\in \left[ \cosh v_{\pm },+\infty \right[ \right\}  \tag{4.4}
\end{equation}
Note that the circle 
$ {\Bbb S}_{1}=\left\{ z=\left( {\rm i}%
y_{0},x_{1}\right) ;\ y_{0}^{2}+x_{1}^{2}=1;\  
y_{0},x_{1}{\rm \ \func{real}}\right\}$ 
of $X_{1}^{\left( c\right) },$ 
represented
by the periodic $u$-axis in $\Pi ,$ is contained in the domain $D.$

{\it An invariant perikernel }${\cal K}\left( z,z'\right) $ on $%
X_{1}^{\left( c\right) }$ is identified with a function ${\cal F}\left(
z\right) ={\cal K}\left( z,z_{0}\right) $ (where $z_{0}=\left( 0,1\right) )$
holomorphic in the domain $D$ of $X_{1}^{\left( c\right) }$ and depending
only on the variable $z^{\left( 1\right) }=\cos \theta .$ Representing ${\cal %
F}$ by the {\it even periodic function }$%
f\left( \theta \right) ={\cal F}\left( z\left( \theta \right) \right),$ holomorphic in
the cut-plane $\Pi ,$ one then defines [25a)] the following
Fourier--Laplace-type transform $\tilde{F}={\cal L}\left( f\right) :$
\begin{equation}
\tilde{F}\left( \lambda \right) =\int\nolimits_{\gamma }{\rm e}^{{\rm i}%
\lambda \theta }f\left( \theta \right) d\theta  \tag{4.5}
\end{equation}
with the prescription of Fig.3 for the contour $\gamma .$ ($\gamma $
``encloses'' the components of $\sigma _{+},\sigma _{-}$ inside the
half-strip: $v>0,-\alpha <u<2\pi -\alpha ,$ with $0<\alpha <\pi ).$

In view of the choice of $\gamma ,$ this transform is well-defined and
holomorphic in a half-plane of the form ${\Bbb C}_{+}^{\left( m\right)
}=\left\{ \lambda \in {\Bbb C};\ \func{Re}\lambda >m\right\} $ provided ${\cal %
F}$ {\it is a perikernel of moderate growth } satisfying the following
bound in $D$:
\begin{equation}
\left| {\cal F}\left( z\right) \right| \leqslant {\rm cst\ }\left(1+\left| z^{(
1) }\right|\right) ^{m},  \tag{4.6}
\end{equation}
or equivalently provided $f$ satisfies the following one in $\Pi$:
\begin{equation}
\left| f\left( u+{\rm i}v\right) \right| \leqslant {\rm cst\ e}^{m|v|}, 
\tag{4.7}
\end{equation}

Let us first assume that $f${\it \ admits continuous boundary values (from
both sides) on the cuts }$\sigma _{+}$ and $\sigma _{-}$ and call $\Delta
f_{+}\left( v\right) ,\Delta f_{-}\left( v\right) $ the corresponding jumps
of ${\rm i}f$, which it is sufficient to consider in the upper half-plane $%
\left( v\geqslant 0\right) :$
\begin{equation}
\Delta f_{+}\left( v\right) ={\rm i}\lim_{\varepsilon \rightarrow 0}\left(
f\left( \varepsilon +{\rm i}v\right) -f\left(- \varepsilon +{\rm i}v\right)
\right),  \tag{4.8}
\end{equation}
\begin{equation}
\Delta f_{-}\left( v\right) ={\rm i}\lim_{\varepsilon \rightarrow 0}\left(
f\left( \pi +\varepsilon +{\rm i}v\right) -f\left( \pi -\varepsilon +{\rm i}%
v\right) \right)  \tag{4.9}
\end{equation}
Assuming that the bound (4.7) is uniform in $\Pi $ and
therefore applies to the discontinuity functions $\Delta
f_{+},\Delta f_{-}$ of ${\rm i}f,$ the Laplace transforms of $%
\Delta f_{+},\Delta f_{-}$ 
\begin{equation}
\tilde{F}_{\pm }\left( \lambda \right) =\int\limits_{0}^{\infty }{\rm e}%
^{-\lambda v}\Delta f_{\pm }\left( v\right) dv  \tag{4.10}
\end{equation}
are holomorphic in ${\Bbb C}_{+}^{\left( m\right) }.$ Then applying a
simple contour-distortion argument to the integral (4.5) yields  
the following relations, valid in ${\Bbb C}_{+}^{\left( m\right) }:$

i)$$\tilde{F}\left( \lambda \right) =\tilde{F}_{+}\left( \lambda
\right) +{\rm e}^{{\rm i}\pi \lambda }\tilde{F}_{-}\left( \lambda \right) .$$

\noindent This follows from replacing the contour $\gamma $ by a pair of
contours $\left( \gamma _{+},\gamma _{-}\right) $ enclosing respectively the
cuts $\sigma _{+},\sigma _{-}$ and then from flattening them (in a folded
way) onto the cuts (see Fig.3).

ii) $$\tilde{F}\left( \ell \right) =f_{\ell },\ \  
{\rm for\  all\  integers}\  \ell\  {\rm  such\  that}\  \ell >m,$$
$${\rm where}\  f_{\ell }=\int\limits_{-\alpha }^{2\pi -\alpha }{\rm e}^{{\rm i}\ell
u}f\left( u\right) du.$$
\noindent This follows from choosing $\gamma =\gamma _{\alpha }$ with
support $\left] -\alpha +{\rm i}\infty ,-\alpha \right] \cup \left[ -\alpha
,2\pi -\alpha \right]$ $ \cup \left[ 2\pi -\alpha ,2\pi -\alpha +{\rm i}\infty
\right[ ,$ and taking into account the $2\pi $-periodicity of the integrand
of (4.5) for $\lambda =\ell $ integer.

We note that the Fourier coefficients $f_{\ell }$ of $f\left( u\right) $ are
associated with the (rotational invariant) kernel ${\bf K}\left( z,z^{\prime
}\right) $ on the ``imaginary circle'' ${\Bbb S}_{1}$ of $X_{1}^{\left(
c\right) }$ which is obtained by taking the restriction of the perikernel $%
{\cal K}\left( z,z^{\prime }\right) ,$ namely ${\bf K}={\cal K}_{\left| 
{\cal S}_{1}\times {\cal S}_{1}\right. }$ and $f\left( u\right) ={\bf K}%
\left( z,z^{\prime }\right) $ with $\cos \!u=-z.z^{\prime
}=y_{0}y_{0}^{\prime }+x_{1}x_{1}^{\prime }.$

We now state in a more detailed form an extension of the previous properties
which applies to the case when $f$ (resp. $F$ or ${\cal K})$ admits {\it %
distribution-like boundary values (and discontinuities) }on the cuts{\it \ }%
which border its domain.

\vskip 0.5cm 
\noindent
\begin{theorem}      
\it Let $f\left( \theta \right) $ be a ($2\pi $-periodic) even holomorphic function
in the cut-plane $\Pi \,$(repre\-senting an invariant perikernel of moderate
growth ${\cal K}\left( z,z^{\prime }\right) $ on $X_{1}^{\left( c\right) })$
satisfying uniform bounds of the following form (with $m$ and $\beta$ fixed,
$m \in {\Bbb R},\ \beta \ge 0$): 
\begin{equation}
\left| f\left( u+{\rm i}v\right) \right| \leqslant C\eta ^{-\beta }{\rm e}%
^{mv},  \tag{4.11}
\end{equation}
in all the corresponding subsets $\Pi _{\eta }^{+}\left( \eta >0\right) $ of
$\Pi ^{+}=\Pi \cap \{\theta \in {\Bbb C}; \ \func{Im} \theta >0 \}: $ 
\begin{eqnarray}
\Pi _{\eta }^{+} &=&\Pi ^{+}\backslash \left\{ \theta \in {\Bbb C};\;\theta
=u+{\rm i}v,\ \left| u-2n\pi \right| <\eta ,\ n\in {\Bbb Z},\ v>v_{+}-\eta
\right\}   \nonumber \\
&&\qquad \backslash \left\{ \theta \in {\Bbb C};\;\theta =u+{\rm i}v,\ \left|
u-\left( 2n-1\right) \pi \right| <\eta ,\ n\in {\Bbb Z},\ v>v_{-}-\eta \right\} 
\nonumber \\
&&  \tag{4.12}
\end{eqnarray}
Then,  

i)\quad The ``discontinuity functions'' $\Delta f_{+},\Delta
f_{-}$ of {\rm i}$f$ across the cuts $\sigma _{+},\sigma _{-}$ are
well-defined in the sense of distributions, and admit Laplace-transforms $%
\tilde{F}_{+}\left( \lambda \right) ,$\noindent $\tilde{F}_{-}\left( \lambda
\right) $ which are holomorphic in ${\Bbb C}_{+}^{\left( m\right) }$ and
satisfy uniform bounds of the following form (for all $\varepsilon,
\varepsilon^{\prime} > 0$): 
\begin{equation}
\left| \tilde{F}_{\pm }\left( \lambda \right) \right| \leqslant C_{\pm
}^{\left( \varepsilon ,\varepsilon ^{\prime }\right) }\left| \lambda-m \right|
^{\beta +\varepsilon ^{\prime }}{\rm e}^{-\left[ \func{Re}\lambda -\left(
m+\varepsilon \right) \right] v_{\pm }}  \tag{4.13}
\end{equation}
in the corresponding half-planes ${\Bbb C}_{+}^{\left( m+\varepsilon
\right) } .$

ii)\quad The transform $%
\tilde{F}={\cal L}\left( f\right) $ of $f,$ namely $\tilde{F}\left( \lambda
\right) =\int\limits_{\gamma }{\rm e}^{{\rm i}\lambda \theta }f\left( \theta
\right) d\theta ,$ is holomorphic in ${\Bbb C}_{+}^{\left( m\right) }$ and
satisfies the following properties:

\noindent a) 
\begin{equation}
\tilde{F}\left( \lambda \right) =\tilde{F}_{+}\left( \lambda \right) +{\rm e}%
^{{\rm i}\pi \lambda }\tilde{F}_{-}\left( \lambda \right)   \tag{4.14}
\end{equation}
b)\ for all integers $\ell $ such that $\ell >m,$ the Fourier coefficients
of $f_{\mid {\Bbb R}},$ namely 
\begin{equation}
f_{\ell }=\int\limits_{-\alpha}^{2\pi -\alpha }{\rm e}^{{\rm i}\ell u}f\left(
u\right) du,  \tag{4.15}
\end{equation}
are given by the following relations: 
\begin{equation}
f_{\ell }=\tilde{F}\left( \ell \right)   \tag{4.16}
\end{equation}
\rm
\end{theorem}

\vskip 0.5cm
\noindent
{\sl Proof}: \quad i) The validity of the bounds (4.11) on the function 
$f$ (which characterize it as a ``function of moderate growth'' near its
boundary set $\sigma _{+}\cup \sigma _{-})$ is equivalent (see e.g. [28]) to the fact
that $f$ admits {\it boundary values in the sense of distributions} on ${\rm %
i}{\Bbb R}$ and $\pi +{\rm i}{\Bbb R}$ (from both sides of each of these
lines) and therefore that the discontinuities $\Delta f_{+},\Delta f_{-}$ are
defined as distributions with respective supports $\sigma _{+},\sigma _{-}.$
We refer the reader to Proposition B.4, for a comprehensive study of
holomorphic functions of this type, considered as {\it derivatives (of
integral or non-integral order) of holomorphic functions with
continuous boundary values.}

In view of the exponential factor in (4.11), the Laplace transforms $%
\tilde{F}_{+}\left( \lambda \right) ,$ $\tilde{F}_{-}\left( \lambda \right) $
of $\Delta f_{+},\Delta f_{-}$ can always be defined as holomorphic
functions in ${\Bbb C}_{+}^{\left( m\right) }$ by the following contour
integrals:
\begin{eqnarray}
\tilde{F}_{+}\left( \lambda \right) &=&\int\limits_{\gamma _{+}}{\rm e}^{%
{\rm i}\lambda \theta }f\left( \theta \right) d\theta ,  \tag{4.17} \\
\tilde{F}_{-}\left( \lambda \right) &=&{\rm e}^{-{\rm i}\pi \lambda
}\int\limits_{\gamma _{-}}{\rm e}^{{\rm i}\lambda \theta }f\left( \theta
\right) d\theta =\int\limits_{\gamma _{+}}{\rm e}^{{\rm i}\lambda \theta
}f\left( \theta +\pi \right) d\theta  \tag{4.18}
\end{eqnarray}
$(\gamma _{+},\gamma _{-}$ being chosen as in Fig.3 with ${\rm \sup \!p}\gamma
_{-}=\left\{ \theta =\pi +\theta ^{\prime },\ \theta ^{\prime }\in {\rm \sup
\!\!p}\gamma _{+}\right\} ).$

When $\gamma _{\pm }$ is flattened onto $\sigma _{\pm },$ the limit of the
r.h.s. of (4.17) (resp. (4.18)) can now be seen as the action of the
distribution $\Delta f_{+}\left( v\right) $ (resp $\Delta f_{-}\left(
v\right) )$ on the test-function ${\rm e}^{-\lambda v}$ (the latter being
admissible for $\func{Re}\lambda >m).$

The derivation of the bounds (4.13) on $\tilde{F}_{\pm }\left( \lambda
\right) ,$ which relies on a technique of Abel transforms (or primitives of
non-integral order) is given in Proposition B.4 .
The latter must be applied to the functions 
$f_{m+}(\theta) = {\rm e}^{{\rm i}m \theta} f(\theta)$ and  
$f_{m-}(\theta) = {\rm e}^{{\rm i}m (\theta+ \pi)} f(\theta + \pi)$, which 
(in view of (4.11) and (4.12)) belong to the class
${\cal O}^{\beta}(B_a^{({\rm cut})})$ of an appropriate domain 
$B_a^{({\rm cut})}$ (e.g. $a= \frac {\pi}{2}$) as described in Appendix B (see Fig.B1). 
The majorization (B.19) then applies to the Laplace transforms
${\tilde F}_{m {\pm}}$ of $f_{m {\pm}},$ which are such that
${\tilde F}_{\pm}(\lambda) =
{\tilde F}_{m {\pm}} (\lambda - m),$ thus yielding the desired result (4.13).   

\vskip 0.3cm

\quad ii) The proof of the relations (4.14) and (4.16) relies on the
contour-distortion argument presented above in the case where $f\left(
\theta \right) $ has continuous boundary values.

\vskip 0.4cm  
\noindent
{\sl Remark:}
\it \quad In view of Eq (4.14), the relations (4.16) yield:
\begin{eqnarray}
{\rm for\ }\ell {\rm \ even,}\quad f_{\ell } &=&\tilde{F}_{+}\left( \ell
\right) +\tilde{F}_{-}\left( \ell \right)  \tag{4.19} \\
{\rm for\ }\ell {\rm \ odd,}\quad f_{\ell } &=&\tilde{F}_{+}\left( \ell
\right) -\tilde{F}_{-}\left( \ell \right)  \tag{4.20}
\end{eqnarray}
Since the holomorphic functions $\tilde{F}_{\pm }\left( \lambda \right) $
satisfy the bounds (4.13), which are in particular dominated by any
exponential function ${\rm e}^{\varepsilon \left| \lambda \right| }\left(
\varepsilon >0\right) $ in ${\Bbb C}_{+}^{\left( m\right) },$ these
functions appear respectively as the (unique) Carlsonian interpolations [29] of
the corresponding sequences $\{ \tilde{F}_{\pm }\left( \ell \right)
;\ell \in {\Bbb N},\ell >m \} .$
However the function $\tilde{F}\left( \lambda \right) 
$ itself (which behaves like ${\rm e}^{-\pi \func{Im}\lambda }$ in ${\Bbb C}%
_{+}^{\left( m\right) })$ {\rm does not satisfy the Carlsonian property }%
with respect to the sequence $\left\{ f_{\ell }\right\} $ which it
interpolates.
\rm

\vskip 0.3cm
This remark suggests the introduction of the following ``symmetrized and
antisymmetrized quantities'':
\begin{equation}
\left( \Delta f^{\left( s\right) }\right) \left( v\right) =\left( \Delta
f_{+}\right) \left( v\right) +\left( \Delta f_{-}\right) \left( v\right)
\;,\;\left( \Delta f^{\left( a\right) }\right) \left( v\right) =\left(
\Delta f_{+}\right) \left( v\right) -\left( \Delta f_{-}\right) \left(
v\right) \;,  \tag{4.21}
\end{equation}
whose respective Laplace transforms are:
\begin{equation}
\tilde{F}^{\left( s\right) }\left( \lambda \right) =\tilde{F}_{+}\left(
\lambda \right) +\tilde{F}_{-}\left( \lambda \right) \;,\;\tilde{F}^{\left(
a\right) }\left( \lambda \right) =\tilde{F}_{+}\left( \lambda \right) -%
\tilde{F}_{-}\left( \lambda \right) \;;  \tag{4.22}
\end{equation}
We can then give the following alternative version of Theorem 2 ii):

\vskip 0.5cm
\noindent
{\sl Proposition 7:}  
\it The transform $\tilde{F}$ of $f$ has the following structure: 
\[
\tilde{F}={\rm e}^{{\rm i}\pi \frac{\lambda }{2}}\left[ \cos \frac{\pi
\lambda }{2}\tilde{F}^{\left( s\right) }-{\rm i}\sin \frac{\pi \lambda }{2}%
\tilde{F}^{\left( a\right) }\right] ,
\]
$\tilde{F}^{\left( s\right) }$ and $\tilde{F}^{\left( a\right) }$ being  
Carlsonian interpolations in the half-plane ${\Bbb C}_{+}^{\left( m\right) }$
of the respective sets of even and odd Fourier coefficients of $f_{\mid 
{\Bbb R}};$ namely, one has: 
\begin{equation}
{\rm for\;}2\ell  >m,\quad \;\quad f_{2\ell }=\tilde{F}^{\left( s\right) }\left( 2\ell
\right)   \tag{4.23} 
\end{equation}
\begin{equation}
{\rm for\;}2\ell +1 >m,\;\ \ f_{2\ell +1}=\tilde{F}^{\left( a\right) }\left(
2\ell +1\right)   \tag{4.24}
\end{equation}
$\tilde{F}_{\lambda }^{\left( s\right) }$ and $\tilde{F}_{\lambda }^{\left(
a\right) }$ satisfying bounds of the form (4.13) in ${\Bbb C}_{+}^{\left(
m\right) }.$
\rm

\vskip 0.5cm
\noindent
{\sl Inversion formulas:}   

\vskip 0.5cm
\noindent a) The discontinuities $\left( \Delta f\right) _{\pm }\left(
v\right) $ (considered as distributions with support in $\left\{ v\geqslant
0\right\} )$ can be recovered from the corresponding functions $%
\tilde{F}_{\pm} \left( \lambda \right) $ by 
the following inverse Fourier formulas (equivalent in view of the Cauchy formula 
applied to $\tilde F_{\pm}(\lambda) {\rm e}^{-\lambda v}$ in ${\Bbb C}_{+}^{(m)}$): 
\begin{equation}
{\rm (for\ }v>0)\quad \left( \Delta f\right) _{\pm }\left( v\right) =\frac{1%
}{\pi }\int\limits_{-\infty }^{+\infty }\tilde{F}_{\pm }\left( m+{\rm i}\nu
\right) \cos \left[\left( \nu -{\rm i}m\right) v\right]\;d\nu ;  \tag{4.25}
\end{equation}
\begin{equation}
{\rm or }\quad \left( \Delta f\right) _{\pm }\left( v\right) =\frac{1%
}{\pi }\int\limits_{-\infty }^{+\infty }\tilde{F}_{\pm }\left( m+{\rm i}\nu
\right) \sin \left[\left( \nu -{\rm i}m\right) v\right]\;d\nu ;  \tag{4.25'}
\end{equation}
(note that under the assumptions of Theorem 2, Eq.(4.25) has to be
understood in the sense of tempered distributions)

On the other hand, there exists a well-defined integral representation of the
holomorphic function $f\left( \theta \right) $ in its domain $\Pi $ in terms of
the Laplace transforms $\tilde{F}_{+}\left( \lambda \right) ,\tilde{F}%
_{-}\left( \lambda \right) $ namely (assuming that $m$ is positive)
\begin{equation}
f\left( \theta \right) =-\frac{1}{2\pi }\int\limits_{-\infty }^{+\infty }%
\frac{\tilde{F}_{+}\left( m+{\rm i}\nu \right) \cos \left[ \left( m+{\rm i}%
\nu \right) \left( \theta \pm \pi \right) \right] }{\sin \pi \left( m+{\rm i}%
\nu \right) }d\nu  \tag{4.26}
\end{equation}
\[
-\frac{1}{2\pi }\int\limits_{-\infty }^{+\infty }\frac{\tilde{F}_{-}\left( m+%
{\rm i}\nu \right) \cos \left[\left( m+{\rm i}\nu \right) \theta\right] }{\sin \pi \left(
m+{\rm i}\nu \right) }d\nu +\frac{1}{2\pi }\sum\limits_{\left| \ell \right|
<m}f_{\ell }\cos \ell \theta 
\]
In fact, the first term at the r.h.s. of (4.26) can be seen to define a pair
of holomorphic functions in the respective strips $0<u<2\pi $ and $-2\pi
<u<0 $ (corresponding to the choice of the sign $-$ or $+$ in the cosine
factor), while the second term defines a holomorphic function in the strip $%
-\pi <u<\pi :$ this follows from the bounds (4.13) on $\tilde{F}_{\pm
}\left( \lambda \right) .$

The proof of (4.26) consists in showing that for $\theta =u$ real, it
reduces to the Fourier series of $f_{\mid {\Bbb R}},$ namely:
\begin{equation}
f\left( u\right) =\dfrac{1}{2\pi }\sum\limits_{\ell \in {\Bbb Z}}f_{\ell
}\cos \ell \theta  \tag{4.27}
\end{equation}
As a matter of fact, by using a standard contour distortion argument and
resummation of residues at integral points inside ${\Bbb C}_{+}^{\left(
m\right) }$ (known as the Sommerfeld-Watson resummation method [30,31]), one
shows that the first two terms at the r.h.s. of Eq.(4.26) are respectively
equal to the sums of the series $\dfrac{1}{\pi }\sum\limits\Sb \ell \in 
{\Bbb N}  \\ l>m  \endSb \tilde{F}_{+}\left( \ell \right) \cos \ell \theta $
and $\dfrac{1}{\pi }\sum\limits\Sb \ell \in {\Bbb N}  \\ l>m  \endSb \left(
-1\right) ^{\ell }\tilde{F}_{-}\left( \ell \right) \cos \ell \theta ,$ which
therefore (in view of Eqs. (4.19), (4.20)) reconstitute the r.h.s. of (4.27).

\vskip 0,5cm
\noindent
{\sl Remarks:}   

\it i)\  Eq.(4.25') for the discontinuities can also be recovered from (4.26) (taken in the
limits $\func{Re}\theta \rightarrow 0$ or $\pi ),$ at first in a formal way,
and more rigorously by using the techniques of primitives, presented in
Appendix B.

ii)\ If Eq.(4.26) is used for $m$ integer, its $r.h.s.$ 
must be understood as the action of the 
distribution $\lim_{\varepsilon \to 0, \varepsilon>0} \frac {1}
{\sin \pi (m-\varepsilon + {\rm i}\nu)}$ 
on the numerator of the integrand.

\rm 

\subsection{Fourier-Laplace transformation on cut-domains of the
complexified hyperboloid $X_{d-1}^{\left( c\right) },d>2$}

\quad We present a geometrical treatment of the $d$-dimensional case $\left(
d>2\right) $ which is very close in its spirit to the one given above for
the case $d=2.$
This treatment (see [25b),c)]) provides the connection, via 
analytic continuation, between Fourier analysis on the sphere $%
{\Bbb S}_{d-1}\approx SO\left( d\right) /SO\left( d-1\right) $ and an appropriate
realization of Fourier-Laplace analysis on the unit one-sheeted hyperboloid $%
X_{d-1}\approx SO_{\circ }\left( 1,d-1\right) /SO_{\circ }\left(
1,d-2\right) .$

Analytic continuation takes place on the complexified unit hyperboloid $%
X_{d-1}^{\left( c\right)} = 
\left\{ z=\left( z^{\left( 0\right) },...,z^{\left( d-1\right) }\right) \in 
{\Bbb C}^{d};z^{2}\equiv z^{\left( 0\right) ^{2}}-z^{\left( 1\right)
^{2}}-\cdots 
-z^{\left( d-1\right) ^{2}}=-1\right\}, $ which contains $%
S_{d-1}$ and $X_{d-1}$ as submanifolds of real type, namely $%
S_{d-1}=X_{d-1}^{\left( c\right) }\cap \left( {\rm i}{\Bbb R}\times {\Bbb R}%
^{d-1}\right) $ and $X_{d-1}=X_{d-1}^{\left( c\right) }\cap {\Bbb R}^{d}.$
One then considers classes of functions which enjoy analyticity, power
boundedness and invariance properties in the ``cut-domain'' $%
D=X_{d-1}^{\left( c\right) }\backslash \left( \Sigma_{+}^{\left( c\right)
}\cup \Sigma_{-}^{\left( c\right) }\right) ,$ where $\Sigma_{+}^{\left( c\right)
},\Sigma_{-}^{\left( c\right) }$ are given (as in Eqs (2.30), (2.31)) by:
\begin{equation}
\Sigma_{\pm }^{\left( c\right) }=\left\{ z=\left( z^{\left( 0\right)
},z^{\left( 1\right) },...,z^{\left( d-1\right) }\right) \in X_{d-1}^{\left(
c\right) };\ \pm z^{\left( d-1\right) }\in \left[ \cosh \!v_{\pm },+\infty
\right[ \right\}  \tag{4.28}
\end{equation}
More specifically, the functions ${\cal F}\left( z\right) $ considered are
supposed to be invariant under the stabilizer $G_{z_{0}}$ (isomorphic to $%
SO_{0}^{\left( c\right) }\left( 1,d-2\right) )$ of the base point $%
z_{0}=\left( 0,...,0,1\right) $ and therefore only depend on $z^{\left(
d-1\right) }=\cos \theta ,$ so that one can again put ${\cal F}\left(
z\right) =f\left( \theta \right) ,$ with $f$ even, $2\pi $-periodic and
holomorphic in the cut-plane $\Pi ={\Bbb C}\backslash \left( \sigma _{+}\cup
\sigma _{-}\right) $ (Fig.3). The analyticity domain $D$ of these
functions ${\cal F}$ is the preimage of $\Pi $ in $%
X_{d-1}^{\left( c\right) }$(through the mapping $z\rightarrow z^{\left(
d-1\right) }=\cos \theta \rightarrow \pm \theta ).$ In particular, the
sphere ${\Bbb S}_{d-1}( z^{\left( 0\right) }={\rm i}y^{\left( 0\right)},  
\ z^{\left( j\right) }=x^{\left( j\right) } $ real for all $j \ne 0,$ 
$\ y^{\left( 0\right)
^{2}}+x^{\left( 1\right) ^{2}}+...+x^{\left( d-1\right) ^{2}}=1 ) $ is
embedded in $D,$ and projects onto the interval $\left[ -1,+1\right] $ in
the $z^{\left( d-1\right) }$-plane. The cuts $\sigma _{+}$ and $\sigma _{-}$
are the images of subsets $\Sigma _{+}$ and $\Sigma _{-}$ of $X_{d-1},$
defined respectively by the conditions $z^{\left( 0\right) }>0,z^{\left(
d-1\right) }\geqslant \cosh \!v_{+}$ and $z^{\left( 0\right) }<0,z^{\left(
d-1\right) }\leqslant -\cosh \!v_{-};$ (see  
Fig.4). The jumps $\Delta f_{+},\Delta f_{-}$ of ${\rm i}f$ across $\sigma
_{+},\sigma _{-}$ can now be considered as functions (or distributions) on $%
X_{d-1}$(depending only on the coordinate $z^{\left( d-1\right) }=\cosh \!v%
{\rm \ or\ }-\cosh \!v)$ with supports contained respectively in $\Sigma_{+},$
$\Sigma_{-}.$

\begin{figure}
\epsfxsize=10truecm
{\centerline{\epsfbox{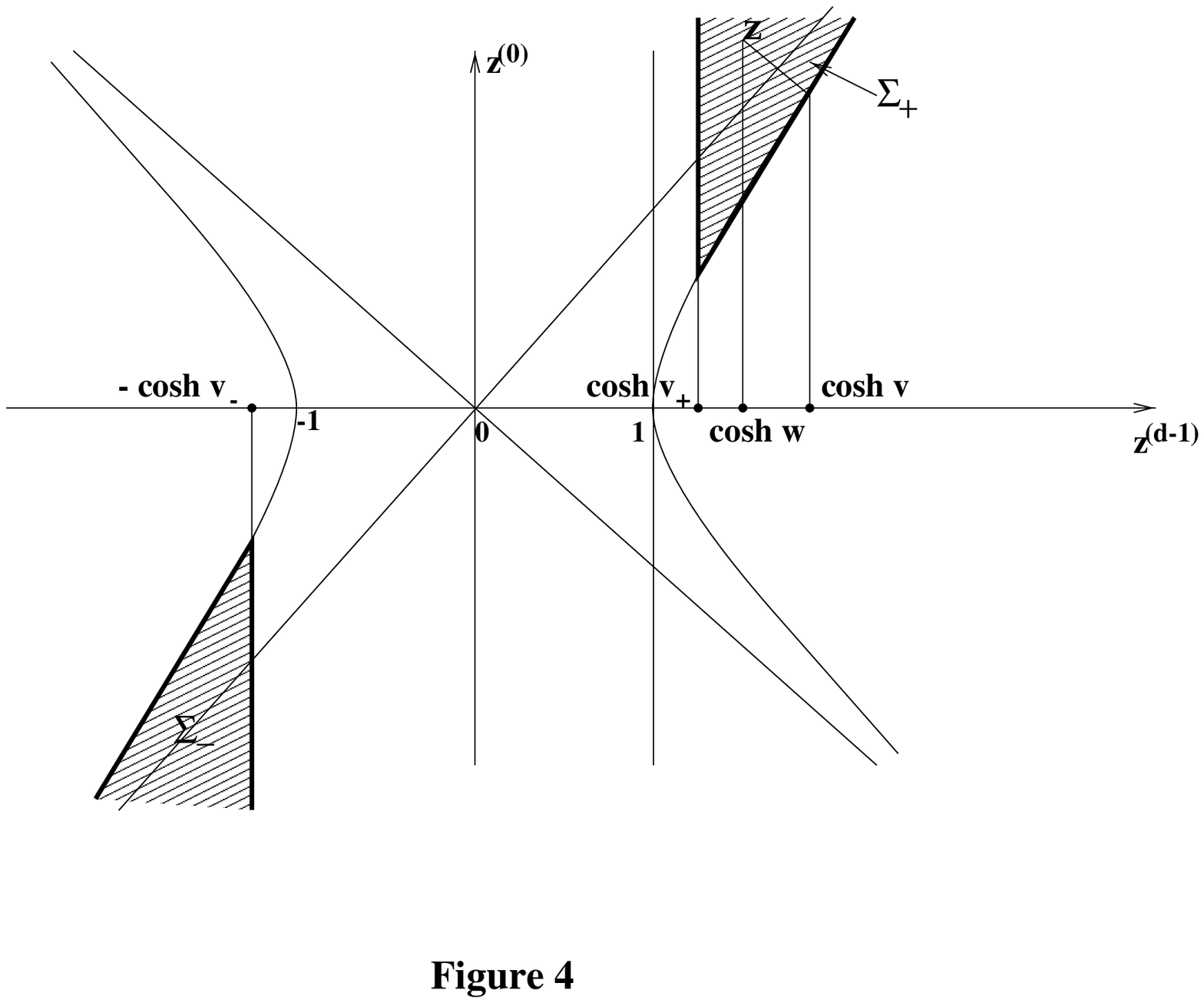}}} 
\end{figure}

Each function ${\cal F}\left( z\right) =f\left( \theta \right) $ also
represents an {\it invariant perikernel} ${\cal K}\left( z,z^{\prime
}\right) $ (such that ${\cal K}\left( gz,gz^{\prime }\right) ={\cal K}\left(
z,z^{\prime }\right) $ for all $g$ in $SO_{_{0}}^{\left( c\right) }\left(
1,d-1\right) ,$ and ${\cal K}\left( z,z_{0}\right) ={\cal F}\left( z\right)
) $ which is holomorphic in $X_{d-1}^{\left( c\right) }\times
X_{d-1}^{\left( c\right) }$ minus the union of the cuts $\hat{\Sigma}%
_{+}^{\left( c\right) }=\left\{ \left( z,z^{\prime }\right) ;z\cdot
z^{\prime }\leqslant -\cosh \!v_{+}\right\} $ and $\hat{\Sigma}_{-}^{\left(
c\right) }=\left\{ \left( z,z^{\prime }\right) ;z\cdot z^{\prime }\geqslant
\cosh v_{-}\right\} $ (these notations being similar to those used in
Theorem 1). The restriction of ${\cal K}$ to the sphere ${\Bbb S}_{d-1},$
namely ${\bf K}={\cal K}_{|{\Bbb S}_{d-1}\times {\Bbb S}_{d-1}}$ is an
analytic invariant kernel on ${\Bbb S}_{d-1},$ represented by ${\cal F}_{_{|%
{\Bbb S}_{d-1}}}=f_{|{\Bbb R}} = {\bf f} . $

While the Fourier analysis of ${\cal K}\left( z,z^{\prime }\right) 
$ on the sphere ${\Bbb S}_{d-1}$ is given [26,27] by the 
following set of coefficients of the generalized
Legendre expansion of ${\bf K}$  
involving the polynomials $P_{\ell}^{(d)}$ (see Eq.(4.3)): 
\begin{equation}
f_{\ell }={\omega _{d-1}}\int\limits_{0}^{\pi }\;P_{\ell
}^{\left( d\right) }\left( \cos \theta \right) {\bf f}\left( \theta \right) \left(
\sin \theta \right)^{d-2} d\theta ,\quad\  \ell \ge {0}  ,  \tag{4.29}
\end{equation}
the introduction of Laplace transforms associated with ${\cal K}$ along the
same line as in the case $d=2$ (see \S 4-1) necessitates a special
geometrical study. Before presenting the latter, we note that the
discontinuities $\left( \Delta f\right) _{+}\left( v\right) ,\left( \Delta
f\right) _{-}\left( v\right) $ of ${\rm i}f$ represent correspondingly the
discontinuities $\left( \Delta {\cal F}\right) _{+}\left( z\right) ,\left(
\Delta {\cal F}\right) _{-}\left( z\right) $ of ${\rm i}{\cal F}\left( z\right) $
on the cuts $\Sigma _{+}^{\left( c\right) },\Sigma _{-}^{\left( c\right) },$
which we can consider (after restriction to the real hyperboloid $X_{d-1})$
as functions (or distributions) with support contained respectively in the regions $%
\Sigma_{+},\Sigma_{-}:$ these functions (depending only on $z_{d-1}=\cosh \!v)$
also represent {\it Volterra kernels\thinspace }$K_{+}\left( z,z^{\prime
}\right) ,K_{-}\left( z,z^{\prime }\right) $ (such that $K_{\pm }\left(
z,z_{0}\right) =\Delta {\cal F}_{\pm }\left( z\right) )$ on the hyperboloid $%
X_{d-1},$ namely kernels with causal support properties on $X_{d-1}\times
X_{d-1}$ which are stable by the composition product [24,32] 
(this structure will be exploited in [14]). 
\vskip 0,5cm 

{\sl Laplace transformation on }$X_{d-1}${\it \ for functions of moderate
growth with support }$\sum_{\pm }.$

\vskip 0.3cm 
Two systems of local coordinates on $X_{d-1},$ are equally valid in a
neighbourhood of the set $\Sigma_{+}=\left\{ z\in X_{d-1};\ z^{\left( d-1\right)
}\geqslant \cosh \!v_{+},\ z^{\left( 0\right) }>0\right\} ,$ namely: 

a) The polar coordinates:

$$\quad\quad\quad   z^{\left( 0\right) }=\sinh w{\rm \;}\cosh \varphi
,\quad z^{\left( d-1\right) }=\cosh w,$$ 

\begin{equation}
\left[ \vec{z}\right] =\left( z^{\left( 1\right) },...,z^{\left( d-2\right)
}\right) =\sinh w\ \sinh \varphi \left[ \vec{\alpha}\right] ,\ \left[ \vec{\alpha%
}\right] \in S_{d-3}  \tag{4.30}
\end{equation}

b) The horocyclic coordinates:
\[
z^{\left( 0\right) }=\sinh \!v+\tfrac{1}{2}\parallel \vec{x}\parallel ^{2}%
{\rm e}^{v},\quad z^{\left( d-1\right) }=\cosh \!v{\rm -}\tfrac{1}{2}%
\parallel \vec{x}\parallel ^{2}{\rm e}^{v}, 
\]
\begin{equation}
\left[ \vec{z}\right] =\vec{x}\;{\rm e}^{v},\qquad \vec{x}\in {\Bbb R}^{d-2}
\tag{4.31}
\end{equation}
The sections $v={\rm cst}$ are paraboloids in the hyperplanes $z^{\left( 0\right)
}+z^{\left( d-1\right) }={\rm e}^{v},$ called horocycles.

For classes of functions $F_{+}\left( z\right) $ with support in $\Sigma
_{+} $ which are invariant under the stabilizer of $z_{0},$ namely $%
F_{+}\left( z\right) \equiv F_{+}\left[ z^{\left( d-1\right) }\right] ={\rm f%
}_{+}\left( w\right) $ (supp. ${\rm f}_{+}\subset \left[ v_{+},+\infty
\right[ ),$ and which moreover satisfy a bound of the form:
\begin{equation}
\left| F_{+}\left( z\right) \right| \leqslant {\rm cst}\left| z^{\left(
d-1\right) }\right| ^{m}  
{\rm or} \left| {\rm f}_{+}\left( w\right) \right| \leqslant {\rm %
cst\;e}^{m|w|},  
\tag{4.32}
\end{equation}
{\it the Laplace transform }$\tilde{F}_{+}\left( \lambda
\right) $ of $F_{+}$ {\it is defined as follows:} 
\begin{equation}
\tilde{F}_{+}\left( \lambda \right) =\int_{\Sigma _{+}}{\rm e}^{-\lambda
v}F_{+}\left[ \cosh w\right] d\vec{x}\;dv  \tag{4.33}
\end{equation}
In the latter we have used ``mixed coordinates'' $v,w,\left[ \vec{x%
}\right] $ (see Fig.4); from (4.30), (4.31) one gets:
\[
\vec{x}={\rm e}^{-v/2}\left[ 2\left( \cosh v-\cosh w\right) \right]
^{1/2}\left[ \vec{\alpha}\right] ,\text{ with }\left[ \vec{\alpha}\right]
\in S_{d-3}, 
\]
which allows one to rewrite Eq.(4.33) as follows $\left( \omega
_{d-2}\text{ being the area of }{\Bbb S}_{d-3}\right): $ 
\begin{equation}
\tilde{F}_{+}\left( \lambda \right) =\omega _{d-2}\int_{v_{+}}^{\infty }{\rm %
e}^{-\lambda v}{\rm e}^{-\left( \frac{d-2}{2}\right) v}{\cal A}_{d}{\rm f}%
_{+}\left( v\right) dv,  \tag{4.34}
\end{equation}
with 
\begin{equation}
{\cal A}_{d}{\rm f}_{+}\left( v\right) =\int_{v_{+}}^{v}{\rm f}_{+}\left( w\right)
\left[ 2\left( \cosh v-\cosh w\right) \right] ^{\frac{d-4}{2}}\sinh w\ dw 
\tag{4.35}
\end{equation}

It has been proved in [25] that {\it under the moderate
growth condition (4.32) the Laplace transform }$\tilde{F}_{+}\left( \lambda
\right) ${\it \ of }$F_{+}\left( z\right) $ {\it is holomorphic in }${\Bbb C}%
_{+}^{\left( m\right) }.$ This follows from the fact that 
{\it provided} $m > -1$ the exponential
bound (4.32) on $F_{+}$ is preserved 
by the transformation ${\rm f}_{+}\left( w\right) \rightarrow 
{\rm e}^{-\frac{d-2}{2}v}\left[ {\cal A}_{d}{\rm f}_{+}\right] \left( v\right) $
(see Proposition II-2 of [25b)] for a precise formulation
of this statement).

On the other hand, by introducing the second-kind function $Q_{\lambda
}^{\left( d\right) }$ via the integral representation (valid for $w\neq 0$
and $\func{Re}\lambda >-1)$
\footnote {We use here a normalization for these functions which is appropriate
to our joint consideration of $P_{\lambda}^{(d)}$ and
$Q_{\lambda}^{(d)}$; for $d=3$, the discrepancy with the standard normalization of 
the second-kind Legendre function [36a)] is a factor $1\over \pi$.} 
\begin{equation}
Q_{\lambda }^{\left( d\right) }\left( \cosh w\right) =\omega _{d-1}^{-1}%
\frac{\omega _{d-2}}{\left( \sinh w\right) ^{d-3}}\int_{w}^{\infty }{\rm e}%
^{-\left( \lambda +\frac{d-2}{2}\right) v}\left[ 2\left( \cosh \!v-\cosh
\!w\right) \right] ^{\frac{d-4}{2}}dv,  \tag{4.36}
\end{equation}
we obtain (by inverting the integrations in (4.34)) the following
alternative expression of $\tilde{F}_{+}(\lambda )\,$in its domain ${\Bbb C}%
_{+}^{(m)}$
\begin{equation}
\tilde{F}_{+}\left( \lambda \right) =\omega _{d-1}\int_{v_{+}}^{\infty
}{\rm f}_{+}(w)\ Q_{\lambda }^{\left( d\right) }\left( \cosh w\right)\  \left( \sinh
\!w\right) ^{d-2}dw.  \tag{4.37}
\end{equation}
(Note that the previous restriction $m > -1$ can be seen to be produced by the pole of the 
function $\lambda \to Q_{\lambda}^{(d)}(\cosh w)$ at $\lambda = -1$.)  

By now replacing $w$ by $w+{\rm i}${\rm $\pi $} and $v$ by $v+{\rm i}\pi $
in (4.30), (4.31), we obtain similar systems of local coordinates which are
valid in a neighbourhood of the set $\Sigma_{-}=\left\{ z\in
X_{d-1};z^{(d-1)}\leqslant -\cosh v_{-},z^{(0)}<0\right\} .$ Then one can
consider similarly the invariant function $F_{-}(z)\equiv F_{-}\left[
z^{(d-1)}\right] ={\rm f}_{-}(w)$ with support contained in $\Sigma_{-}$ (supp. $%
{\rm f}_{-}\subset \left[ v_{-},+\infty \right[ )$ and satisfying the growth
condition (4.32). This function $F_{-}$ admits the following Laplace
transform which is also holomorphic in ${\Bbb C}_{+}^{(m)}:$
\begin{equation}
\tilde{F}_{-}(\lambda )\  =\ \int\limits_{\Sigma_{-}}{\rm e}^{-\lambda
v}F_{-}\left[ -\cosh w\right] d\vec{x}dv  
\ =\ \omega _{d-2}\int_{v_{-}}^{\infty }{\rm e}^{-\lambda v}{\rm e}^{-(\frac{%
d-2}{2})v}{\cal A}_{d}{\rm f}_{-}(v)dv,  
\tag{4.38}
\end{equation} 
with
\begin{equation}
{\cal A}_{d}{\rm f}_{-}(v)=\int_{v-}^{v}{\rm f}_{-}(w)\left[ 2\left( \cosh v-\cosh w 
\right) \right] ^{\frac{d-4}{2}}\sinh w\ dw,  \tag{4.39}
\end{equation}
or equivalently in view of Eq.(4.36):
\begin{equation}
\tilde{F}_{-}(\lambda )=\omega _{d-1}\int_{v_{-}}^{\infty
}{\rm f}_{-}(w)\ \ Q_{\lambda }^{(d)}(\cosh w)\ (\sinh w)^{d-2}dw  \tag{4.40}
\end{equation}

\vskip 0.5cm

\noindent
{\sl Laplace transformation on }$X_{d-1}^{(c)}${\it \ for
holomorphic functions in } $D$    
{\sl with  
continuous boundary values}

\vskip 0.3cm

With every $G_{z_{0}}$-invariant holomorphic function ${\cal F}(z)=f(\theta
) $ defined in the cut-domain $D$ of $X_{d-1}^{(c)}$ and satisfying moderate
growth condition of the form $\left| {\cal F}(z)\right| \leqslant {\rm cst}%
\left| z_{d-1}\right| ^{m}$ (or $\left| f(u+{\rm i}v)\right| \leqslant {\rm %
cst}\ {\rm e}^{m|v|}),$ we shall now associate a Fourier--Laplace-type transform $%
\tilde{F}(\lambda ),$ holomorphic in the half-plane ${\Bbb C}_{+}^{(m)},$ by
a formula similar to (4.34) and (4.38),
except that a complex integration contour is used, namely
\begin{equation}
\tilde{F}(\lambda )=\omega _{d-2}\int_{\gamma }{\rm e}^{{\rm i}\left(
\lambda +\frac{d-2}{2}\right) \theta }\left( {\cal A}_{d}^{(c)}f\right)
\left( \theta \right) d\theta ;  \tag{4.41}
\end{equation}
here $\gamma $ is the same contour as for the case $d=2$ (see Fig.3 and
Eq.(4.5)), and the definition of ${\cal A}_{d}^{\left( c\right) }$ requires
the following procedure.
We introduce a decomposition of $f\left( \theta \right) $ of the form $%
f\left( \theta \right) =f_{+}\left( \theta \right) +f_{-}\left( \theta
\right) $ where $f_{+}$ and $f_{-}$ have the same analyticity and symmetry
properties as $f,\,$but enjoy the following additional property: $f_{+}$
(resp. $f_{-})$ admits {\it a single cut,} namely $\sigma _{+}$ (resp. $%
\sigma _{-})$ across which its discontinuity coincides with the
corresponding one of $f,$ denoted unambiguously by $\Delta f_{+}\left(
v\right) $ (resp. $\Delta f_{-}\left( v\right) ).$ Such a decomposition can
be done by considering the representation $\underline{f}\left( \cos \theta
\right) =f\left( \theta \right) $ of $f$ as a holomorphic function $%
\underline{f}$ in ${\Bbb C}\backslash \left\{ \left[ \cosh \!v_{+},+\infty
\left[ \cup \right] -\infty ,-\cosh \!v_{-}\right] \right\} $ bounded by $%
{\rm cst}\left| \cos \theta \right| ^{m}$ and defining $\underline{f}_{\pm
}\left( \cos \theta \right) =f_{\pm }\left( \theta \right) $ through
appropriate Cauchy integrals involving the respective weights
${\left( \Delta \underline{f}\right) _{+}\left( {\rm \cosh }%
v\right) }/ {\left( {\rm \cosh }v\right) ^{{\rm E}(m)+1}}$ and ${\left( \Delta 
\underline{f}\right) _{-}\left( {\rm \cosh }v\right) }/ {\left( {\rm \cosh }%
v\right) ^{{\rm E}(m)+1}}$ on the corresponding cuts of $\underline{f}$ (i.e. in
physical terms by the method of \it ``subtracted dispersion relations''\rm ).
The decomposition is non-unique, but defined up to a polynomial in 
$\cos \theta $ with degree ${\rm E}(m).$ 
We then define:
\begin{equation}
\left( {\cal A}_{d}^{(c)}f\right) \left( \theta \right) =\left( {\cal A}%
_{d+}^{(c)}f_{+}\right) \left( \theta \right) +\left( {\cal A}%
_{d-}^{(c)}f_{-}\right) \left( \theta \right) ,  \tag{4.42}
\end{equation}
where ${\cal A}_{d+}^{(c)}f_{+}$ and ${\cal A}_{d-}^{(c)}f_{-}$ are
respectively defined as holomorphic functions in the periodic cut-planes $%
{\Bbb C}\backslash \sigma _{+}$ and ${\Bbb C}\backslash \sigma _{-}$ by the
following integrals:

\begin{equation}
\left( {\cal A}_{d+}^{(c)}f_{+}\right) \left( \theta \right) =-\int_{\gamma
\left( \pi ,\theta \right) }f_{+}\left( \tau \right) \left[ 2\left( \cos
\theta -\cos \tau \right) \right] ^{\frac{d-4}{2}}\sin \tau d\tau  \tag{4.43}
\end{equation}

\begin{equation}
\left( {\cal A}_{d-}^{(c)}f_{-}\right) \left( \theta \right) =-\int_{\gamma
\left( 0,\theta \right) }f_{-}\left( \tau \right) \left[ 2\left( \cos \theta
-\cos \tau \right) \right] ^{\frac{d-4}{2}}\sin \tau \;d\tau .  \tag{4.44}
\end{equation}
In the latter, the path $\gamma \left( \pi ,\theta \right) $ (resp. $\gamma
\left( 0,\theta \right) )$ with end-points $\pi $ and $\theta $ (resp. $0$
and $\theta )$ has to belong to the domain ${\Bbb C}\backslash \sigma _{+}$
(resp. ${\Bbb C}\backslash {\Bbb \sigma }_{-})$ and the function $\left[
2\left( \cos \theta -\cos \tau \right) \right] ^{\frac{d-4}{2}}$ is
determined by the condition that it is positive for $\theta ={\rm i}v,\tau =%
{\rm i}w,0<w<v.$

Let us first assume that the boundary values of ${\cal F}(z)=f(\theta )$ on
the cuts $\sum_{\pm }^{(c)}$ (resp.$\sigma _{\pm })$ are continuous (from
both sides). One then checks that the jumps of ${\rm i}{\cal A}%
_{d}^{(c)}f(\theta )$ across the cuts $\sigma _{+}$ and $\sigma _{-}$ are
respectively equal to ${\cal A}_{d}\Delta f_{+}(v)$ and {\rm $e$}$^{-{\rm i}%
\pi \left( \frac{d-2}{2}\right) }{\cal A}_{d}\Delta f_{-}(v)$ (in view of
Eq.(4.35) and (4.39)). By using the same contour distortion argument as for
the case $d=2,$ namely by replacing $\gamma $ by $\gamma _{+}+\gamma _{-},$
and then flattening $\gamma _{+},\gamma _{-}$ on the respective cuts $\sigma
_{+},\sigma _{-},$ one can then rewrite the integral at the r.h.s. of (4.41)
as:

\begin{eqnarray}
&& 
\begin{array}{ll}
\int_{\gamma _{+}+\gamma _{-}}{\rm e}^{{\rm i}\left( \lambda +\frac{d-2}{2}%
\right) \theta }\left( {\cal A}_{d}^{(c)}f\right) (\theta )d\theta \\ 
= \int\nolimits_{\gamma _{+}}{\rm e}^{{\rm i}\left( \lambda +\frac{d-2}{2}%
\right) \theta }\left( {\cal A}_{d+}^{\left( c\right) }f_{+}\right) \left(
\theta \right) d\theta +  
\int\nolimits_{\gamma _{-}}{\rm e}^{{\rm i}\left( \lambda +\frac{d-2}{2}%
\right) \theta }\left( {\cal A}_{d-}^{\left( c\right) }f_{-}\right) \left(
\theta \right) d\theta \\ 
=\int\limits_{v_{+}}^{+\infty }{\rm e}^{-\left( \lambda +\frac{d-2}{2}\right)
v}\left( {\cal A}_{d}\Delta f_{+}\right) \left( v\right) dv+{\rm e}^{{\rm i}%
\pi \lambda } \int\limits_{v_{-}}^{+\infty }{\rm e}^{-\left( \lambda +\frac{%
d-2}{2}\right) v}\left( {\cal A}_{d}\Delta f_{-}\right) \left( v\right) dv
\end{array}
\nonumber \\
&&  \tag{4.45}
\end{eqnarray}
In view of Eqs.(4.34), (4.38), the latter can be rewritten (as for $%
d=2):$ 
\begin{equation}
\tilde{F}\left( \lambda \right) =\tilde{F}_{+}\left( \lambda \right) +{\rm e}%
^{i\pi \lambda }\tilde{F}_{-}\left( \lambda \right)  \tag{4.45'} 
\end{equation}
where the functions $\tilde{F}_{+},\tilde{F}_{-},$ holomorphic in ${\Bbb C}%
_{+}^{(m)}$ now denote the Laplace transforms of the discontinuities $\Delta
f_{+\mid \Sigma _{+}},\Delta f_{-\mid \Sigma _{-}}$ taken on the
corresponding sets $\Sigma _{+},\Sigma _{-}$ according to formulas (4.33),
(4.38) (with $F_{\pm }=\Delta f_{\pm \mid \Sigma _{\pm }}).$

\vskip 0.5cm
\noindent
{\sl Remark:}  
\it The Laplace transform $\tilde{F}\left( \lambda \right) $ that we have
introduced only depends on the function ${\cal F}\left( z\right) $ through
its discontinuities on $\Sigma _{+},\Sigma _{-};$ it therefore does {\it not}
depend on the particular decomposition $f=f_{+}+f_{-},$ in spite of the fact
that ${\cal A}_{d}^{\left( c\right) }f$ actually depends on the latter.
\rm

\vskip 1cm
\noindent
{\sl Link with the Fourier expansion on the sphere }${\Bbb S}%
_{d-1}$  {\sl (Froissart--Gribov
-type equalities):}

For $\lambda =\ell $ integer (with $\ell >m),$ we rewrite Eq.(4.41) with the
choice of contour $\gamma =\gamma _{\alpha }$ (as in the case $d=2,$ see
Fig.3); by taking the periodicity of $\left( {\cal A}_{d}^{\left( c\right)
}f\right) \left( \theta \right) .$ ${\rm e}^{{\rm i}\left( \frac{d-2}{2}%
\right) \theta }$ into account, this yields:
\begin{equation}
\tilde{F}\left( \ell \right) ={\omega _{d-2}}\int_{-\alpha
}^{2\pi -\alpha }{\rm e}^{{\rm i}\left( \ell +\frac{d-2}{2}\right) u}\left( 
{\cal A}_{d}^{(c)}f\right) (u)du  \tag{4.46}
\end{equation}
By choosing $\alpha =\pi $, the latter can be rewritten in
view of Eq.(4.42):
\begin{equation}
\tilde{F}\left( \ell \right) ={\omega _{d-2}}%
\int\nolimits_{-\pi }^{\pi }{\rm e}^{{\rm i}\left( \ell +\frac{d-2}{2}%
\right) u}\left[ \left( {\cal A}_{d+}^{\left( c\right) }f_{+}\right) \left(
u\right) +\left( {\cal A}_{d-}^{\left( c\right) }f_{-}\right) \left(
u\right) \right] du  \tag{4.47}
\end{equation}
Then by applying Eqs. (4.43), (4.44), inverting the order of integrations
and using obvious symmetries in the double integrals, one obtains:
\begin{equation}
\begin{array}{l}
{\omega _{d-2}}\int_{-\pi }^{\pi }{\rm e}^{{\rm i}\left(
\ell +\frac{d-2}{2}\right) u}\left( {\cal A}_{d+}^{\left( c\right)
}f_{+}\right) \left( u\right) du = \\ 
\\ 
{\omega _{d-2}}\int_{0}^{\pi }f_{+}\left( t\right) \sin
\!tdt\int_{-t}^{t}{\rm e}^{{\rm i}\left( \ell +\frac{d-2}{2}\right) u}\left[
2\left( \cos u - \cos t\right) \right] ^{\frac{d-4}{2}}du
\end{array}
\tag{4.48}
\end{equation}
and
\begin{equation}
\begin{array}{l}
{\omega _{d-2}}\int_{-\pi }^{\pi }{\rm e}^{{\rm i}\left(
\ell +\frac{d-2}{2}\right) u}\left( {\cal A}_{d-}^{\left( c\right)
}f_{-}\right) \left( u\right) du= \\ 
\\ 
{\omega _{d-2}}\int_{0}^{\pi }f_{-}\left( t\right) \sin
\!tdt\int_{t}^{2\pi -t}( -{\rm i}) ^{d-2}{\rm e}^{{\rm i}\left(
\ell +\frac{d-2}{2}\right) u}\left[ 2\left( \cos t- \cos u\right)
\right] ^{\frac{d-4}{2}}du
\end{array}
\tag{4.48'}
\end{equation}
We now use the following integral representations of the ultraspherical
Legendre polynomials, 
which are consequences of the representation (4.3) (see in [25c)] the derivation  of 
Eqs. (III-25) and (III-25') from (III-18)): 
\begin{equation}
\begin{array}{l}
P_{\ell }^{\left( d\right) }\left( \cos t \right) =
\left(\frac{\omega
_{d-2}}{\omega _{d-1}}\right)\left( \sin t \right) ^{-\left( d-3\right)
}\int_{-t}^{t} {\rm e}^{{\rm i} ( \ell +{(d-2)/ {2}}) u } 
[ 2( \cos u -\cos t ) ] ^{\frac{d-4}{2}} du = \\ 
\\   
(-{\rm i})^{d-2} \left(\frac{\omega
_{d-2}}{\omega _{d-1}}\right)\left( \sin t \right) ^{-\left( d-3\right)
}\int_{t}^{2\pi -t} {\rm e}^{{\rm i} ( \ell +{(d-2)/ {2}}) u } 
[ 2( \cos t -\cos u ) ] ^{\frac{d-4}{2}} du.  
\tag{4.49}
\end{array}
\end{equation}
The latter imply that Eqs (4.48) and (4.48') can be rewritten as follows: 
\begin{equation}
{\omega _{d-2}}\int\limits_{-\pi }^{\pi }{\rm e}^{{\rm i}%
\left( \ell +\frac{d-2}{2}\right) u}\left( {\cal A}_{d\pm }^{\left( c\right)
}f_{\pm }\right) \left( u\right) du=  
{\omega _{d-1}}\int\limits_{0}^{\pi }f_{\pm }\left(
t\right) P_{\ell }^{\left( d\right) }( \cos t)\ ( \sin t)^{d-2}dt.  
\tag{4.50}
\end{equation}
In view of Eq.(4.50), Eq.(4.47) can now be rewritten
\begin{equation}
\tilde{F}\left( \ell \right) ={\omega _{d-1}}
\int\limits_{0}^{\pi }\left( f_{+}\left( t\right) +f_{-}\left( t\right)
\right) P_{\ell }^{\left( d\right) }\left( \cos \!t\right) \left( \sin
\!t\right) ^{d-2}dt,  \tag{4.51}
\end{equation}
and since $f = f_{+} + f_{-},$ by comparing to Eq.(4.29):
\begin{equation}
{\rm (for\ }\ell >m){\rm \;\;\;}\quad \tilde{F}\left( \ell \right) =f\left( \ell
\right).  \tag{4.52}
\end{equation}
These {\it Froissart--Gribov--type equalities} can also be given the
following more precise form (in view of Eq.(4.45')):
\begin{equation}
f_{\ell }=\tilde{F}_{+}\left( \ell \right) +\left( -1\right) ^{\ell }\tilde{F%
}_{-}\left( \ell \right)  \tag{4.53}
\end{equation}
Note that if one calls $f_{\ell \pm }$ the Legendre coefficients of the
corresponding functions $f_{\pm }\left( \theta \right) ={\cal F}_{\pm
}\left( z\right) $ on the sphere ${\Bbb S}_{d-1},$ it can be easily checked
that $\tilde{F}_{+}\left( \ell \right) =f_{\ell +}$ and $\left( -1\right)
^{\ell }\tilde{F}_{-}\left( \ell \right) =f_{\ell -}.$

\vskip 0.5cm 
\noindent
{\sl The case of perikernels with distribution-like boundary
values:}

As in \S\ 4.1 (Theorem 2), we shall now give a detailed version of
the previous properties under the assumption that ${\cal F}\left( z\right)
=f\left( \theta \right) $ admits distribution-like boundary values (and
discontinuities) on the cuts $\Sigma _{\pm }^{\left( c\right) }$ (resp. $%
\sigma _{\pm }).$
\begin{theorem}
\it Let ${\cal F}\left( z\right) =f\left( \theta \right) $ represent an
invariant perikernel of moderate growth on $X_{d-1}^{(c)},$ satisfying
uniform bounds of the following form 
\begin{equation}
\left| f\left( u+{\rm i}v\right) \right| \leqslant c\eta ^{-\beta }{\rm e}%
^{mv}  \tag{4.54}
\end{equation}
in all the subsets $\Pi _{\eta }^{+}\left( \eta >0\right) $ of $\Pi ^{+}$
(see Eq.(4.12)), or equivalently: 
\begin{equation}
\left| {\cal F}\left( z\right) \right| \leqslant C\eta ^{-\beta }\left|
z^{(d-1)}\right| ^{m}  \tag{4.54'}
\end{equation}
in the subsets $D_{\eta },$ defined as the preimages of $\Pi _{\eta }^{+}$
in $D.$ In (4.54), (4.54'), we assume that $m>-1$ and $\beta \ge 0.$ 

\noindent Then

\noindent i)\quad The discontinuities $\left( \Delta {\cal F}\right) _{\pm
}(z)=\left( \Delta f\right) _{\pm }(v)$ of ${\rm i}{\cal F}$ (resp. ${\rm i}%
f)$ across the cuts $\Sigma_{\pm }^{(c)}$ (resp. $\sigma _{\pm })$ are
well-defined as distributions. They admit Laplace-transforms $\tilde{F}_{\pm
}\left( \lambda \right) $ on the hyperboloid $X_{d-1}$ defined for $\func{Re}\lambda > m$ by:
\begin{equation}
\tilde{F}_{\pm }\left( \lambda \right) =\omega _{d-1}\int_{v_{\pm}}^{+\infty
}\Delta f_{\pm }(v)\quad Q_{\lambda }^{(d)}\left( \cosh \!v\right) \ \left( \sinh
\!v\right) ^{d-2}dv,  
{\rm \tag{4.55}} 
\end{equation}
where these integrals are understood as the action of the distributions $%
\Delta f_{\pm }$ on the (admissible) test-functions $Q_{\lambda
}^{(d)}\left( \cosh \!v\right) \left( \sinh \!v\right) ^{d-2}.$ $\tilde{F}%
_{\pm }\left( \lambda \right) $ are holomorphic in ${\Bbb C}_{+}^{(m)}$ and
satisfy uniform bounds of the following form (for all $\varepsilon,
\varepsilon^{\prime} > 0$):
\begin{equation}
\left| \tilde{F}_{\pm }(\lambda )\right| \leqslant C_{\pm
}^{(\varepsilon ,\varepsilon ^{\prime })}\left| \lambda-m  \right| ^{\beta -%
\frac{d-2}{2}+\varepsilon ^{\prime }}{\rm e}^{-\left[ \func{Re}\lambda
-\left( m+\varepsilon \right) \right] v_{\pm }}  \tag{4.56}
\end{equation}
in all the corresponding half-planes ${\Bbb C}_{+}^{\left( m+\varepsilon
\right) }.$

\noindent ii)\quad The Laplace-transform $\tilde{F}={\cal L}_{d}\left( {\cal %
F}\right) $ of ${\cal F}$ is defined as $\tilde{F}={\cal L}\left( \hat{f}%
\right) ,$ where 
$\hat{f}\left( \theta \right) =\omega _{d-2}e^{{\rm i}\left( \frac{d-2}{2}%
\right) \theta }( {\cal A}_{d}^{\left( c\right) }f) \left( \theta
\right) $ 
and ${\cal L}$ is the Fourier-Laplace transformation (4.5); 
$( {\cal A}_{d}^{\left( c\right) }f) \left(
\theta \right) $ is defined by means of formulae (4.42), (4.43), (4.44)
involving a decomposition $f=f_{+}+f_{-}$ of $f$ into ``single-cut
functions'' $f_{+},f_{-}.$ This transform $\tilde{F}\left( \lambda \right) $
is holomorphic in ${\Bbb C}_{+}^{(m)}$ and satisfies the following
properties:

a)
\begin{equation}
\tilde{F}\left( \lambda \right) =\tilde{F}_{+}\left( \lambda \right) +{\rm e}%
^{i\pi \lambda }\tilde{F}_{-}\left( \lambda \right)  \tag{4.57}
\end{equation}

b) for all integers $\ell $ such that $\ell >m,$ the Legendre coefficients $%
f_{\ell }$ of ${\cal F}_{\left| {\Bbb S}_{d-1}\right. },$ defined by
Eq.(4.29), are given by the following (Froissart--Gribov-type) relations:

\begin{equation}
f_{\ell }=\tilde{F}\left( \ell \right) =\tilde{F}_{+}\left( \ell \right)
+\left( -1\right) ^{\ell }\tilde{F}_{-}\left( \ell \right)  \tag{4.58}
\end{equation}

c) 
\begin{equation}
\tilde{F}(\lambda )=\ (-{\rm i})^{d-2}\omega _{d-1}\int_{\gamma} 
f(\theta )\ Q_{\lambda }^{(d)}(\cos \theta )\ (\sin \theta )^{d-2}d\theta 
\tag{4.59}
\end{equation}

\rm
\end{theorem}
\noindent
{\sl Proof:} 

As in Theorem 2, the validity of bounds of the form (4.54') (resp (4.54)) is
equivalent to the existence of distribution boundary values and  
discontinuities on $\Sigma_{\pm }^{(c)}$ (resp. $\sigma _{\pm }).$ The
corresponding discontinuities $\Delta f_{\pm },$ now defined as
distributions with support $\sigma _{\pm },$ can still be used for
introducing a decomposition $f=f_{+}+f_{-}$ of $f$ into ``single-cut
functions'' $f_{\pm }$ by means of Cauchy integrals in the $\cos \theta $%
-plane; in the latter, the ``weights'' ${\Delta \underline{f}_{\pm
}(\cosh v)}/ {(\cosh v)^{{\rm E}(m) +1}}$ act as distributions on the Cauchy
kernel considered as a test-function. The
expressions (4.43), (4.44) of $\left( {\cal A}_{d\pm }^{\left( c\right)
}f_{\pm }\right) \left( \theta \right) $ then remain well-defined in the
corresponding cut-planes ${\Bbb C}\backslash \sigma _{\pm }$ and we can
introduce the functions 
\begin{equation}
{\hat f}_{\pm }\left( \theta \right) =\omega
_{d-2}e^{{\rm i}\left( \frac{d-2}{2}\right)\theta }\left( {\cal A}%
_{d\pm}^{\left( c\right) }f_{\pm }\right) \left( \theta \right)  
\tag{4.60}
\end{equation}
which allow
us to write $\tilde{F}={\cal L}\left( \hat {f}\right) ,$ with $\hat{f}=\hat{%
f}_{+}+\hat{f}_{-}.\ $ The functions $\hat{f}_{+},\hat{f}_{-}$ and $\hat{f}$ are 2$\pi $%
-periodic and holomorphic in $\Pi ^{+}$ 
and we claim that the assumed bounds (4.54) on $f$ (with $m > -1$ and $\beta \ge 0 $)  
imply that $\hat f$ satisfies the assumptions of Theorem 2, namely  bounds of the form (4.11)
with the same value of $m$ (although not the same value of $\beta$). 
This fact is fully justified in the detailed analysis given below for proving 
the bounds (4.56);
it relies on the interpretation of the transformation ${\cal A}%
_{d}^{(c)}$ as {\it a primitive of non-integral order with respect to the
variable }$\cos \theta $ (see Appendix B, Proposition B.6). 
The conclusions of Theorem 2 then imply the expression (4.57) of $\tilde{%
F}\left( \lambda \right) ,$ with
\begin{equation}
\tilde{F}_{\pm }\left( \lambda \right) =\int_{v_{_{\pm }}}^{+\infty }\Delta 
\hat{f}_{\pm }(v){\rm e}^{-\lambda v}dv,  \tag{4.61}
\end{equation}
these integrals being (if necessary) understood as the action of the
distributions $\Delta \hat{f}_{\pm }$ on the exponential function ${\rm e}%
^{-\lambda v}$ (as specified in the proof of Theorem 2). The proof of
Eq.(4.58) has already been given above in full generality (see the
computation after (4.46)) which is independent of the continuous or
distribution-like character of the boundary values of $f.$

\vskip 0.3cm
\noindent
{\sl Proof of the bounds (4.56):} 
One applies the results of Proposition B.6 with the following specifications.
The holomorphic function $f(\theta)$ of Proposition B.6 plays the role 
respectively of $f_+(\theta)$ and $f_-(\theta + \pi)$. One then considers
the function ${\hat f}_m^{(\alpha)}$ studied in Proposition B.6 for the value
$\alpha = \frac {d-2}{2}.$ Then, in view of Eq.(4.60),  
${\hat f}_m^{(\alpha)}(\theta)$ coincides respectively (up to a constant 
factor) with ${\hat f}_{m+}(\theta) = {\rm e}^{{\rm i}m\theta} {\hat f}_+(\theta)$ and 
${\hat f}_{m-}(\theta) = {\rm e}^{{\rm i}m(\theta+\pi)} {\hat f}_-(\theta+ \pi)$  
and the corresponding Laplace transforms 
$\tilde{F}_{m{\pm}}( \lambda)  $ of ${\hat f}_{m\pm}(\theta)$    
are such that
$\tilde{F}_{\pm}( \lambda)  
=\tilde{F}_{m{\pm}}( \lambda-m). $   
According to the results of Proposition B.6, one is then led to
distinguish three cases: 

a) $\beta >\frac{d-2}{2}:$ in this case, 
${\hat f}_{m+} $ and ${\hat f}_{m-}$ 
belong to the class ${\cal O}^{\beta - \frac {d-2}{2}}(B_{\pi}^{({\rm cut})}).$ 
(Note that ${\hat f} ={\hat f}_+ +{\hat f}_-$ then
satisfies uniform bounds of the form
$\left| \hat{f}\left( u+{\rm i}v\right) \right| \leqslant C\eta ^{-\left(
\beta -\frac{d-2}{2}\right) }{\rm e}^{mv} $  
in all the corresponding subsets $\Pi _{\eta }^{+}\left( \eta >0\right) $ of 
$\Pi ^{+}).$

As in Theorem 2 i), 
the corresponding majorization (4.56) of $\tilde{F}_{\pm}\left( \lambda
\right) $    
(namely of $\tilde{F}_{m{\pm}}( \lambda-m) $)   
then follows from Proposition B.4 iii), formula (B.19),  with (in the present case)   
$\beta$ replaced by $\beta -\frac{d-2}{2}.$ 

b) $\beta =\frac{d-2}{2}:$  the functions  
${\hat f}_{m+} $ and ${\hat f}_{m-}$ 
belong to ${\cal O}^{0*}(B_{\pi}^{({\rm cut})}),$ 
(which then implies that
${\hat f} ={\hat f}_+ +{\hat f}_-$ 
is bounded by $C\left| \log \eta \right| {\rm e}^{mv}$ in each $\Pi
_{\eta }^{+}>0 $). Proposition B.4 iii) still applies and yields again 
the corresponding majorization (4.56) (involving the power 
$\left| \lambda-m \right| ^{\varepsilon ^{\prime }})$. 

c) $\beta <\frac{d-2}{2}:$ in this case, ${\hat f}_{m{\pm}}( \theta ) $
admit continuous boundary values; more precisely, Proposition
B.5 shows that ${\hat f}_{m{\pm}}( \theta ) $   
belongs to the class ${\cal O}_{\frac {d-2}{2}- \beta}(B_{\pi}^{({\rm cut})});$ 
therefore, in view of Proposition B.3, 
${\tilde F}_{\pm}( \lambda) $ again satisfy the bound (4.56). 

\begin{figure}
\epsfxsize=8truecm
{\centerline{\epsfbox{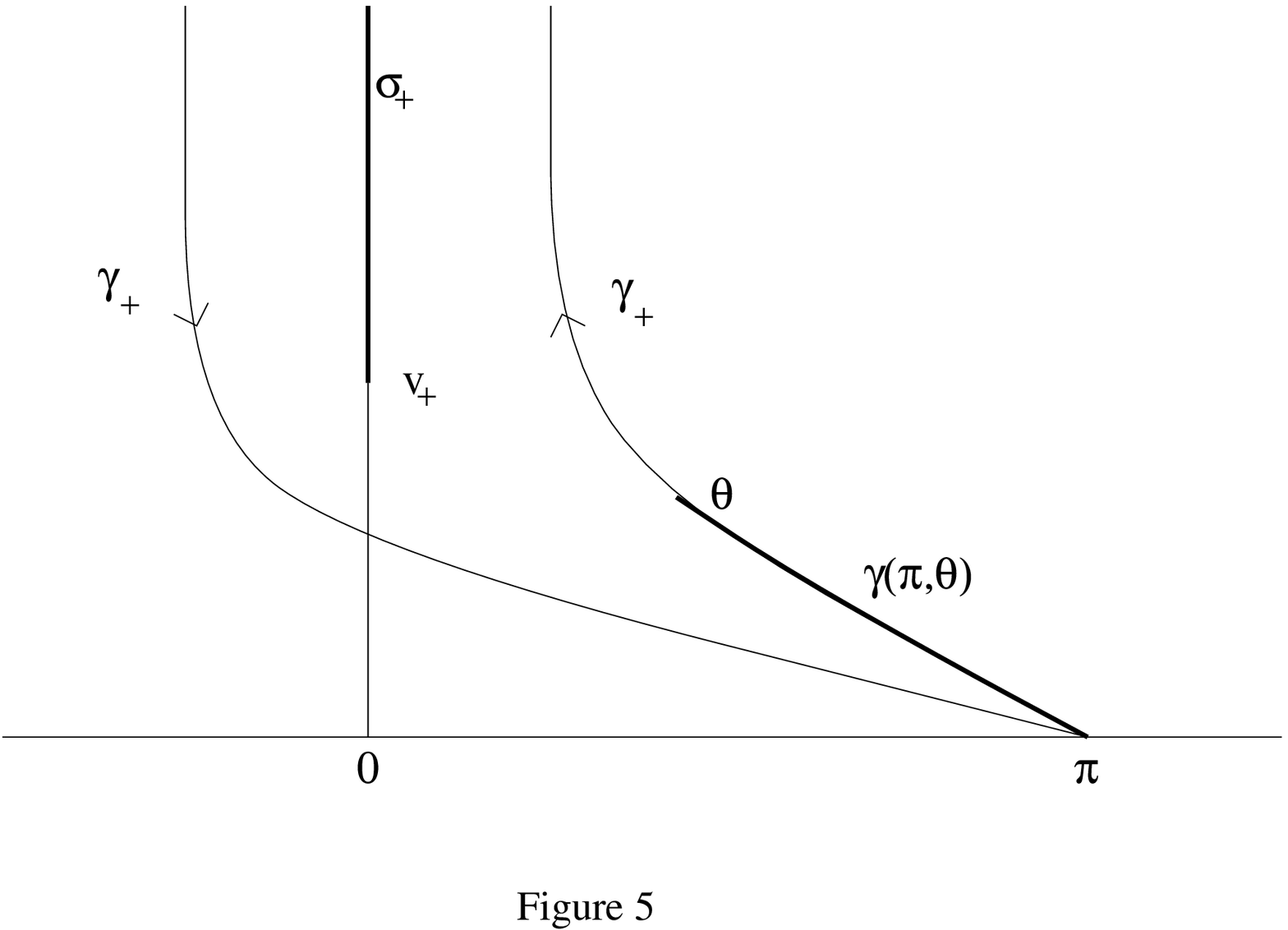}}} 
\end{figure}

It remains to show that the expressions (4.61) of $\tilde{F}_{\pm }(\lambda
) $ imply the corresponding alternative form (4.55). Considering $\tilde{F}_ 
{+}(\lambda ),$ one rewrites (4.61) (as in Theorem 2) as

\begin{eqnarray}
\tilde{F}_{+}(\lambda ) &=&\int_{\gamma _{+}}{\rm e}^{{\rm i}\lambda \theta }%
\hat{f}\left( \theta \right) d\theta =\int_{\gamma _{+}}{\rm e}^{{\rm i}%
\lambda \theta }\hat{f}_{+}\left( \theta \right) d\theta ,{\rm \ i.e.} 
\nonumber \\
\tilde{F}_{+}(\lambda ) &=&\omega _{d-2}\int_{\gamma _{+}}{\rm e}^{{\rm i}%
\left( \lambda +\frac{d-2}{2}\right) \theta }\left( {\cal A}%
_{d+}^{(c)}f_{+}\right) \left( \theta \right) d\theta  \tag{4.62}
\end{eqnarray}
with $\left( {\cal A}_{d+}^{(c)}f_{+}\right) \left( \theta \right) $
expressed by Eq.(4.43). 
By choosing $\gamma _{+}$ and $\gamma \left( \pi,  
\theta \right) $ such that, for all $\theta ,$ supp. $\gamma \left( \pi,  
\theta \right) \subset $ supp $\gamma _{+}$ (as e.g. in Fig.5), one can
treat the resulting expression for $\tilde{F}_{+}(\lambda )$ as a double
integral (convergent for $\lambda $ in ${\Bbb C}_{+}^{\left( m\right) })$ in
which the order of the integrations can be inverted.
This yields:
\begin{equation}
\tilde{F}_{+}(\lambda )=\ (-{\rm i})^{d-2}\omega _{d-1}\int_{\gamma
_{+}}f_{+}(\tau )\ Q_{\lambda }^{(d)}(\cos \tau )\ (\sin \tau )^{d-2}d\tau 
\tag{4.63}
\end{equation}

Now, by the very definition of boundary values of holomorphic functions in
the sense of distributions, the expression (4.63) of $\tilde{F}_{+}(\lambda
) $ can be rewritten in the distribution form (4.55) (by flattening the
folded contour $\gamma _{+}$ onto the cut $\sigma _{+}).$ A similar argument
holds for $\tilde{F}_{-}(\lambda ).$   
Moreover, by plugging the expression (4.63) of  
$\tilde{F}_{+}(\lambda)$ and the analogous one for  
$\tilde{F}_{-}(\lambda)$ into Eq.(4.57) and then noticing that the corresponding  
integration paths $\gamma_+$ and $\gamma_-$ can be replaced by $\gamma$, one obtains 
the expression (4.59) of ${\tilde F}(\lambda) $ in terms of $f = f_+ + f_-$.

\vskip 0.6cm
As in the case $d=2$,  one still defines
the quantities $\tilde{F}^{\left( s\right) }(\lambda ),\tilde{F}^{\left(
a\right) }(\lambda )$ by formula (4.22): they are respectively  
the Laplace transforms on $X_{d-1}$ of the distributions $\Delta
f^{\left( s\right) },\Delta f^{\left( a\right) }$ defined (on $
X_{d-1}$) by Eqs.(4.21). 
One can thus complete the second part of 
Theorem 3 by the 

\vskip 0.5cm
\noindent
{\sl Proposition 7bis:} 

\it The statement of Proposition 7 is valid without modification 
in the $d-$ dimensional case ($d \ge 3$) 
apart from the bounds on $\tilde{F}^{\left( s\right) }(\lambda ),\tilde{F}^{\left(
a\right) }(\lambda )$ which are now given by the r.h.s. of (4.56). 
\rm

\vskip 3cm

\noindent
{\sl Inversion formulas:}   

\vskip 0.1cm

We shall give formulas which express ${\cal F}\left( z\right) =f\left(
\theta \right) $ and its discontinuities $\left( \Delta {\cal F}\right)
_{\pm }\left( z\right) =\left( \Delta f\right) _{\pm }\left( v\right) $ in
terms of the Laplace transforms $\tilde{F}_{\pm }\left( \lambda \right) $ of
the latter. The formulas exactly parallel the inversion formulas (4.25),
(4.26), (4.27) of the two-dimensional case; they only differ from the latter
by the fact that the trigonometric kernel $\cos \lambda \theta $ is replaced
by $h_{d}\left( \lambda \right) P_{\lambda }^{\left( d\right) }\left( \cos
\theta \right) ,$ where $P_{\lambda }^{\left( d\right) }$ is the $d$%
-dimensional first-kind Legendre function
\begin{equation}
P_{\lambda }^{\left( d\right) }\left( \cos \theta \right) =2\frac{\omega
_{d-2}}{\omega _{d-1}}\left( \sin \theta \right) ^{-\left( d-3\right)
}\int_{0}^{\theta }\cos [ ( \lambda +{(d-2)/ {2}}) \tau ] 
[ 2( \cos \tau -\cos \theta ) ] ^{\frac{d-4}{2}%
}d\tau   \tag{4.64}
\end{equation}
and
\begin{equation}
h_{d}\left( \lambda \right) =\frac{\left( 2\lambda +d-2\right) }{\left(
d-2\right) !}. \frac{\Gamma \left( \lambda +d-2\right) }{\Gamma \left(
\lambda +1\right) }.  \tag{4.65}
\end{equation}
$P_{\lambda }^{\left( d\right) }\left( \cos \theta
\right) $ is defined as a holomorphic function in the cut plane ${\Bbb C}%
\backslash \left] -\infty ,-1\right] .$ 

The following formula is shown to hold in the open set  
$0<\left| \func{Re}\theta \right| <\pi :$
with the specification
$\varepsilon _{\theta }={\rm sgn}\left( \func{Re}\theta \right) $ 
\begin{eqnarray}
{\cal F}\left( z\right) &\equiv &f\left( \theta \right) =-\frac{1}{2\omega
_{d}}\int_{-\infty }^{+\infty }\frac{\tilde{F}_{+}\left( m+{\rm i}\nu
\right) h_{d}\left( m+{\rm i}\nu \right) P_{m+{\rm i}\nu }^{\left( d\right)
}\left( \cos \theta -\varepsilon _{\theta }\pi \right) }{\sin \pi \left( m+%
{\rm i}\nu \right) }d\nu  \nonumber \\
&&-\frac{1}{2\omega _{d}}\int_{-\infty }^{+\infty }\frac{\tilde{F}%
_{-}\left( m+{\rm i}\nu \right) h_{d}\left( m+{\rm i}\nu \right) P_{m+{\rm i}%
\nu }^{\left( d\right) }\left( \cos \theta \right) }{\sin \pi \left( m+{\rm i%
}\nu \right) }d\nu  \nonumber \\
&&+\frac{1}{\omega _{d}}\sum_{0\leqslant \ell <m}f_{\ell }h_{d}\left( \ell
\right) P_{\ell }^{\left( d\right) }\left( \cos \theta \right) .  \tag{4.66}
\end{eqnarray}

As in Eq.(4.26), the first term at the r.h.s. of (4.66) defines a pair of
holomorphic functions in the respective strips $0<u<2\pi $ and $-2\pi <u<0$
(corresponding to the choice $\varepsilon _{\theta }=+$ or $-$ in the
argument of the cosine), while the second term defines a holomorphic
function in the strip $-\pi <u<\pi :$ this follows from the bounds (4.57) on 
$\tilde{F}_{\pm }\left( \lambda \right) $ and the power behaviour of $P_{m+%
{\rm i}\nu }^{\left( d\right) }\left( \cos \theta \right) $ as $\left| \nu
\right| ^{\frac{d-1}{2}}$ (easily derived from the representation (4.64) of $%
P_{\lambda }^{\left( d\right) }).$

The restriction of ${\cal F}$ to the sphere ${\Bbb S}_{d-1},$ namely ${\cal F}%
_{\mid {\Bbb S}_{d-1}}\left( z\right) =f_{\mid {\Bbb R}}\left( \theta
\right), $ is also expressed by the generalized Legendre (or
``partial-wave'') expansion:
\begin{equation}
f_{\mid {\Bbb R}}\left( \theta \right) =\frac{1}{\omega _{d}}\sum_{\ell \in 
{\Bbb N}}f_{\ell }h_{d}\left( \ell \right) P_{\ell }^{\left( d\right)
}\left( \cos \theta \right)  \tag{4.67}
\end{equation}
Finally, the discontinuities $\Delta f_{+}\left( v\right) $ and $\Delta
f_{-}\left( v\right) $ of $f$ across the cuts $\sigma _{+}$ and $\sigma _{-}$
are given by the following (identical) formulas:
\begin{equation}
\left( \Delta f\right) _{\pm }\left( v \right) =\frac{1}{\omega _{d}}%
\int_{-\infty }^{+\infty }\tilde{F}_{\pm }\left( m+{\rm i}\nu \right) 
\ h_d(m+{\rm i}\nu)\ P_{m+%
{\rm i}\nu }^{\left( d\right) }\left( \cosh \!v\right) d\nu  \tag{4.68}
\end{equation}
which (in view of the polynomial increase in $\nu $ of all factors of the
integrand) must be understood in the sense of distributions according to  
Appendix B.

\vskip 0.2cm
All these formulas have been established in [25c)] (under assumptions of
continuity for the boundary values of ${\cal F})$ in the case where a single
cut, namely $\sigma _{+},$ is present. The proof given in [25c)] applies
equally well to the derivation of Eq.(4.66) under the present assumptions;
however, for tutorial reasons, we will sketch the derivation of this result
which relies on the inversion of the two transformations (4.41) and (4.42),\ldots 
(4.44). We must treat separately the cases of even and odd dimensions $d.$

\noindent 
a) $d${\it \ even }$\left( d\geqslant 4\right) $: the Abel-type
transformations (4.43), (4.44) can be inverted as follows:

\begin{equation}
f_{\pm }\left( \theta \right) =\omega _{d-2}\left( -\frac{1}{2\pi }\frac{1}{%
\sin \theta }\frac{d}{d\theta }\right) ^{\frac{d-2}{2}}\left[ \left( {\cal A}%
_{d\pm }^{\left( c\right) }f_{\pm }\right) \left( \theta \right) \right]\quad 
\tag{4.69}
\end{equation}
and the inversion of the Fourier integrals over $\gamma _{\pm }$ in (4.45)
yields (by taking into account that $f_{\ell }=f_{\ell +}+f_{\ell -}):$
\begin{eqnarray}
\omega _{d-2}\left( {\cal A}_{d}^{\left( c\right) }f\right) \left( \theta
\right) &=&\omega _{d-2}\left( {\cal A}_{d+}^{\left( c\right) }f_{+}\right)
\left( \theta \right) +\omega _{d-2}\left( {\cal A}_{d-}^{\left( c\right)
}f_{-}\right) \left( \theta \right)  \nonumber \\
&=&\frac{\left( -1\right) ^{\frac{d}{2}}}{2\pi }\int_{-\infty }^{\infty }%
\frac{\tilde{F}_{+}\left( m+{\rm i}\nu \right) \cos \left[ \left( m+{\rm i}%
\nu +\frac{d-2}{2}\right) \left( \theta -\varepsilon _{\theta }\pi \right)
\right] }{\sin \pi \left( m+{\rm i}\nu \right) }d\nu  \nonumber \\
&&-\int_{-\infty }^{\infty }\frac{\tilde{F}_{-}\left( m+{\rm i}\nu \right)
\cos \left( m+{\rm i}\nu +\frac{d-2}{2}\right) \theta }{\sin \pi \left( m+%
{\rm i}\nu \right) }d\nu +  \nonumber \\
&&+\sum_{-m-d+2<\ell <m}f_{\ell }\cos ( \ell +{(d-2)/{2}})
\theta \quad \tag{4.70}
\end{eqnarray}
By applying the differential operator at the r.h.s. of (4.69) to both sides
of Eq.(4.70), we then directly obtain Eq.(4.66), thanks to the integral
representation (see [25b)] Eq.(II.85)):
\begin{equation}
h_{d}\left( \lambda \right) P_{\lambda }^{\left( d\right) }\left( \cos
\theta \right) =\frac{2\omega _{d}}{\left( 2\pi \right) ^{d/2}}\left( -\frac{%
1}{\sin \theta }\frac{d}{d\theta }\right) ^{\frac{d-2}{2}} [ \cos 
(\lambda + {(d-2)/2}) \theta ] \quad \tag{4.71}
\end{equation}

\vskip 1cm

\noindent
b) $d$ {\it odd} $\left( d\geqslant 3\right) $: the Abel
inversion formulae (4.69) are replaced by

\begin{eqnarray}
f_{+}\left( \theta \right) &=&-2\omega _{d-2}\left( -\frac{1}{2\pi }\frac{1}{%
\sin \theta }\frac{d}{d\theta }\right) ^{\frac{d-1}{2}}\int_{\varepsilon
_{\theta }\pi }^{\theta }\left( {\cal A}_{d+}^{\left( c\right) }f_{+}\right)
\left( \tau \right) \left[ 2\left( \cos \theta -\cos \tau \right) \right]
^{-1/2}\sin \tau d\tau  \nonumber \\
&&\;  \tag{4.72} \\
f_{-}\left( \theta \right) &=&-2\omega _{d-2}\left( -\frac{1}{2\pi }\frac{1}{%
\sin \theta }\frac{d}{d\theta}\right) ^{\frac{d-1}{2}}\int_{0}^{\theta }\left( 
{\cal A}_{d-}^{\left( c\right) }f_{-}\right) \left( \tau \right) \left[
2\left( \cos \theta -\cos \tau \right) \right] ^{-1/2}\sin \tau d\tau 
\nonumber \\
&&\;  \tag{4.73}
\end{eqnarray}
and correspondingly the inversion of the Fourier integrals in (4.45) yields
the following expressions for ${\cal A}_{d\pm }^{\left( c\right) }f_{\pm }$
in the open set $\left\{ \theta ;0<\left| \func{Re}\theta \right| <\pi
\right\} :$
\[
\omega _{d-2}\left( {\cal A}_{d+}^{\left( c\right) }f_{+}\right) \left(
\theta \right)\  =  
\]
\begin{eqnarray}
\frac{{\rm i} \varepsilon_{\theta}\ (-1)^
{\frac{d-1}{2}}}{2\pi }\   
\left\{ -\int_{-\infty }^{\infty }\frac{F_{+}\left( m+{\rm i}\nu \right)
\sin \left[ \left( m+{\rm i}\nu +\frac{d-2}{2}\right) \left( \theta
-\varepsilon _{\theta }\pi \right) \right] }{\sin \pi \left( m+{\rm i}\nu
\right) }d\nu 
\right.  \nonumber \\
\left. +\ \sum_{-m-d+2< \ell <m}f_{\ell +}\sin [ ( \ell +
(d-2)/{2}) \theta -\varepsilon_{\theta }( (d-2)/{2}) \pi] 
\right\}\quad   
\tag{4.74}
\end{eqnarray}
\begin{eqnarray}
\omega _{d-2}\left( {\cal A}_{d-}^{\left( c\right) }f_{-}\right) \left(
\theta \right) =
\frac{{\rm i}}{2\pi }\left\{ -\int_{-\infty }^{\infty }%
\frac{\tilde{F}_{-}\left( m+{\rm i}\nu \right) \sin \left( m+{\rm i}\nu +%
\frac{d-2}{2}\right) \theta }{\sin \pi \left( m+{\rm i}\nu \right) }d\nu
\right.  \nonumber \\
\left. +\sum_{-m-d+2<\ell <m}f_{\ell -}\sin ( \ell +(d-2)/{2}) 
\theta \right\}\quad   \tag{4.75}
\end{eqnarray}
By applying the integro-differential operator at the r.h.s. of (4.72)
(resp.(4.73)) to both sides of Eq.(4.74) (resp.(4.75), we then directly
obtain Eq.(4.66), thanks to the integral representation (see [25b)]
Eq.(II-86)):
\begin{equation}
h_d\left( \lambda \right) P_{\lambda }^{\left( d\right) }\left( \cos \theta
\right) =\frac{4{\rm i}\omega _{d}}{\left( 2\pi \right) ^{\frac{d+1}{2}}}%
\left( \frac{-1}{\sin \theta }\;\frac{d}{d\theta }\right) ^{\frac{d-1}{2}%
}\int_{0}^{\theta }\frac{\sin \left( \lambda +\frac{d-2}{2}\right) \tau \sin
\tau }{\left[ 2\left( \cos \theta -\cos \tau \right) \right] ^{1/2}}d\tau 
\tag{4.76}
\end{equation}

For $\theta =u$ real (i.e. $z\in {\Bbb S}_{d-1}),$ formula (4.66) can be
seen to reduce to the expansion (4.67) by using the same contour distortion
argument [30,31] in the $\lambda $-plane as in the case $d=2$ (see Eqs.(4.26),
(4.27)); this can be done in two equivalent ways:

i) by proceeding directly with the r.h.s. of (4.66) thanks to the
properties of $P_{\lambda }^{\left( d\right) }\left( \cos \theta \right) $ 
{\it as a holomorphic function of }$\lambda $ in the right-hand plane.

ii) by proceeding with the r.h.s. of (4.70) (resp. (4.74), (4.75))
which only involves trigonometric functions and then applying the inverse
Abel operator of Eq.(4.69) (resp. (4.72), (4.73)) which will restore the
ultraspherical polynomials $h_{d}\left( \ell \right) P_{\ell }^{\left(
d\right) }\left( \cos \theta \right) $ term by term in the expansion.

Finally, the formulas (4.68) for the discontinuities $\Delta f_{\pm }$
are obtainable from (4.66) either directly (thanks to relevant
discontinuity formulas for the $P_{\lambda }^{\left( d\right) }\left( \cos
\theta \right) $ on $\left] -\infty ,-1\right] )$ or by computing the
corresponding discontinuities of $\left( {\cal A}_{d\pm }^{\left( c\right)
}f_{\pm }\right) \left( \theta \right) $ from their representations (4.70)
(resp. (4.74), (4.75)) and then applying the inverse Abel operators of
Eq.(4.69) (resp. (4.72) (4.73)). The case of
distribution-like discontinuities requires a suitable regularization
corresponding to the application of a ``cut-off'' to $\tilde{F}_{\pm }\left(
m+{\rm i}\nu \right) .$

\vskip 0.3cm
\noindent {\sl Remarks:}

\it i)\ If Eq.(4.66) is used for $m$ integer, its $r.h.s.$  
must be understood as the action of the 
distribution $\lim_{\varepsilon \to 0, \varepsilon>0} \frac {1}{\sin \pi (m-\varepsilon + {\rm i}\nu)}$ 
on the numerator of the integrand.

ii)\ We refer to Theorem 4 of [25c)] for the possible use of formula (4.66) in a precise range  
of negative values of $m$. However, it is only for $m > -1$ that Eq.(4.66) is exactly the inverse 
of the transformation (4.55) defined under the assumptions of Theorem 3 (for $m \le -1$, this 
transformation can generate poles of $\tilde F(\lambda)$ located as those of $Q_{\lambda}$ at all the 
negative integers).
\rm 

\vskip 0.3cm
The following complement of Theorem 3 which emphasizes the reciprocal 
property of the transformation follows from 
the previous study of the inversion used conjointly with the principle of uniqueness of 
analytic continuation (it is the adaptation of Theorem 3 of [25c)] to the case with two cuts) 

\begin{theorem}
Let ${\bf K}(z,z')$ be an $SO(d)-$invariant kernel on the sphere ${\Bbb S}_{d-1}$ 
with set of Legendre coefficients $f_{\ell}$ (defined by Eq.(4.29) in terms of the function  
${\bf f}(\theta) ={\bf F}(z)= {\bf K}(z,z_0)$,\   $z.z_0 = - \cos \theta$).
Let us then assume that the sets of even and odd coefficients $f_{\ell}$ 
admit respectively analytic interpolations $\tilde F^{(s)}(\lambda)$ and $\tilde F^{(a)}(\lambda)$ 
in ${\Bbb C}^{(m)}_{+}$ satisfying uniform bounds of the form (4.56) with $m > -1.$  

Then there exists an invariant perikernel ${\cal K}(z,z')$ of moderate growth on $X_{d-1}^{(c)}$ 
represented by a holomorphic function ${\cal F}(z) = {\cal K}(z,z_0) = f(\theta)$ satisfying 
all the assumptions of Theorem 3 such that 
${\cal K}_{|{\Bbb S}_{d-1}\times {\Bbb S}_{d-1}} = {\bf K}$ and correspondingly  
$f_{|{\Bbb R}} = {\bf f}$; moreover, the functions   
$\tilde F^{(s)}(\lambda)$ and $\tilde F^{(a)}(\lambda)$ 
appear respectively as the symmetric and antisymmetric combinations of the Laplace 
transforms $\tilde F_{\pm}$ of the discontinuities $\Delta f_{\pm}$ of $f$ defined by
formula (4.55). 
\end{theorem}

\subsection{Complex angular momentum analysis of the four-point functions}
\quad Starting from the basic postulates of Q.F.T., we have established in  
Theorem 1 that the four-point function $H\left([ k]\right) $ of
any set of scalar fields enjoys a structure of {\it invariant perikernel} in
each submanifold $\hat{\Omega}_{\left( \zeta ,\zeta ^{\prime },K\right) }$
of the set ${\hat \Omega}_{K}$ associated with any space-like 
energy-momentum vector $K=\left( 0,...,0,\sqrt{-t}\right)$, with  
$t\leqslant 0.$ In particular, we have shown that the temperateness assumption 
expressed by the bounds (3.1) results in the properties of moderate growth 
(3.25), (3.26) of these perikernels:
\begin{equation}
{\cal K}_{\left( \zeta ,\zeta ^{\prime },K\right) }\left( z,z^{\prime
}\right) \equiv H\left([k]_{\left( \zeta ,\zeta ^{\prime },K\right)}  
\left(z,z'\right)\right)   
= {\underline H}_{(\zeta,\zeta',K)}(\cos \Theta_t),  
\tag{4.77}
\end{equation}
\[
{\rm with}\ 
\cos \Theta _{t}= -z.z^{\prime}.  
\]
One can therefore apply the results of Theorem 3 to the latter,  
for which 
the notations of \S 4.2 and  
the identification (4.2) 
can also be used:  
\begin{equation}
{\underline H}_{(\zeta,\zeta',K)}(\cos \Theta_t) 
= f_{\left( \zeta ,\zeta ^{\prime },K\right) }(\Theta_t) 
=F\left( \zeta ,\zeta ^{\prime };t,\cos \Theta _{t}\right)  
\tag{4.78}
\end{equation}

As in \S 4.2 (for $d > 2$) or \S 4.1 (for $d=2$), 
we introduce the discontinuities $\left( \Delta f_{\left(
\zeta ,\zeta ^{\prime },K\right) }\right) _{\pm }\left( v\right) $ of the
function ${\rm i}f_{\left( \zeta ,\zeta ^{\prime },K\right) }$ across the
respective cuts $\sigma _{+}\left( v_{s}\right) ,\sigma _{-}\left(
v_{u}\right) $ with thresholds $v_{s}=v_{s}\left( \zeta ,\zeta ^{\prime
},t\right) ,v_{u}=v_{u}\left( \zeta ,\zeta ^{\prime },t\right) $ given by
Eqs.(2.23), (2.24). These discontinuities are interpreted as the $s$- and $u$%
-channel ``absorptive parts'' of $F$ in the submanifold $\hat{\Omega}%
_{\left( \zeta ,\zeta ^{\prime },K\right) }:$
\begin{equation}
\left( \Delta f_{\left( \zeta ,\zeta ^{\prime },K\right) }\right) _{+}\left(
v\right) =\Delta _{s}F\left( \zeta ,\zeta ^{\prime };t,\cosh \!v\right) 
\tag{4.79}  
\end{equation}
\begin{equation}
\left( \Delta f_{\left( \zeta ,\zeta ^{\prime },K\right) }\right) _{-}\left(
v\right) =\Delta _{u}F\left( \zeta ,\zeta ^{\prime };t,\cosh \!v\right) 
\tag{4.80}
\end{equation}

In view of the two possible dimensions of the manifolds $\hat \Omega_{(\zeta,
\zeta^{\prime},K)}$, namely $2(d-1)$ for $K \neq 0$ and $2d$ for $K=0$
(see \S 2.2 and Theorem 1), the application of Theorem 3 
(resp. Theorem 2 for the case $d=2$) allows one to define 
correspondingly two different Laplace transforms of $H([k])$.
However, by considering the case $K \neq 0$ (i.e. $t <0$), one obtains
\it the generic complex angular momentum analysis of \rm $H([k])$
whose results are specified in the following theorem; the peculiarities of
the case $K=0$ will be briefly commented at the end.

\begin{theorem}
Let $F\left( \zeta ,\zeta ^{\prime };t,\cos \Theta _{t}\right) \equiv
H\left( \left[ k\right] \right) $ be any four-point function of local scalar
fields satisfying bounds of the form (3.25), (3.26) in each section of maximal analyticity  
(or cut-submanifold) $\hat{\Omega}_{\left( \zeta ,\zeta ^{\prime },K\right) }^{\left( {\rm %
cut}\right) },$ with $K^2 = t <0$, $(\zeta,\zeta^{\prime}) \in
\Delta_t \times \Delta_t .$ Then there exists a function $\tilde F(\zeta,\zeta^{\prime};t,
\lambda_t)$ which is holomorphic 
with respect to $\lambda_t $ in ${\Bbb C}_{+}^{\left( m_* \right)
}$ and satisfies the following properties: 

a) 
\begin{equation}
\tilde{F}=\tilde{F}_{s}+e^{{\rm i}\pi \lambda_t }\tilde{F}_{u},  \tag{4.81}
\end{equation}
where, in the general case $ d> 2$:   
\begin{eqnarray}
\tilde{F}_{s}\left( \zeta ,\zeta ^{\prime };t,\lambda_t \right)  &=&\omega
_{d-1}\int_{v_{s}\left( \zeta ,\zeta ^{\prime },t\right) }^{+\infty }\Delta
_{s}F\left( \zeta ,\zeta ^{\prime };t,\cosh \!v\right)   \nonumber \\
&&Q_{\lambda_t }^{\left( d\right) }\left( \cosh \!v\right) \left( \sinh
\!v\right) ^{d-2}dv,  \tag{4.82}
\end{eqnarray}
\begin{eqnarray}
\tilde{F}_{u}\left( \zeta ,\zeta ^{\prime };t,\lambda_t \right)  &=&\omega
_{d-1}\int_{v_{u}\left( \zeta ,\zeta ^{\prime },t\right) }^{+\infty }\Delta
_{u}F\left( \zeta ,\zeta ^{\prime };t,\cosh \!v\right)   \nonumber \\
&&Q_{\lambda_t }^{\left( d\right) }\left( \cosh \!v\right) \left( \sinh
\!v\right) ^{d-2}dv . \tag{4.83}
\end{eqnarray}

b) $\tilde{F}_{s}$ and $\tilde{F}_{u}$ are holomorphic functions of $\lambda_t 
$ in ${\Bbb C}_{+}^{\left( m_*\right) }$ which satisfy uniform bounds of the
following form: 
\begin{equation}
\left| \tilde{F}_{s,u}\left( \zeta ,\zeta ^{\prime };t,\lambda_t \right)
\right| \leqslant C_{s,u}^{\varepsilon ,\varepsilon ^{\prime }}\left|
\lambda_t -m_* \right| ^{n -\frac{d-2}{2}+\varepsilon ^{\prime }}e^{-\left[ 
\func{Re}\lambda_t -\left( m_*+\varepsilon \right) \right] v_{s,u}}  \tag{4.84}
\end{equation}
in the corresponding half-planes ${\Bbb C}_{+}^{\left( m_* +\varepsilon \right)}.$ 

c) for $\ell >m_*,$ the off-shell partial-wave functions $f_{\ell }\left(
\zeta ,\zeta ^{\prime },t\right) $ of $F$, defined for $\zeta ,\zeta ^{\prime }\in
\Delta _{t}\times \Delta _{t},\ t<0$ by Eq.(4.2), are given by the following
(Froissart-Gribov-type) relations: 
\begin{equation}
f_{\ell }\left( \zeta ,\zeta ^{\prime },t\right) =\tilde{F}\left( \zeta
,\zeta ^{\prime };t,\ell \right)   \tag{4.85}
\end{equation}
\quad\quad Moreover, the ``symmetric and antisymmetric Laplace transforms'' 
\footnote{ Note that when $F$ is the four-point function of a {\it single } scalar field $\Phi$ 
or of two fields $\Phi, \Phi'$ in such a way that the $t-$channel is $(\Phi,\Phi) \to 
(\Phi',\Phi')$,  one has $\Delta_s F = \Delta_u F $; in such cases ${\tilde F}^{(a)} = 0,
f_{2 \ell +1}=0,$ and only Eq.(4.87) survives.}  

\noindent $\tilde{F}^{(s)}\left( \zeta
,\zeta ^{\prime };t,\lambda_t \right)$ and    
$\tilde{F}^{(a)}\left( \zeta
,\zeta ^{\prime };t,\lambda_t \right)$  
defined by  
\begin{equation}
\tilde{F}^{\left( s\right) } =\tilde{F}_{s}
+\tilde{F}_{u} \;,\;
\tilde{F}^{\left(
a\right) } =
\tilde{F}_{s}-
\tilde{F}_{u}   \tag{4.86}
\end{equation}
are {\rm Carlsonian} interpolations in  
${\Bbb C}_{+}^{\left( m_* +\varepsilon \right)}$ for the respective sets of even and odd 
partial-waves of $F$, namely one has: 
\begin{equation}
{\rm for\;}2\ell  >m,\quad \;\quad  
f_{2 \ell }\left( \zeta ,\zeta ^{\prime },t\right) =
\tilde{F}^{(s)}\left( \zeta
,\zeta ^{\prime };t,2 \ell \right)   
\tag{4.87} 
\end{equation}
\begin{equation}
{\rm for\;}2\ell +1 >m,\;\ \ 
f_{2 \ell +1}\left( \zeta ,\zeta ^{\prime },t\right) =
\tilde{F}^{(a)}\left( \zeta
,\zeta ^{\prime };t,2 \ell +1\right).  
\tag{4.88}
\end{equation}

d) The four-point function $F$ and the absorptive parts $\Delta _{s}F,\Delta
_{u}F$ are reobtained in terms of $\tilde{F}_{s}$ and $\tilde{F}_{u}$ by the
following formulas: 
\[
F\left( \zeta ,\zeta ^{\prime };t,\cos \Theta _{t}\right) =
\]
\[
-\frac{1}{2\omega _{d}}\int_{-\infty }^{+\infty }\frac{\tilde{F}_{s}\left(
\zeta ,\zeta ^{\prime };t,m_* +{\rm i}\nu \right) 
h_{d}\left( m_* +{\rm i}\nu
\right) P_{m_*+{\rm i}\nu }^{\left( d\right) }\left( \cos \left( \Theta
_{t}-\varepsilon _{\Theta _{t}}\pi \right) \right) }{\sin \pi \left( m_*+{\rm i%
}\nu \right) }d\nu 
\]
\begin{equation}
-\frac{1}{2\omega _{d}}\int_{-\infty }^{+\infty }\frac{\tilde{F}_{u}\left(
\zeta ,\zeta ^{\prime };t,m_* +{\rm i}\nu \right) 
h_{d}\left( m_* +{\rm i}\nu
\right) P_{m_* +{\rm i}\nu }^{\left( d\right) }\left( \cos \Theta _{t}\right) }{%
\sin \pi \left( m_* +{\rm i}\nu \right) }d\nu   \tag{4.89}
\end{equation}
\[
+\frac{1}{\omega _{d}}\sum_{0\leqslant \ell <m_*}f_{\ell }\left( \zeta ,\zeta
^{\prime },t\right) h_{d}\left( \ell \right) P_{\ell }^{\left( d\right)
}\left( \cos \Theta _{t}\right) 
\]
\begin{eqnarray}
\Delta _{s,u}F\left( \zeta ,\zeta ^{\prime };t,\cosh \!v\right)  &=&\frac{1}{%
\omega _{d}}\int_{-\infty }^{+\infty }\tilde{F}_{s,u}\left( \zeta ,\zeta
^{\prime };t,m_* +{\rm i}\nu \right) h_{d}\left( m_* +{\rm i}\nu \right)  
\nonumber \\
&&P_{m_* +{\rm i}\nu }^{\left( d\right) }\left( \cosh \!v\right) d\nu . 
\tag{4.90}
\end{eqnarray}

e) When $[k]$ belongs to the Euclidean region ${\hat {\cal E}}_{K},$  
(i.e. for 
$\cos \Theta _{t} \in \left[ -1,+1\right]$), formula (4.89) can be replaced by the   
partial wave expansion of 
$H_{|{\hat {\cal E}}_{K}},$ namely: 
\begin{equation}
H([k])  =F\left( \zeta ,\zeta ^{\prime };t,\cos
\Theta _{t}\right) =\frac{1}{\omega _{d}}\sum_{0\leqslant \ell <\infty} 
f_{\ell }\left( \zeta ,\zeta ^{\prime },t\right) h_{d}\left( \ell \right)
P_{\ell }^{\left( d\right) }\left( \cos \Theta _{t}\right),  \tag{4.91} 
\end{equation}

f) In the case $d=2$, all the previous results are valid provided the following 
replacements are performed:

In Eqs (4.82),(4.83):\quad  $\omega_{d-1} Q_{\lambda_t}^{(d)}(\cosh v)$\quad  by\quad  
${\rm e}^{-\lambda_t v}$, 

In Eq.(4.89):\quad  
$\frac{1}{\omega_{d}}
h_{d}\left( m_* +{\rm i}\nu \right) 
P_{m_* +{\rm i}\nu }^{\left( d\right) }\left( \cos \Theta_{t}\right) $ \quad by 
\quad $\frac{1}{ \pi} \cos [(m+ {\rm i}\nu) \Theta_{t}]$,   

\noindent
and similar ones in Eqs (4.90),(4.91).  

\end{theorem}
The proof of properties a) and b) is a direct application of
Theorem 3, since the bounds (3.25), (3.26) established in
Sec.3 are equivalent in an obvious way to (4.54), (4.54') with the pair $(m, \beta)$  
replaced by $(m_*, n)$. Property c) follows from Eq(4.58) in Theorem 3  
completed by Proposition 7bis. Properties d) and e)
directly follow from the inversion formulas (4.66), (4.68) and (4.67) derived
after Theorem 3.
The case $d=2$ (property f)) is obtained similarly as an application of Theorem 2 
and of the inversion formulas (4.25),(4.26),(4.27).

\vskip 0.5cm
In the case $K=0$, it is possible to define \it two \rm Laplace transforms,  
namely:

\noindent
a) the function $\tilde F(\zeta,\zeta^{\prime};0,\lambda_t)$ obtained as the limit
of  $\tilde F(\zeta,\zeta^{\prime};t,\lambda_t)$ 
(for $ (\zeta,\zeta^{\prime})
\in \Delta_0 \times \Delta_0 $) when $t$ tends to zero (and satisfying exactly
all the formulas of Theorem 3, with $t=0$),

\noindent
b) the function $\tilde F_0 (\zeta,\zeta^{\prime};\lambda_t)$ obtained 
by formulas similar to Eqs (4.81),...,(4.83), with $d$ replaced by $d+1$
(a complete substitute to Theorem 5 being equally valid).

The connection between these two functions, both defined  
for $ (\zeta,\zeta^{\prime})
\in \Delta_0 \times \Delta_0 $ and $\lambda_t \in {\Bbb C}_+^{(m)}$,
and the possible exploitation of this connection for the structure of
the four-point function $H([k])$ will be treated elsewhere.

\vskip 0.3cm

\noindent
{\sl Remark}

{\it In view of the field-theoretical interpretation of $m_* = {\rm max}(m, n)$ and $n$ as {\rm degrees of 
temperateness of the four-point function}, the properties obtained in Theorem 5 
have only made use of the results of harmonic analysis of Theorem 3 for the case 
$ m \ge 0$. However it may occur that negative values of this parameter,  
corresponding to extended analyticity or meromophy properties 
of the four-point functions in the complex angular momentum plane, be of relevant use 
in a second step where     
the ``Bethe-Salpeter structure'' will be taken into account [14].}   

\subsection{Analytic continuation of Euclidean field theory and complex angular momentum analysis}   
In this last subsection, let us forget about the Minkowskian framework 
of Q.F.T. considered throughout this paper and adopt the viewpoint of Euclidean field theory. 
The starting point of such a theory is the set of $n-$point ``Schwinger functions'', considered as 
tempered distributions on the corresponding $n-$point Euclidean 
($(d+1)-$dimensional) space-time, whose Fourier transforms 
are the $n-$point Green functions taken in the Euclidean energy-momentum space.
In particular, the data which concern the four-point Green function are encoded in the properties of 
$H_{{\cal E}}$=  
$H_{|{\cal E}}$, 
where ${\cal E}$ is the Euclidean subspace of $M^{(c)}$ introduced in \S 3.1  
(see the Theorem at the end of \S 3.1). The crucial problem of Euclidean field theory, which is 
the problem of reconstructing Minkowskian Q.F.T. by analytic continuation of the Schwinger functions, 
has been solved under a certain set of sufficient conditions called the
Osterwalder-Schrader axioms [33,34]. Here we wish to stress the special and unexpected
relationship which exists between this problem and the validity of complex angular momentum 
analysis, as a direct corollary of our Theorem 4.

In fact, from the assumed $SO(d+1)-$invariance of the theory it follows that 
the Euclidean four-point function 
$H_{\cal E}$ is represented by the function  
$ F(\zeta,\zeta';t,\cos \Theta_t)$ considered as given on the set 
$\{(\zeta,\zeta',t,\cos \Theta_t );\  t<0,\ (\zeta,\zeta') \in\  
\Delta_t \times \Delta_t,\   \cos \Theta_t \in\  [-1,+1] \}.  $ 
It is therefore equivalent to give oneself  
$H_{{\cal E}}$   
or the corresponding set   
of ``partial-wave-functions'' $\{ f_{\ell}(\zeta,\zeta',t),\  \ell\ge 0 \}$ 
defined by Eq.(4.2) (for $t <0,
(\zeta,\zeta') \in \Delta_t \times \Delta_t$). 
We can then state:

\begin{theorem}
Let $H_{\cal E}([k]) =   
F(\zeta,\zeta';t,\cos \Theta_t)$  be any $SO(d+1)-$invariant Euclidean four-point 
function whose sets of even and odd partial-wave functions $f_{\ell}(\zeta,\zeta',t)$ 
admit Carlsonian analytic interpolations in a given half-plane $\{ \lambda_t \in {\Bbb C}^{(m)}_+ \},$ 
denoted respectively by 
$\tilde{F}^{(s)}\left( \zeta ,\zeta ^{\prime };t,\lambda_t \right),$  
$\tilde{F}^{(a)}\left( \zeta ,\zeta ^{\prime };t,\lambda_t \right)$  
and satisfying Eqs (4.87), (4.88) and bounds of the form (4.84) with 
\footnote{for the sake of simplicity} $v_s = v_u.$   

Then for each vector $K \in {\Bbb R}^{d+1},$  
and each unit vector ${\rm e}_0$ orthogonal to $K$ fixing the time-direction, 
the restriction of 
$H_{\cal E}$ to    
${\hat {\cal E}}_{K},$  
admits an analytic continuation in the corresponding cut-manifold 
${\hat \Omega}_K^{(cut)} = {\hat \Omega}_K \backslash (\Sigma_s \cup \Sigma_u)$  
(represented by the cut $\cos \Theta _{t}$-plane, with $\zeta
,\zeta ^{\prime }$ varying in $\Delta _{t}\times \Delta _{t}).$ 
This analytic continuation is defined explicitly via formula (4.89), which 
reduces in 
${\hat {\cal E}}_{K},$  
to the
partial wave expansion (4.91) of  
$H_{\cal E}.$   
\end{theorem}

The analytic continuation of  
$ H_{\cal E}$  
in the cut-manifolds 
${\hat \Omega}_K^{(cut)} $ (for all $K$ orthogonal to a given ${\rm e}_0$) 
is indeed interpretable as  
{\it an analytic continuation which reaches the Minkowskian momentum space} 
(spanned by the hyperplane orthogonal to 
${\rm e}_0$ and the vector ${-\rm i} {\rm e}_0$) 
{\it and moreover generates the ``absorptive part structure''
with specified mass thresholds}.

\vskip 1cm 
\begin{center}
{\bf APPENDIX A:\   
Analytic completion and propagation of bounds}
\end{center}

\vskip 0.3cm
\noindent
{\sl A property of analytic completion}

\vskip 0.2cm
We recall the following result, first obtained by V. Glaser by a
method based on the Cauchy integral, then extended to the case of two
polydisks in arbitrary situations by the tube method presented below [35].

\vskip 0.5cm

\noindent
{\sl Proposition A.1}
\quad \it Let ${\cal T}_{+}=\left\{ \left( \eta ,\eta ^{\prime }\right) \in {\Bbb C}%
^{2};\ \ \func{Im}\eta >0,\ \func{Im}\eta ^{\prime }>0\right\} $ and ${\cal T}%
_{-}=\left\{ \left( \eta ,\eta ^{\prime }\right) \in {\Bbb C}^{2};\func{Im}%
\eta <0,\func{Im}\eta ^{\prime }<0\right\} $ and let ${\cal R}$ be the
``coincidence region'' 

\[
{\cal R}=\left\{ \left( \eta ,\eta ^{\prime }\right) \in {\Bbb R}^{2};\ a<\eta
<b,\ a^{\prime }<\eta^{\prime} <b^{\prime }\right\} .
\]

Then the holomorphy envelope of the ``edge-of-the-wedge domain'' $\Delta =%
{\cal T}_{+}\cup {\cal T}_{-}\cup {\cal R}$ can be defined as follows: $%
{\cal H}\left( \Delta \right) =\stackunder{0\leqslant \alpha \leqslant \pi }{%
\cup }{\cal T}_{\alpha },$ where each domain ${\cal T}_{\alpha }$ is the
following polydisk:

\[
{\cal T}_{\alpha }=\left\{ \left( \eta ,\eta ^{\prime }\right) \in {\Bbb C}%
^{2};\ \eta \in \Gamma _{ab}\left( \alpha \right) ,\ \eta ^{\prime }\in \Gamma
_{a^{\prime }b^{\prime }}\left( \alpha \right) \right\} ; 
\]
$\Gamma _{ab}\left( \alpha \right) $ and $\Gamma _{a^{\prime
}b^{\prime }}\left( \alpha \right) $ respectively denote the disks whose
bordering circles make the angle $\alpha $ with the real axis and intersect
the latter respectively at $a,b$ and $a^{\prime },b^{\prime }$ (see Fig A.1);
\rm 

\noindent
{\sl Proof:}\quad  Let $\chi =\log \dfrac{\eta -b}{\eta -a},\ \chi
^{\prime }=\log \dfrac{\eta ^{\prime }-b^{\prime }}{\eta ^{\prime
}-a^{\prime }}.$ One easily checks that the images of ${\cal T}_{+},{\cal T}%
_{-}$ in the space of variables $\chi ,\chi ^{\prime }$ are the respective
tubes:

\[
T_{+}=\left\{ \left( \chi ,\chi ^{\prime }\right) \in {\Bbb C}^{2};\ 0<\func{Im%
}\chi <\pi ,\;0<\func{Im}\chi ^{\prime }<\pi \right\} , 
\]

\[
T_{-}=\left\{ 
\left( \chi ,\chi ^{\prime }\right) \in {\Bbb C}^{2};\ \pi <\func{Im}\chi
<2\pi ,\ \pi <\func{Im}\chi ^{\prime }<2\pi  
\right\} , 
\]
while the image of  ${\cal R}$  is the set    
$R = \left\{ \left( \chi ,\chi ^{\prime }\right) \in {\Bbb C}^{2},\ \func{%
Im}\chi =\pi ,\ \ \ \func{Im}\chi ^{\prime }=\pi \right\} ,$ which is the common
edge to $T_{+}$ and $T_{-}$.

\begin{figure}
\epsfxsize=8truecm
{\centerline{\epsfbox{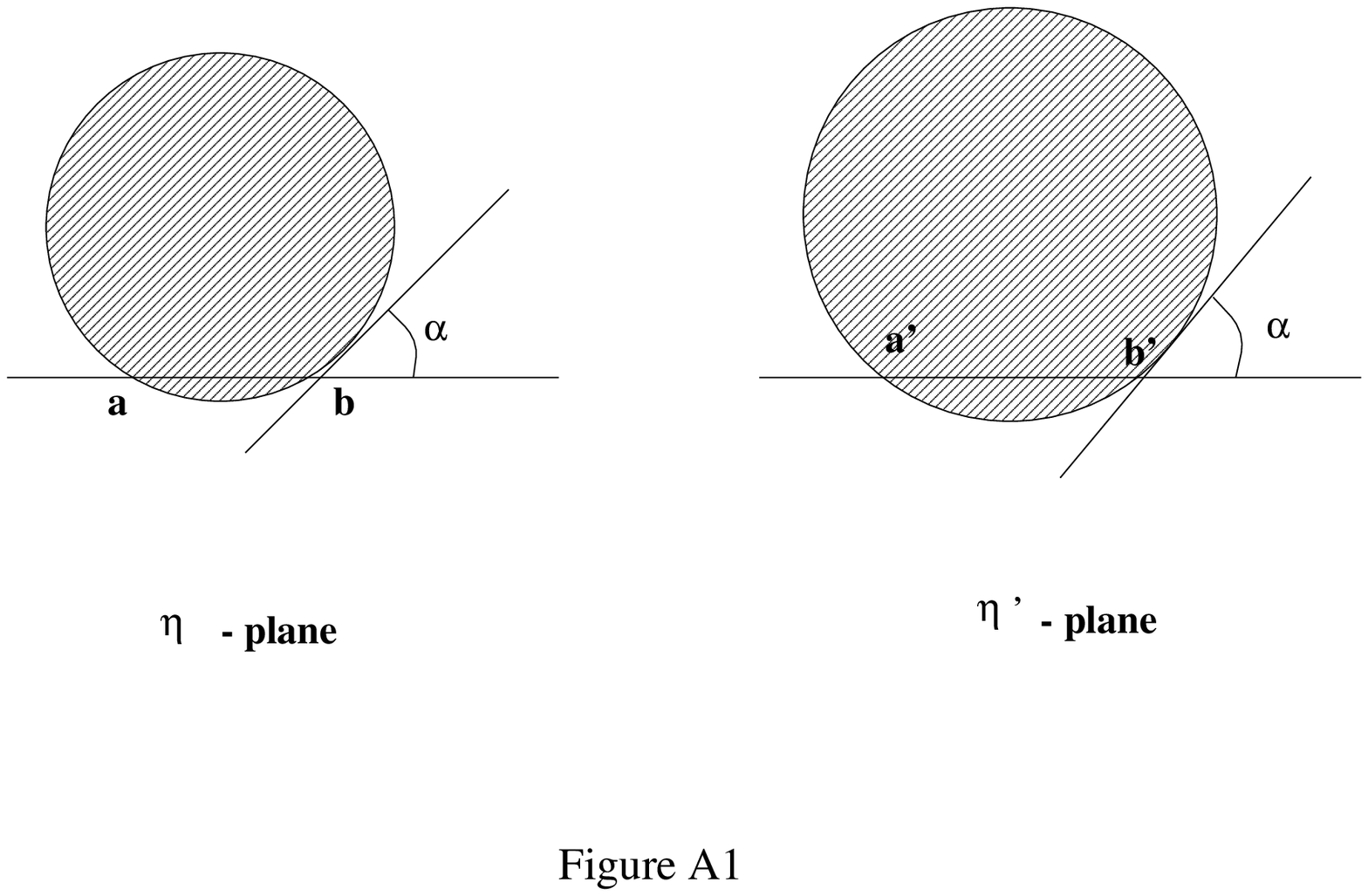}}}
\end{figure}

\begin{figure}
\epsfxsize=8truecm
{\centerline{\epsfbox{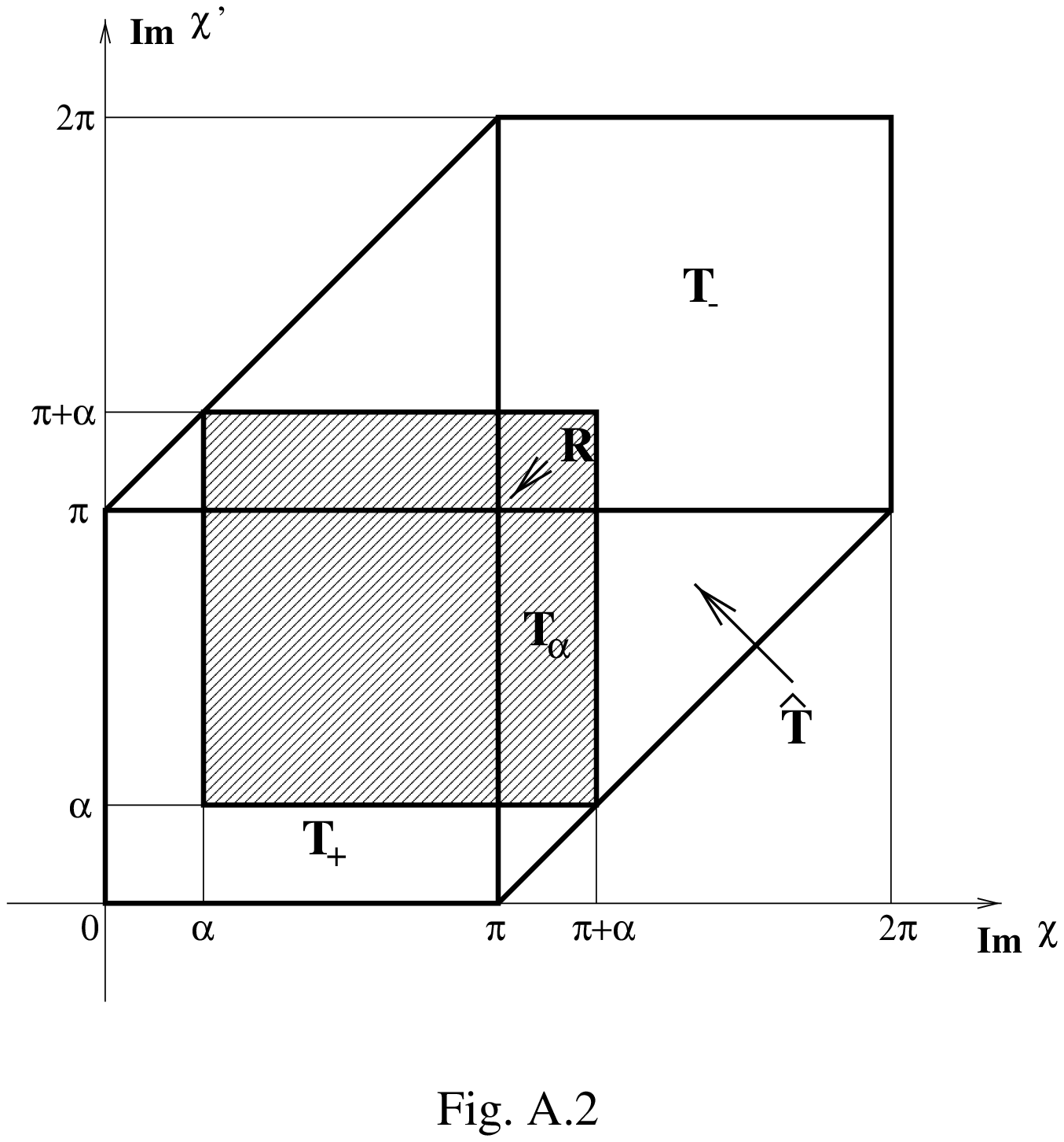}}}
\end{figure}

Then, in view of the tube theorem (applied in the limiting ``edge of the
wedge situation'', illustrated by Fig A.2), the holomorphy envelope ${\cal H}%
\left( T_{+}\cup T_{-}\cup R\right) $ is the convex tube $\hat{T}=%
\stackunder{0\leqslant \alpha \leqslant \pi }{\cup }T_{\alpha }$ whose basis
in $\left( \func{Im}\chi ,\;\func{Im}\chi ^{\prime }\right) $-space is
represented on Fig A.2. The result of proposition A.1 is readily obtained by
taking the inverse image ${\cal T}_{\alpha }$ of each tube $ T_{\alpha }$
in the variables $\left( \eta ,\eta ^{\prime }\right) .$

\vskip 0.5cm
\noindent
{\sl Corollary A.2:}\quad \it  The holomorphy envelope ${\cal H}\left(
\Delta \right) $ contains all the points of the form $\left( \eta ,0\right)
, $ with $\eta $ varying in the cut-plane ${\Bbb C} \backslash 
\{\eta\  {\rm real} ;\  \eta \notin ]a,b[ \ \}. $  \rm  

\vskip 0.5cm

\noindent
{\sl Proof:} \quad These points $\left( \eta ,\eta ^{\prime }=0\right) $
have images $\left( \chi ,\chi ^{\prime }\right) $ such that: $0<\func{Im}%
\chi <2 \pi $ and Im$\chi ^{\prime }=\pi $ and therefore belong to the convex
tube $\hat{T}$ (see Fig.A.2).

\vskip 1cm
\noindent
{\sl Propagation of bounds in the analytic completion procedure}

\vskip 0.5cm

We need the following extension of the ``maximum modulus principle''.

\vskip 0.5cm
\noindent
{\sl Proposition A.3:}

\it Let $f\left( \eta \right) $ be holomorphic in the domain $\Delta
_{b}=\left\{ \eta \in {\Bbb C};\left| \func{Re}\eta \right| <b,\left| \func{%
Im}\eta \right| <b\right\} ,$ and continuous in the closure $\bar{\Delta}%
_{b} $ of $\Delta _{b}.$ Let the following majorization hold in $\bar{\Delta}%
_{b}: $ \rm  

\begin{equation}
\left| f\left( \eta \right) \right| \leqslant M\left| \func{Im}\eta \right|
^{-n},  \tag{A.1}
\end{equation}

\noindent
\it where $n$ is a given positive number and $M$ is a constant.

Then for all $\beta$ with $\ -b  \le \beta \le b,$ one has: 
\begin{equation}
\left| f\left( {\rm i}\beta\right) \right| \leqslant {\sqrt 5}^n Mb^{-n}  \tag{A.2}
\end{equation}

\vskip 0.5cm
\noindent
Proof: \rm

One considers the function $g(\eta) = (b^2 -{\eta}^2)^n f(\eta)$, which is also holomorphic in 
$\Delta_b$ and continuous in $\bar{\Delta}_b$. One directly deduces from (A.1) the following 
uniform majorization for $g$ on the boundary of $\Delta_b$:
$$ |g(\eta)| \le 5^{n \over 2} M b^n.$$
In view of the maximum modulus principle, this majorization extends to all points in $\Delta_b$; 
by writing it at $\eta={\rm i} \beta$, one then obtains: 
$$ |f({\rm i}\beta)| = |g({\rm i}\beta)|( b^2 + \beta^2)^{-n} 
\le 5^{n \over 2} M 
\left(\frac{b}{b^2+ \beta^2}\right)^n
\le 5^{n \over 2} M b^{-n}. $$

\vskip 2.5cm

\begin{center}
{\bf APPENDIX B}
\end{center}

\vskip 0.5cm

\noindent
{\bf Primitives and derivatives of non-integral order in a complex domain and
Laplace transformation}

\vskip 0.3cm
$a$ being a given positive number, we define in ${\Bbb C}$ 
the following subset
$ B_{a}^{({\rm cut})} =B_{a}\backslash \sigma ,$
where $B_{a}=\left\{ \theta \in {\Bbb C};\ \theta =u+{\rm i}v,\left| u\right|
< a,v\geqslant 0 \right\} $ and $\sigma =\left\{ \theta \in 
{\Bbb C};\ \theta ={\rm i}v,v\geqslant v_{0}\right\}$,  $v_0 >0$ 
(see Fig.B1). 

We then introduce the space of holomorphic functions denoted 
${\cal O}^{\infty }( B_{a}^{({\rm cut}%
)}) $ which is generated by  

\noindent
i) all functions $f\left( \theta \right) $ holomorphic
in $ B_{a}^{({\rm cut})} $
and 
satisfying bounds of the form  
\begin{equation}
\left| f\left( u+{\rm i}v\right) \right| \leqslant C_{\eta }(v) 
\tag{B.1}
\end{equation}
in the corresponding subsets
\begin{equation}
B_{a}^{(\eta )}=B_{a}\backslash \left\{ \theta \in {\Bbb C;\ \theta =}u{\Bbb +}%
{\rm i}v,\ \left| u\right| <\eta ,\ v>v_{0}-\eta \right\} 
\tag{B.2}
\end{equation}
of ${\Bbb C}$, for all $\eta >0.$ In (B.1),  
$C_{\eta }(v)$ denotes an increasing and
locally bounded function \it with at most power-like behaviour \rm for $v$ tending
to infinity.

\vskip 0.3cm
\noindent
ii) the products of functions of the previous type by $\theta^{\rho}$, with
$\rho \ real\ > 0$.

\vskip 0.5cm
\noindent
{\sl Laplace transforms} 

\vskip 0.3cm
Let $\gamma _{0}$ and $\gamma _{0}^{\prime }$ be two infinite paths with
origin $0$ in the respective half-strips $u>0$ and $u<0$
of $B_{a},$ and whose infinite branches are asymptotically parallel to the
imaginary axis of the $\theta $-plane.
We associate with each function $f(\theta ) 
\in {\cal O}^{\infty }( B_{a}^{({\rm cut}%
)}) $ 
the Laplace-type transforms:
\begin{eqnarray}
{\cal L}_{0}(f)\left( \lambda \right) &=&\int_{\gamma _{0}}{\rm e}^{{\rm i}%
\lambda \theta }f\left( \theta \right) d\theta,  \tag{B.3} \\
{\cal L}_{0}^{\prime }(f)(\lambda ) &=&\int_{\gamma _{0}^{\prime }}{\rm e}^{%
{\rm i}\lambda \theta }f\left( \theta \right) d\theta.  \tag{B.3'}
\end{eqnarray}
In view of (B.1), the latter are holomorphic in the half-plane
${\Bbb C}_+ = \{\lambda \in {\Bbb C};\  \func{Re} \lambda >0 \}$  
and admit bounds of the form  
$c_{\varepsilon, \eta }
{\rm e}^{\eta 
| \func{Im}\lambda |},$ 
for all $\varepsilon >0, \eta >0,$ 
in the corresponding half-planes 
${\Bbb C}%
_{+}^{(\varepsilon )} = 
\{\lambda \in {\Bbb C};\  \func{Re} \lambda > \varepsilon \}$  
(as it results from a suitable distortion of the
paths $\gamma _{0},\gamma _{0}^{\prime }$ in the integrals (B.3), (B.3')).

\begin{figure}
\epsfxsize=5truecm
{\centerline{\epsfbox{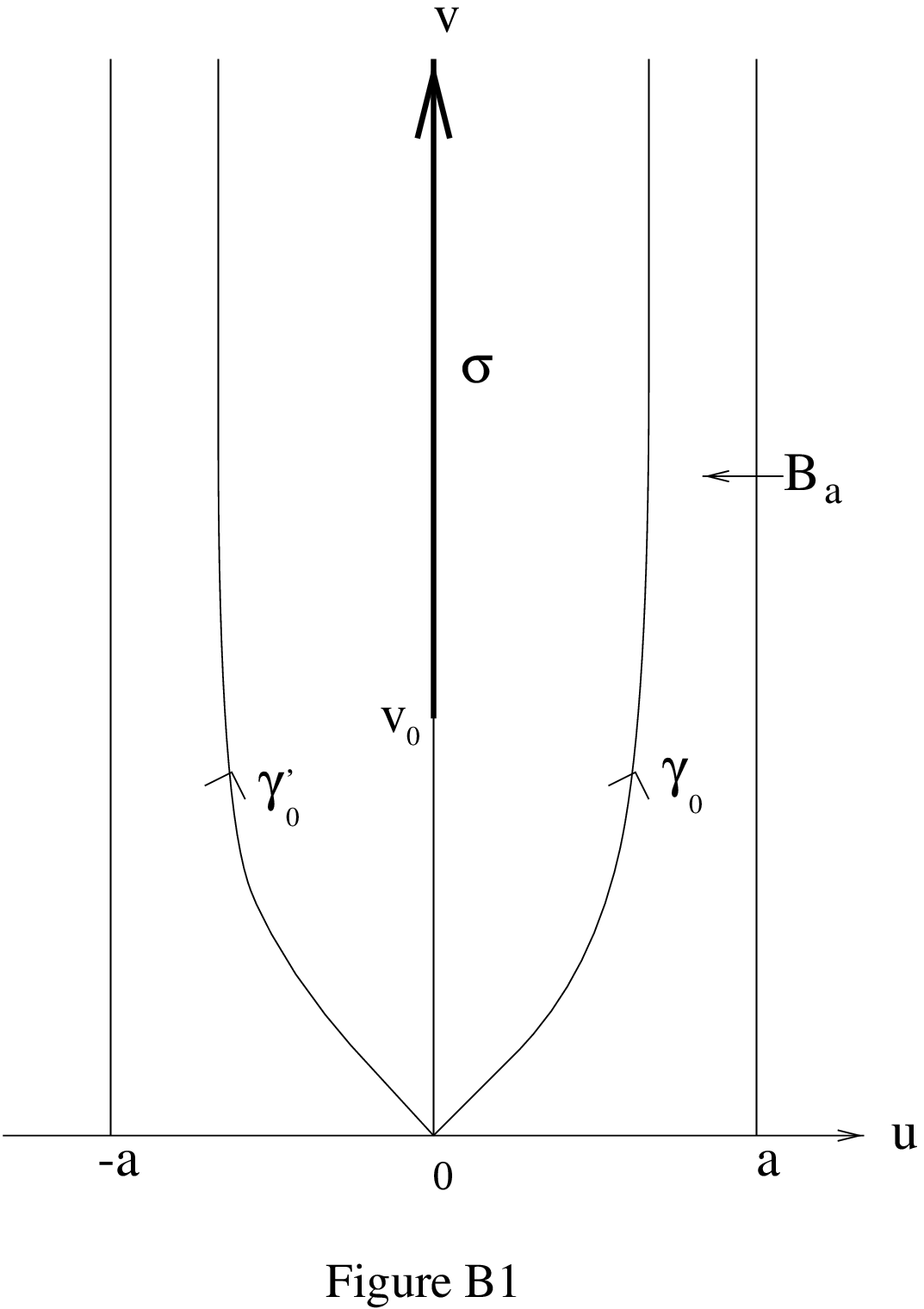}}}
\end{figure}
If $f$ has continuous boundary values (from both sides) on $\sigma ,$ the
corresponding discontinuity function $\Delta f(v)={\rm i}
\displaystyle {\lim_{\eta \rightarrow 0, \eta >0}
\left[ \left( f(\eta +{\rm i}v)\right) -f(-\eta +{\rm i%
}v)\right]} $ admits the Laplace transform:
\begin{equation}
L\left( \Delta f\right) \left( \lambda \right) ={\cal L}_{0}(f)-{\cal L}%
_{0}^{\prime }(f)
=\int_{\gamma_0 - \gamma'_0} {\rm e}^{{\rm i}\lambda \theta} f(\theta) d\theta
=\int_{0}^{\infty }{\rm e}^{-\lambda v}\Delta f(v)dv 
\tag{B.4}
\end{equation}
In the general case, we still say that $L(\Delta f)(\lambda )={\cal L}%
_{0}(f)-{\cal L}_{0}^{\prime }(f)$ represents the Laplace-transform of the
discontinuity $\Delta f$ of $f,$ now considered as a hyperfunction with
support $\sigma ;\ L(\Delta f)(\lambda )$ is holomorphic in ${\Bbb C}_{+}$
and such that:
\begin{equation}
\left| L(\Delta f)(\lambda )\right| \leqslant 2c_{\varepsilon, \eta }{\rm e}%
^{-\left( \func{Re}\lambda -\varepsilon \right) (v_{0}-\eta) }{\rm e%
}^{\eta \left| \func{Im}\lambda \right| }  \tag{B.5}
\end{equation}
in each subset ${\Bbb C}_{+}^{(\varepsilon )}$ of ${\Bbb C}_{+},$
for all $\eta >0.$
\vskip 1cm

\noindent
{\sl Primitives of non integral order in the complex domain }

\vskip 0.2cm

For every {\it real positive }$\alpha ,$ we associate with any function $%
f\left( \theta \right) $ of the previous class 
${\cal O}^{\infty }
(B_{a}^{({\rm cut})}) $ 
the following function
\begin{equation}
[P_{\alpha }f]
\left( \theta \right) =\frac{1}{\Gamma \left( \alpha \right) }%
\int_{\gamma _{\left( 0,\theta \right) }}\left( \theta -\theta ^{\prime
}\right) ^{\alpha -1}f\left( \theta ^{\prime }\right) d\theta ^{\prime }, 
\tag{B.6}
\end{equation}
where $\theta $ varies in $B_{a}^{\left( {\rm cut}\right) }$
and $\gamma _{\left( 0,\theta \right) }$ is a path with end-points $0,\theta
,$ (homotopous to the linear segment $\left[ 0,\theta \right] )$ whose
support is contained in $B_{a}^{\left( {\rm cut}\right) }.$

By choosing for $\gamma _{\left( 0,\theta \right) }$ the linear segment $%
\left[ 0,\theta \right] $ and  
making the change of variables $\theta' = \theta t$, with $t \in [0,1]$ in
(B.6), one checks that each function
$P_{\alpha }f$ is the product of the ramified function $\theta^{\alpha}$ (case i))
or more generally $\theta^{\alpha + \rho}$ (case ii)) by a function
holomorphic in $%
B_{a}^{\left( {\rm cut}\right) }$ and that it
also satisfies bounds of the form (B.1) (with functions $C_{\eta }^{\alpha
}(v)=v^{\alpha }C_{\eta }(v))$ and therefore belongs to the class ${\cal O}%
^{\infty }( B_{a}^{\left( {\rm cut}\right) }) .$
The same change of variables also shows that the integral (B.6) reduces to a
Riemann--Liouville integral. Therefore, by using the standard properties of
the latter (see [36b)], p 181-182), we obtain the following properties of the
operators $P_{\alpha }:$ for all $\alpha ,\beta >0,$
\begin{equation}
P_{\alpha }\circ P_{\beta}=
P_{\beta }\circ P_{\alpha}=
P_{\alpha +\beta }  \tag{B.7}
\end{equation}
and for all positive integers $n:$
\begin{equation}
\left( \frac{d}{d\theta }\right) ^{n}\left[ P_{n}f\right] \left( \theta
\right) =f\left( \theta \right),  \tag{B.8}
\end{equation}
from which it follows that, for all $\alpha > n$:
\begin{equation}
\left( \frac{d}{d\theta }\right) ^{n}\left[ P_{\alpha }f\right] \left(
\theta \right) =  [P_{\alpha -n }  
f] \left( \theta \right).  \tag{B.8a}
\end{equation}
These equations lead one to call $P_{\alpha }f$ (for general $\alpha )$
a ``primitive of order $\alpha $ of $f$ in the complex domain''.  
\footnote{Note that all the primitives $P_{\alpha}f$ are still well-defined 
(via Eq. (B.6)) for functions $f$ such that, for some $\varepsilon >0$, 
$\theta^{1-\varepsilon} f(\theta) \in 
{\cal O}%
^{\infty }( B_{a}^{\left( {\rm cut}\right) }). $}

\vskip 0.5cm 

\noindent
{\sl Derivatives }

\vskip 0.2cm 
Since the operations 
$D_n= \left( \dfrac{d}{d\theta }\right) ^{n}$ act on the class 
${\cal O}^{\infty }
(B_{a}^{({\rm cut})}) $ 
for all integers $n,$ it is natural to extend Eq.(B.8a)
to the case $\alpha < n$ and to introduce 
\it the derivation $D_{\nu}$ of non-integral order $\nu =  n-\alpha$\rm \  
by the following formula:
\begin{equation}
D_{\nu}f  \equiv D_n P_{\alpha}f  = D_{n+r} P_{\alpha + r}f,    
\tag{B.8b}
\end{equation}
in which the last equality holds, in view of (B.7),(B.8), for every integer $r$ 
such that $\alpha + r >0$. We then have:

\vskip 0.8cm
\noindent
{\sl Proposition B.1}

\it a) For any function $f(\theta)$ in  
${\cal O}%
^{\infty }( B_{a}^{\left( {\rm cut}\right) }) $ 
and any real positive number $\nu$, the derivative $D_{\nu}f$ 
is the product of a function in 
$\displaystyle \bigcap_{\delta;\ \delta >0}$ $ {\cal O}%
^{\infty }( B_{a- \delta}^{\left( {\rm cut}\right) }) $ 
by $\theta^{-\nu}$.  

b) Eqs (B.8a),(B.8b) admit the following generalizations, valid for all positive numbers 
$\beta$ and $\nu$:
\begin{equation}
D_{\beta}P_{\beta}f  = f,  \tag{B.8c} 
\end{equation}
\begin{equation}
{\rm if}\quad \beta > \nu, \ \ D_{\nu}P_{\beta} = P_{\beta - \nu},   \tag{B.8d}  
\end{equation}
\begin{equation}
{\rm if} \quad \beta < \nu, \ \ D_{\nu}P_{\beta} = D_{\nu - \beta}.  \tag{B.8e} 
\end{equation}

c) If the function $f\left( \theta \right) $ is holomorphic in $B_a^{({\rm cut})}$, then 
for all $\alpha >0$ and for all positive integers $n$, one has:  
\begin{equation}
[D_{n}P_{\alpha}f](\theta) - [P_{\alpha}D_n f](\theta) = \theta^{\alpha -n}  
\sum_{p=0}^{n-1} [D_p f](0) {\theta^p \over {\Gamma(\alpha -n+p+1)}} 
\tag{B.9}
\end{equation}
and for all $\rho >0$:
\begin{equation}
D P_{\alpha}(\theta^{\rho} f) = P_{\alpha}D (\theta^{\rho} f).   
\tag{B.9'}
\end{equation}
\rm

\noindent
{\sl Proof:}

a) Since $D_{\nu}f = D_n P_{\alpha} f$, with $\nu = n- \alpha$, and since $P_{\alpha} f$ 
is the product of $\theta^{\alpha}$ by a function in 
${\cal O}%
^{\infty }( B_{a}^{\left( {\rm cut}\right) }) $
(for any $\alpha >0$ and $f$ in  
${\cal O}%
^{\infty }( B_{a}^{\left( {\rm cut}\right) }) $), 
the usual derivation $D_n$ yields the analytic structure with the factor $\theta^{-\nu}$ and   
the Cauchy inequalities imply bounds of the form 
(B.1) in $B_{a-\delta}^{({\rm cut})}$, for all $\delta >0$. 

b) In view of (B.8b) and (B.7), one can always write: $D_{\nu}P_{\beta} =  D_n P_{\alpha + \beta}$
with $\nu = n - \alpha,\ n$ integer, Applying Eqs (B.8) or (B.8a) or the definition (B.8b)  
according to whether $n = \alpha + \beta$ or $n < \alpha + \beta$  or $n > \alpha + \beta$ yields  
respectively Eqs (B.8c),(B.8d) and (B.8e). 

c)  If $f$ is holomorphic at the origin, one can apply  
integration by parts to Eq.(B.6) with $f$ replaced by any derivative  
$D_{p} f$ (with $ 0 \le p \le n-1$); one gets:   
\[
[P_{\alpha}D_{p}f](\theta) = [P_{\alpha +1}D_{p+1}f](\theta) + [D_{p}f](0)\frac{\theta^{\alpha}}
{\Gamma(\alpha +1)},  
\]
and therefore in view of (B.8a):
\[
[D_{n-p} P_{\alpha} D_{p}f](\theta) = [D_{n-p-1}P_{\alpha} D_{p+1}f](\theta) 
+ [D_{p}f](0) \frac{\theta^{\alpha-n +p}}{\Gamma(\alpha-n+p+1)}.
\]
Using the latter recursively with $0 \le p \le n-1 $ then yields Eq.(B.9).
Eq.(B.9') is obtained as (B.9) for $n = 1$, the r.h.s. of (B.9') being still meaningful 
(see footnote 11) since $ \theta^{1-\rho} D (\theta^{\rho} f)$ belongs to  
${\cal O}%
^{\infty }( B_{a - \delta}^{\left( {\rm cut}\right) })$ (for all $\delta >0$). 

\vskip 0.3cm
\noindent
{\sl Remark:}\quad  
{\it Property c) extends the usual Taylor expansion (obtained for $\alpha = n$ in Eq.(B.9)).
In particular, the holomorphic (ramified) function at the r.h.s. of Eq. (B.9)   
has no discontinuity across $\sigma$; therefore,   
$ D_n P_{\alpha} f$
and $ P_{\alpha} D_n f$
have ``the same discontinuity'' 
across $\sigma$ 
(i.e. represent the same hyperfunction with support $\sigma$)   
which we denote $P_{\alpha-n}\Delta f$
when $n < \alpha$ and $D_{n- \alpha}\Delta f$ when $n > \alpha$. 
Since $D_n f$ is holomorphic at the origin, $ P_{\alpha} D_n f$
belongs to 
$ {\cal O}^{\infty }( B_{a-\delta}^{({\rm cut}%
)})$ (for all $\delta > 0$).  
In other words, if $f$ (in 
${\cal O}%
^{\infty }( B_{a}^{( {\rm cut}) }) $) 
is holomorphic at the origin, all the hyperfunctions 
$D_{\nu}\Delta f$ admit a representative in  
$\displaystyle \bigcap_{\delta;\ \delta >0} {\cal O}^{\infty }( B_{a-\delta}^{({\rm cut}%
)}).$} 

\vskip 0.3cm
\noindent
{\sl Laplace transforms of the primitives $%
P_{\alpha }f$ and derivatives $D_{\nu}f$:} 

We first notice that since all the primitives $P_{\alpha} f$ of a function $f$ in  
${\cal O}%
^{\infty }( B_{a}^{( {\rm cut}) }) $ 
remain in the same space, they all admit well-defined Laplace-type transforms
${\cal L}_0 (P_{\alpha} f)$, 
${\cal L}'_0 ( P_{\alpha} f)$ (defined via Eqs (B.3), (B.3')). 
On the contrary, the operations ${\cal L}_0$ and ${\cal L}'_0$ do not act in general on  
the corresponding derivatives $D_{\nu} f$, since the latter may contain non-integrable 
factors $\theta^{\varrho}$ (with $\varrho \le -1$). However, the Laplace transforms 
$ L(\Delta D_{\nu}f) \equiv L(D_{\nu} \Delta f)$ are always well-defined via the following procedure. 
One uses the fact that for functions $f$ (and $P_{\alpha} f$) in  
${\cal O}%
^{\infty }( B_{a}^{( {\rm cut}) }), $ 
the defining formula (B.4) can be alternatively replaced by 
$ L(\Delta f)(\lambda) = \int_{\gamma} {\rm e}^{{\rm i}\lambda \theta} f(\theta) d\theta,$  
where $\gamma$ is a cycle homotopous to $\gamma_0 -\gamma'_0 $ in $B_a^{({\rm cut})}$  
whose support avoids the origin (i.e. lies in the interior of $B_a^{({\rm cut})}$).
Since each derivative $D_{\nu} f = D_n P_{\alpha} f$ is holomorphic and of 
power-like growth at infinity in 
the interior of $B_a^{({\rm cut})},$ the previous formula applies  and defines  
\begin{equation}
L(D_{\nu}\Delta f)(\lambda) = \int_{\gamma} {\rm e}^{{\rm i}\lambda \theta} [D_{\nu}f](\theta) d\theta  
\tag {B.10} 
\end{equation}
as a holomorphic function in ${\Bbb C}_+$. 

The following statement extends to the primitives $P_{\alpha}$ and derivatives $D_{\nu}$ the 
usual property of Laplace transforms. 

\vskip 0.3cm
\noindent
{\sl Proposition B.2}
\quad \it For any holomorphic function $f\left( \theta \right) $ in the space  
${\cal O}%
^{\infty }( B_{a}^{( {\rm cut}) }) $ 
and for the
corresponding (hyperfunction) discontinuity $\Delta f,$ there holds the
following property of the Laplace transforms in the half-plane ${\Bbb C}%
_{+} :$    

\noindent
a)\ for all $\alpha >0$,   
\[
{\cal L}_{0}\left( P_{\alpha }f\right)(\lambda) ={\rm e}^{\frac{{\rm i}\pi }{2}%
\alpha }\lambda ^{-\alpha }{\cal L}_{0}\left( f\right)(\lambda), \quad  
{\cal L}_{0}^{\prime }\left( P_{\alpha }f\right)(\lambda) ={\rm e}^{^{\frac{{\rm i}%
\pi }{2}\alpha }}\lambda ^{-\alpha }{\cal L}_{0}^{\prime }\left( f\right)(\lambda),  
\]
\begin{equation}
L\left( P_{\alpha }\Delta f\right)(\lambda) ={\rm e}^{^{\frac{{\rm i}\pi }{2}\alpha
}}\lambda ^{-\alpha} L\left( \Delta f\right)(\lambda),  \tag{B.11}
\end{equation}
b)\  for all $\nu >0, $ 
\begin{equation}
L\left(D_{\nu}\Delta f\right)(\lambda) = {\rm e}^{-\frac{{\rm i}\pi}{2} \nu} \lambda^{\nu}
L\left(\Delta f \right)(\lambda).  \tag{B.12} 
\end{equation}
\rm
\vskip 0.3cm

\noindent
{\sl Proof:}\quad a) It is sufficient to prove the first equation in (B.11); 
the r.h.s.. of this equation can be written for
all $\alpha >0:$
\begin{equation}
{\cal L}_{0}\left( P_{\alpha }f\right)(\lambda) =\frac{1}{\Gamma \left( \alpha
\right) }\int\limits_{\gamma _{0}}{\rm e}^{{\rm i}\lambda \theta }d\theta
\int\limits_{\gamma _{\left( 0,\theta \right) }}\left( \theta -\theta
^{\prime }\right) ^{\alpha -1}f\left( \theta ^{\prime }\right) d\theta
^{\prime }  \tag{B.13}
\end{equation}
For simplicity, we choose $\gamma _{0}$ such that its support is a {\it %
convex} (infinite) curve (see Fig.B1) and we specify $\gamma _{\left(
0,\theta \right) }$ by the condition that its support is contained in 
the support of $\gamma_{0}.$ 
For $\lambda $ in ${\Bbb C}_{+},$ the integral in  
(B.13) is absolutely convergent and can be rewritten (by
inverting the integrations and putting $\theta'' =\theta
-\theta ^{\prime }):$
\begin{equation}
{\cal L}_{0}\left( P_{\alpha }f\right)(\lambda) =\frac{1}{\Gamma \left( \alpha
\right) }\int\limits_{\gamma _{0}}{\rm e}^{{\rm i}\lambda \theta ^{\prime
}}f\left( \theta ^{\prime }\right) d\theta ^{\prime }\int\limits_{\gamma
_{0}\left( \theta ^{\prime }\right) }{\rm e}^{{\rm i}\lambda \theta''} 
\left( \theta'' 
\right) ^{\alpha
-1}d\theta'' ,  \tag{B.14}
\end{equation}
where the support of $\gamma _{0}\left( \theta ^{\prime }\right) $ is the
set $\left\{ \theta'' \in {\Bbb C};\ \theta'' 
+\theta ^{\prime }\in {\rm \sup \!\!p}\ \gamma _{0}\setminus {\rm \sup \!\!p}%
\ \gamma _{\left( 0,\theta ^{\prime }\right) }\right\} .$ Since this
(infinite) path $\gamma _{0}\left( \theta ^{\prime }\right) $ is homotopous
to $\left[ 0,{\rm i}\infty \right[ ,$ the subintegral of (B.14) is
independent of $\theta ^{\prime }$ and equal to ${\rm e}^{{\rm i}\frac{\pi }{%
2}\alpha }\int_{0}^{\infty }{\rm e}^{-\lambda v}v^{\alpha -1}dv=\frac{{\rm e}%
^{\frac{{\rm i}\pi}{2}\alpha }\Gamma \left( \alpha \right) }{\lambda ^{\alpha }%
}.$ 

\vskip 0.3cm
b) Let $\nu = n- \alpha$, with $\alpha < n$; in view of (B.10),
we have:
\[
L\left(D_{\nu}\Delta f \right)(\lambda) = L\left(D_nP_{\alpha} \Delta f \right)(\lambda) 
= \int_{\gamma} {\rm e}^{{\rm i}\lambda \theta} [D_n (P_{\alpha} f)](\theta) d\theta  
\]
\[
=(-{\rm i}\lambda)^n \int_{\gamma} {\rm e}^{{\rm i}\lambda \theta} [P_{\alpha} f](\theta) d\theta  
=(-{\rm i}\lambda)^n   
L\left(P_{\alpha} \Delta f \right)(\lambda).   
\]
Eq.(B.12) then readily follows from the latter and from (B.11).  

\vskip 0.6cm
\noindent
{\sl The case of distribution-like boundary values on $\sigma$}
\vskip 0.3cm

We shall now restrict our attention to functions of the class 
${\cal O}^{\infty }( B_{a}^{( {\rm cut}) }) $ 
which
are ``of moderate growth'' near the cut ${\rm \sigma .}$ More precisely, we
introduce for each real number $\beta ,$ with $\beta \geqslant 0,$ the class 
${\cal O}^{\beta }( B_{a}^{( {\rm cut}) }) $ by the same definition as  
${\cal O}^{\infty }( B_{a}^{\left( {\rm cut}\right) }), $ 
except for the uniform bounds (B.1) which are replaced by:   
\begin{equation}
\left| f\left( u+{\rm i}v\right) \right| \leqslant \frac{C\left( v 
\right) }{\eta ^{\beta }},  \tag{B.15}
\end{equation}
in the corresponding subsets $B_{a}^{\left( \eta \right) }$ of $B_{a},$ $%
(C\left( v\right) $ being again a locally bounded function with power-like
behaviour for $v$ tending to infinity).

When the bounds (B.15) are
replaced by logarithmic bounds of the form:
\begin{equation}
\left| f\left( u+{\rm i}v\right) \right| \leqslant C\left( v\right) \left|
\ln \eta \right| ;  \tag{B.16}
\end{equation}
the corresponding class of holomorphic functions $f\left( \theta \right) $
is called ${\cal O}^{0*}( B_{a}^{( {\rm cut})
}) .$

We also need to consider functions $f$ of the class  
${\cal O}^{0 }( B_{a}^{\left( {\rm cut}\right) }) $ 
which have {\it %
continuous boundary values on $\sigma ,$ as well as all their derivatives} $%
D_{\nu' }f$ for all $\nu' <\nu ,\ \nu $ being a given
positive number. If moreover each of these 
derivatives $D_{\nu' }f$ 
is the product of a 
function in 
$\displaystyle \bigcap_{\delta;\ \delta >0}$ $ {\cal O}%
^{0}( B_{a- \delta}^{\left( {\rm cut}\right) }) $ 
by $\theta^{-\nu'}$,   
we say that $f$ belongs to
the class ${\cal O}_{\nu }( B_{a}^{\left( {\rm cut}\right) }) .$
Functions in these classes satisfy the 

\vskip 0.3cm

\noindent
{\sl Proposition B.3}\quad \it If $f$ belongs to ${\cal O}_{\nu} 
( B_{a}^{\left( {\rm cut}\right) }) ,$ then the Laplace transform 
$L\left( \Delta f\right) $ of the discontinuity $\Delta f$ of $f$ satisfies
uniform bounds of the following form 
\rm 
\begin{equation}
\left| L\left( \Delta f\right) \left( \lambda \right) \right| \leqslant
c_{\varepsilon \varepsilon ^{\prime }}\left| \lambda \right| ^{-\nu
+\varepsilon ^{\prime }}{\rm e}^{-\left( \func{Re}\lambda -
\varepsilon \right)  v_{0}}  \tag{B.17}
\end{equation}
\it in the corresponding half-planes ${\Bbb C}_{+}^{(\varepsilon )},$ 
for all $\varepsilon >0,\varepsilon
^{\prime }>0.$
\rm

\vskip 0.3cm 
\noindent
{\sl Proof:}\quad In view of proposition B.2 b), we can write for
any $\varepsilon ^{\prime }>0:$
\begin{equation}
L\left( D_{\nu -\varepsilon ^{\prime }}\Delta f\right)(\lambda) 
={\rm e}^{{\rm i}\frac{\pi }{2%
}\left( \varepsilon ^{\prime }-\nu \right) }\lambda ^{\nu -\varepsilon
^{\prime }}L\left( \Delta f\right)(\lambda) .  \tag{B.18}
\end{equation}
Since $\left| [D_{\nu -\varepsilon ^{\prime }}f] \left(
u+ {\rm i}v\right) \right| \leqslant C \left( v\right) ,$ 
for $|u| < a- \delta,\  v > \delta$ (with $ 0 < \delta < v_0$),  
the expression of $L( D_{\nu -\varepsilon ^{\prime }}\Delta
f) ( \lambda ) $ given by (B.10) 
(with $\gamma$ flattened onto $\sigma$ from both sides) 
can be uniformly bounded in modulus by 
$c_{\varepsilon \varepsilon ^{\prime }}{\rm e}^{-\left( \func{Re}\lambda
-\varepsilon \right)  v_{0}}$ in any half-plane ${\Bbb C}%
_{+}^{\left( \varepsilon \right) }\left( \varepsilon >0\right) .$ This implies the 
bound (B.17) in view of Eq.(B.18).

\vskip 0.3cm
We now study the properties of the functions in the classes ${\cal O}%
^{\beta }( B_{a}^{( {\rm cut}) }) ,\beta \geqslant
0,$ and ${\cal O}^{0*}( B_{a}^{( {\rm cut}) }) $ and
characterize in a precise way their distribution-like boundary values on the
cut $\sigma $ and their Laplace transforms.  

\vskip 0.3cm
\noindent
{\sl Proposition B.4}
\quad \it Let $f\left( \theta \right) $ belong to a class ${\cal O}^{\beta }
(B_{a}^{( {\rm cut}) }) ,$ with $\beta \geqslant 0,$ 
or (for $\beta = 0$) to 
${\cal O}^{0*}(B_a^{({\rm cut})} ).$ 
Then  

\noindent i)\quad The various ``primitives'' $P_{\alpha }f 
\ \left( \alpha >0\right) $ satisfy the following properties:

a) if $\alpha <\beta ,\ P_{\alpha }f $ belongs to the
class ${\cal O}^{\beta -\alpha }( B_{a}^{( {\rm cut})
}), $

b) if $\alpha =\beta ,\ P_{\alpha }f $ belongs to the
class ${\cal O}^{0*} ( B_{a}^{( {\rm cut}) 
}), $

c) if $\alpha >\beta ,\ P_{\alpha }f $ belongs to the
class ${\cal O}_{\alpha -\beta }( B_{a}^{( {\rm cut})
}); $

\vskip 0.3cm
\noindent ii)\quad The boundary values $f_{+},f_{-}$ of $f$ on ${\rm i}{\Bbb %
R}$ from the respective sides $ \func{Re}\theta >0$,  
$ \func{Re}\theta <0$,  
and the corresponding discontinuity $\Delta f={\rm i}\left(
f_{+}-f_{-}\right) $ (with support contained in $\sigma )$ are defined in
the sense of distributions and such that

\[
f_{\pm }=D_{p}F_{\pm },\qquad \Delta f=D_{p}\Delta F,
\]

\noindent with $F_{\pm }$ continuous on ${\rm i}{\Bbb R},$  supp $\Delta
F\subset \sigma $ and $p=E(\beta )+1;$

\vskip 0.3cm
\noindent iii)\quad The Laplace transform $L(\Delta f)$ of the distribution $%
\Delta f$ satisfies uniform bounds of the following form (for all $%
\varepsilon >0,\varepsilon ^{\prime }>0)$ \rm  

\begin{equation}
\left| L\left( \Delta f\right) \left( \lambda \right) \right| \leqslant
c_{\varepsilon \varepsilon ^{\prime }}\left| \lambda \right| ^{\beta
+\varepsilon ^{\prime }}{\rm e}^{-\left( \func{Re}\lambda -
\varepsilon \right)  v_{0}}  \tag{B.19}
\end{equation}
\it in the corresponding half-planes ${\Bbb C}_{+}^{(\varepsilon )}.$
\rm

\vskip 0.3cm
\noindent
{\sl Proof:}

\noindent i)\quad For all $\alpha\  (\alpha >0),$ the expression (B.6) of $%
[P_{\alpha }f]( \theta ) $ can be rewritten with the following
choice: supp $\gamma_{(0,\theta)} 
=\left[ 0,b\right] \cup \left[
b,b+{\rm i}v\right] \cup \left[ b+{\rm i}v,u+{\rm i}v\right] ,$ where 
$\theta =u+{\rm i}v$ and  
$b$ is
a fixed number such that $0<\left| b\right| <a.$
As seen below, this choice is suitable for showing that 
$[P_{\alpha }f]( \theta ) $ satisfies  
bounds of the form 
(B.15) or (B.16) on the part $u= \pm\eta, \ v \ge v_0-\eta$ of the border of a given region 
$B_a^{(\eta)}$ (estimates on the remaining ``small'' part $ |u| < \eta,$ $ v=v_0 -\eta$
are similar  
\footnote 
{For $|u|<\eta,\ v=v_0 - \eta$, one chooses the path with support 
$\left[0,u \right] \cup \left[u, u+{\rm i}(v_0 - \eta) \right] $, which yields two
contributions to (B.6): while the first one is bounded by a constant, the second one
is bounded (up to a constant factor) by the same integral as in (B.20) or (B.20') 
whose dependence on $\eta$ yields the desired result.}).

Let us first assume that $f$ belongs to 
${\cal O}^{\beta }( B_{a}^{({\rm cut})%
}), $ with $\beta \ge 0.$  
In view of (B.15), 
one readily obtains
that the first two contributions to $[P_{\alpha }f]( \theta ) $
(given by the integrations on $\left[ 0,b\right] $ and $\left[ b,b+%
{\rm i}v\right] $) admit uniform bounds of the form $c(v)$, 
where $c(v)$ is locally bounded and power-like behaved for $v$ tending
to infinity. The third contribution (given by the interval $\left[ b+%
{\rm i}v,u+{\rm i}v\right] $) can be majorized by the following expression
(written for the case $0<u=\eta <b$):

\begin{equation}
\frac{1}{\Gamma \left( \alpha \right) }C\left( v\right) 
\int_{\eta}^{b}\left( u^{\prime }-\eta\right) ^{\alpha -1}\left( u^{\prime
}\right) ^{-\beta }du^{\prime }.  \tag{B.20}
\end{equation}

a)\ If $\alpha < \beta,$ the integral in (B.20) is bounded by 
${\rm cst\;}\eta^{-(\beta-\alpha)}$  
and therefore 
$P_{\alpha }f$  
belongs to ${\cal O}^{\beta -\alpha }( B_{a}^{({\rm cut})%
}). $

b)\ If $\alpha = \beta,$ the integral in (B.20) is  
bounded by ${\rm cst}\left| \ln \eta\right| .$ 
This shows  
that $P_{\beta }f$ belongs to the class ${\cal O}%
^{0*}( B_{a}^{( {\rm cut}) }) .$

c)\  If $\alpha > \beta ,$ the integral in (B.20) is bounded by a constant and
therefore $P_{\alpha }f $ belongs to ${\cal O}%
^{0}( B_{a}^{({\rm cut})}) .$ In order to show that $P_{\alpha
}f $ admits {\it continuous} boundary values on $\sigma
,$ one writes $P_{\alpha }f =P_{\varepsilon }P_{\alpha
-\varepsilon }f $ for a given $\varepsilon >0$ such
that $\alpha -\varepsilon >\beta .$
\noindent Since $g =P_{\alpha -\varepsilon }f$  
is then itself in ${\cal O}^{0}( B_{a}^{({\rm cut})%
}) ,$ one is led to apply directly the following result to the
expression (B.6) of $[P_{\varepsilon }g]( \theta ) $ (with the choice of the 
linear segment $\left[ 0,\theta \right] $  
for supp $\gamma (0,\theta )$, $\theta$  
being either in 
$ B_{a}^{({\rm cut})} $ 
or on the cut $\sigma$ ):
for every $\varepsilon >0,$ the Abel transform $%
g_{\varepsilon }(x)=\int_{0}^{x}f(y)\left( x-y\right) ^{\varepsilon -1}dy$
of a locally bounded function $f$ is continuous.
Moreover the previous argument holds for every derivative $D_{\nu' }
(P_{\alpha }f) $ such that $\nu' <\alpha -\beta,$ since in this case (in view of (B.8d)) 
$D_{\nu' }
(P_{\alpha }f) =P_{\alpha -\nu' }f$. 
We have thus proved that $P_{\alpha }f$ belongs to the class ${\cal O}%
_{\alpha -\beta }( B_{a}^{( {\rm cut}) }) .$

In order to complete the study of the case c), let us now assume that $f$ 
belongs to ${\cal O}^{0*} ( B_{a}^{( {\rm cut})
});$ in view of (B.16),  
the majorization (B.20) on the third contribution to $[P_{\alpha}f](\theta)$ 
is now replaced by 
\begin{equation}
\frac{1}{\Gamma \left( \alpha \right) }C\left( v\right) 
\int_{\eta}^{b}\left( u^{\prime }-\eta\right) ^{\alpha -1}\left| \ln  
u^{\prime }\right| du^{\prime }  \tag{B.20'} 
\end{equation} 
which is bounded by ${\rm cst\;}C(v).$
This proves that $P_{\alpha }f $ belongs to ${\cal O}%
^{0}( B_{a}^{( {\rm cut}) })
, $ and since the result holds for every $\alpha ^{\prime },$ with $0<\alpha
^{\prime }<\alpha ,$ the same argument as above in c) 
shows that $P_{\alpha }f$ belongs to the class ${\cal %
O}_{\alpha }( B_{a}^{( {\rm cut}) }) $ for all $%
\alpha >0.$ 

\vskip 0.2cm
\noindent ii)\quad If $f$ belongs to a class ${\cal O}^{\beta }
(B_{a}^{( {\rm cut}) }) ,$ or also (for $\beta=0,$) to  
${\cal O}^{0*}( B_{a}^{( {\rm cut}) }) ,$
let $p=E(\beta )+1;\ $ it follows
from i)c) that the function $F=P_{p}f$ admits continuous boundary values $%
F_{+},F_{-}$ on $\sigma .$ Then it results from the standard definition of
distribution-like boundary values of holomorphic functions that the function 
$f\left( \theta \right) =\left( \dfrac{d}{d\theta }\right) ^{p}F\left(
\theta \right) $ admits boundary values on $\sigma $ which are the
corresponding derivatives {\it in the sense of distributions }$f_{\pm
}=\left(\dfrac{1}{\rm i} \dfrac{d}{dv }\right) ^{p}F_{\pm }$ denoted $D_{p}F_{\pm }.$
Since $F_{+\left| \left[ 0,v_{0}\right[ \right. }=F_{-\left| \left[ 
0,v_{0}\right[ \right. },$ the discontinuity $\Delta F={\rm i}\left(
F_{+}-F_{-}\right) $ of $F$ is a continuous function with support contained
in $\sigma ,$ which yields the desired structure for the distribution $%
\Delta f=D_{p}\Delta F.$

\vskip 0.2cm

\noindent iii)\quad Let us consider, for any $\varepsilon ^{\prime }>0,$ the
Laplace transform $L\left( P_{\beta +\varepsilon ^{\prime }}\Delta f\right)
; $ in view of i)c), $P_{\beta +\varepsilon ^{\prime }}\Delta f$ is a
continuous function with support contained in $\sigma $ and satisfying a
bound of the following form:
\[
\left| [ P_{\beta +\varepsilon ^{\prime }}\Delta f] 
(v) \right| \leqslant C_{\varepsilon'}\left( v\right) ,  
\]
where $C_{\varepsilon'}\left( v\right) $ has power-like behaviour at infinity.

Therefore the corresponding expression of $L\left( P_{\beta +\varepsilon
^{\prime }}\Delta f\right)(\lambda) $ given by (B.4) can be uniformly bounded by $%
c_{\varepsilon \varepsilon ^{\prime }}{\rm e}^{-\left( \func{Re}\lambda
-\varepsilon \right)  v_{0}}$ in any half-plane ${\Bbb C}%
_{+}^{\left( \varepsilon \right) }\left( \varepsilon >0\right) .$ Since we
have (in view of Proposition B.2, Eq.(B.11)):

\[
L\left( \Delta f\right) \left( \lambda \right) ={\rm e}^{-\frac{{\rm i}\pi }{%
2}\left( \beta +\varepsilon ^{\prime }\right) }\lambda ^{\beta +\varepsilon
^{\prime }}L\left( P_{\beta +\varepsilon ^{\prime }}\Delta f\right) \left(
\lambda \right), 
\]
the majorization (B.19) follows from the
previous bound on $L\left( P_{\beta +\varepsilon ^{\prime }}\Delta f\right)
. $

\vskip 0.5cm
We now complete the statements of Proposition B.4 i) by considering
the action of derivatives $D_{\nu }$ of arbitrary order $\nu .$

\vskip 0.3cm
\noindent
{\sl Proposition B.5}

\noindent
\it  Let $f(\theta)$ belong to a class
${\cal O}^{\beta }
(B_{a}^{( {\rm cut}) }) ,$ 
with $\beta \ge 0.$ Then, for all $\nu >0,$  
the product $\theta^{\nu}\ D_{\nu}f(\theta)$ belongs to the class 
${\cal O}^{\beta +\nu} 
( B_{a-\delta }^{( {\rm cut}) }) $ for any $\delta
>0.$ \rm  

\vskip 0.3cm
\noindent
{\sl Proof}\quad Putting $\nu = n- \alpha$, with $n$ integer and $0< \alpha <1$, 
we can write in view of Eqs (B.9),(B.9'):
\[ 
[D_{\nu}f](\theta) = [D_nP_{\alpha}f](\theta) = [D_{n-1}(P_{\alpha}Df)](\theta)
+ f(0) \frac{\theta^{-\nu}}{\Gamma(1-\nu)}.  
\]
Since the second term 
at the r.h.s. of this equation has no discontinuity across $\sigma$, we are
led to prove that   
if $f$ belongs to ${\cal O}^{\beta }
(B_{a}^{( {\rm cut}) }) ,$ then  
$D_{n-1}(P_{\alpha}Df)$ belongs to  
${\cal O}^{\beta +\nu} 
( B_{a-\delta }^{( {\rm cut}) }) $ for any $\delta >0.$  
At first, simple estimates based on the
Cauchy formula for the derivative of a holomorphic function show that $%
Df$ belongs to ${\cal O}^{\beta +1} 
( B_{a-\delta }^{( {\rm cut}) }) $ for all $\delta
>0.$ Then, since $\alpha < 1< \beta +1$, the case a) of Proposition B.4 i) applies to $%
P_{\alpha}Df$ and implies that this function belongs to the corresponding classes 
${\cal O}^{\beta +1 - \alpha}$; applying now again the Cauchy formula to the derivative $D_{n-1}$    
of the previous function implies that  
$D_{n-1}(P_{\alpha}Df)$ belongs to  
${\cal O}^{\beta -\alpha +n} 
( B_{a-\delta }^{( {\rm cut}) }) $ for all $\delta >0$.

\vskip 0.8cm 
The rest of this Appendix is devoted to proving the following result which is of direct use
for our Theorem 3 (see Sec 4). Although very close to 
Proposition B.4 i) in its form, this result is technically 
more sophisticated since its statement involves  
conjointly 
primitives ${\bf P}_{\alpha}$ with respect to the complex variable
$z = \cos\theta $,   
together with the previous derivatives $D_{\nu}$ with respect to $\theta$  
(involved in the definition of the classes 
${\cal O}_{\alpha-\beta}(B_{\pi}^{({\rm cut})})$ used again below).  

\vskip 0.5cm
\noindent
{\sl Proposition B.6}

\noindent
\it Let $\alpha, \beta $ and $m$ be fixed real numbers such that $\alpha >0$,
$\beta \ge 0$ and $m > -1$.   
With each function $ f$ holomorphic in $B_{\pi}^{({\rm cut})}$ and such that:

\noindent
i) $f$ satisfies uniform bounds of the following form   
\[
|f(\theta)| \le {C {\rm e}^{m {\rm Im}\theta} \over \eta^{\beta}} 
\]
in all the corresponding subsets $B_{\pi}^{(\eta)}$ of $B_{\pi}^{({\rm cut})},$    

\noindent
ii) $f(u)=f(-u)$ for $0 \le u \le \pi$, 

\noindent
one associates the following function:

\begin{equation}
{\hat f}_m^{(\alpha)}(\theta)  = {\rm e}^{{\rm i}(m+\alpha)\ {\theta}}  
\dfrac{1}{\Gamma(\alpha)}  
\ \int_{-1}^{\cos \theta} {(\cos \theta -\cos \tau)}^{\alpha-1} f(\tau) \ d \cos \tau,  
\tag{B.21}  
\end{equation}
where $\theta$ varies in $B_{\pi}^{({\rm cut})}.$ 

\vskip 0.5cm 
\noindent
Then 

\noindent
a) If $\alpha < \beta,\quad    
{\hat f}_m^{(\alpha)}$ belongs to the class   
${\cal O}^{\beta-\alpha}(B_{\pi}^{({\rm cut})})$,

\noindent
b) If $\alpha = \beta,\quad      
{\hat f}_m^{(\alpha)}$ belongs to the class   
${\cal O}^{0*}(B_{\pi}^{({\rm cut})})$,

\noindent
c) If $\alpha > \beta,\quad      
{\hat f}_m^{(\alpha)}$  belongs to the class   
${\cal O}_{\alpha-\beta}(B_{\pi}^{({\rm cut})})$. \rm

\vskip 0.5cm
The proof of the latter relies on two auxiliary lemmas, for which 
we need the following notations. 
Let ${\Bbb C}_{A}={\Bbb C}%
\backslash \{z\  {\rm real}
; z\ge A \} 
\backslash \{z\  {\rm real}
; z \le -1 \}$   
with $A \ge 1.$
For every function $\underline f\left( z\right) ,$ holomorphic in ${\Bbb C}_{A}$ and
continuous on the cut $z\le -1$, and for every $\alpha >0,$ we put  
\begin{equation}
[{\underline P}_{\alpha }{\underline f}]( z) =  \frac{1}{\Gamma(\alpha)}  
\int_{-1}^{z}{\underline f}\left( z^{\prime }\right)
\left( z-z^{\prime }\right)  ^{\alpha -1}dz^{\prime }.  
\tag{B.22}
\end{equation}
By choosing $A= \cosh v_0$,   
the cut-plane ${\Bbb C}_{A}$ 
appears as the image 
of the set  
$B_{\pi}^{({\rm cut})}$ by the mapping $z = \cos\theta $.    
Considering the function $f(\theta) = {\underline f}(\cos\theta)$, holomorphic in 
$B_{\pi}^{({\rm cut})}$ (and such that $f(u) = f(-u)$ for $-\pi \le u \le \pi$), we then also put:    
\begin{equation}
[{\bf P}_{\alpha }{f}]( \theta) =   
[{\underline P}_{\alpha }{\underline f}]( \cos \theta) = - \frac{1}{\Gamma(\alpha)}  
\int_{\pi}^{\theta}
\left( \cos\theta-\cos\tau \right)  ^{\alpha -1}\ f(\tau) \ \sin\tau d\tau 
\tag{B.23}
\end{equation}
Lemma B.7 closely parallels the results of Proposition B.4 i) but 
it involves primitives  
${ \underline P}_{\alpha}$ taken 
in the cut-plane ${\Bbb C}_{A}$ 
and a corresponding new 
specification of the increase properties of the holomorphic functions considered. 

\vskip 0.5cm
\noindent
{\sl Lemma B.7}

\noindent
\it Let ${\underline f}(z)$, holomorphic in ${\Bbb C}_{A}$ and continuous on the cut $z \le -1$, 
satisfy uniform bounds of the following form     
\begin{equation}
\left| {\underline f}\left( z\right) \right| \leqslant \frac{C\left(1+\left| z\right| \right)^{m}}{\phi 
^{\beta }},\ \ \   {\rm with}\ \  m > -1  \ \ {\rm and}  \ 
\beta \ge 0
\tag{B.24}
\end{equation}
in the corresponding regions  (see Fig.B2)   
\[{\Bbb C}_{A}^{(\phi )}={\Bbb C}_{A}\  
\backslash 
\left\{ z\in {\Bbb C}%
; \ z =\rho {\rm e}^{{\rm i}\psi},\ \rho > A(1-\frac{\phi}{\pi}),\ |\psi|< \phi 
\right\}  
\]
\noindent for all $\phi\  \ (0 < \phi < \pi).$

\vskip 0.5cm 
\noindent  Then  
$[{\underline P}_{\alpha }{\underline f}](z)$ 
is holomorphic in 
$ {\Bbb C}_{A}$ and   
satisfies uniform bounds of the
following form  
for $z\in {\Bbb C}_{A}^{(\phi )}:$

\noindent a)\quad  If  $\alpha <\beta ,$

\begin{equation}
\left| 
[{\underline P}_{\alpha }{\underline f}](z) 
\right| 
\leqslant \dfrac
{C_{\alpha }
\left(1 + \left| z\right|\right)^{m+\alpha}} 
{\phi^{^{\beta -\alpha }}}  \tag{B.25}
\end{equation}

\noindent b)\quad If $\alpha =\beta ,$

\begin{equation}
\left| 
[{\underline P}_{\alpha }{\underline f}](z) 
\right| \leqslant 
C_{\alpha }
\left(1 + \left| z\right|\right)^{m+\alpha} 
\left| \ln \phi \right|   \tag{B.26}
\end{equation}

\noindent c)\quad  If  $\alpha >\beta ,$

\begin{equation}
\left| 
[{\underline P}_{\alpha }{\underline f}](z) 
\right| \leqslant C_{\alpha }
\left(1 + \left| z\right|\right) 
^{m+\alpha }  \tag{B.27}
\end{equation}

\noindent  and 
$[{\underline P}_{\alpha }{\underline f}](z)$  
is continuous in the
closure of ${\Bbb C}_{A}$ (from both sides of the cuts). \rm  

\vskip 0.5cm

\noindent
{\sl Proof:}
In order to obtain the bounds (B.25)...(B.27), it is sufficient to consider two typical
geometrical situations :

i) $z$ is of the form $z = A(1-\frac{\phi}{\pi}) {\rm e}^{{\rm i}\psi},$ with $0 \le |\psi| \le \phi; $ 
the integration path in (B.22) is then chosen as the union of two linear paths with supports
$\{ z'\  {\rm real}; \ -1 \le z' \le 0\}$ and 
$\{ z' \in {\Bbb C};\  z' = A(1-\frac{\phi'}{\pi}) {\rm e}^{{\rm i} \psi},\ \phi \le \phi' \le \pi \}.$ 
By using the assumption (B.24), one checks that the first contribution to
$[{\underline P}_{\alpha }{\underline f}](z)$  
is bounded by a constant, while the second one is majorized 
(up to a constant factor) by
$ \int_{\phi}^{\pi} (\phi' -\phi)^{\alpha-1} (\phi')^{-\beta} d\phi', $ which is of the same
form as the integral in (B.20).
In view of the analysis after (B.20) (cases a), b), c)), we then obtain the
corresponding bounds (B.25)...(B.27)(with $|z|^{m+{\alpha}}$ replaced by a constant). 

\begin{figure}
\epsfxsize=10truecm 
{\centerline{\epsfbox{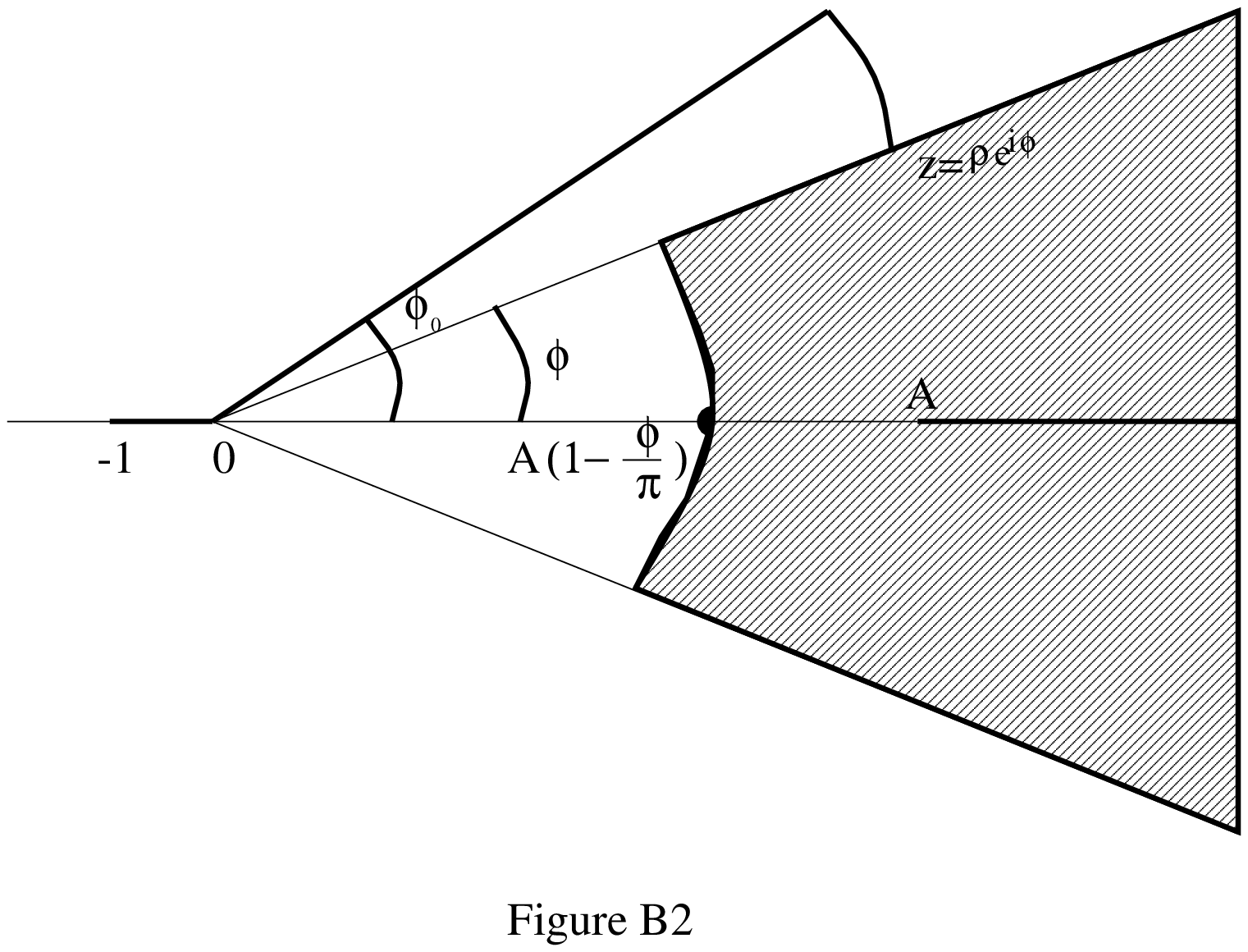}}}
\end{figure}

ii) $z$ is of the form $z = \rho {\rm e}^{{\rm i}\phi},$ with $\rho > A(1-\frac{\phi}{\pi})$; the integration
path in (B.22) is then chosen as the union of three paths (see Fig.B2) with respective supports
$\{ z'\  {\rm real};\ -1 \le z' \le 0 \}$, $\{ z' = \rho' {\rm e}^{{\rm i}\phi_o};\ 0\le \rho' \le \rho \}$
and $\{ z' = \rho {\rm e}^{{\rm i}\phi'};\ \phi \le \phi' \le \phi_0 \}$ ($\phi_0$ being a 
fixed angle with $0 < \phi_0 \le {\pi}$). In view of (B.24), the corresponding first two
contributions to  
${\underline P}_{\alpha }{\underline f}(z)$  
are majorized respectively by $cst |z|^{\alpha -1}$ and
$cst |z|^{m + \alpha}$ and therefore (since $m > -1$) both by $cst |z|^{m+\alpha}$.  
The contribution given by the third path is majorized by
$cst |z|^{m+\alpha} \int_{\phi}^{\phi_0} (\phi' - \phi)^{\alpha -1}\ (\phi')^{-\beta} d\phi'.$
By applying again the results described after Eq.(B.20), we then obtain the majorizations  
(B.25)...(B.27) in the corresponding cases a), b) and c). Finally the continuity of 
${\underline P}_{\alpha }{\underline f}$ on $\sigma$ (from both sides)   
in case c) is again justified as in Proposition B.4.

\vskip 0.5cm
\noindent
Lemma B.8 shows that the identity 
$h(\theta) ={ d\over d\theta}\left[P_{1-\nu}(P_{\nu}h)\right](\theta)$
is replaced by an equally regular operation when $P_{\nu}$ is replaced by ${\bf P}_{\nu}$ and 
intertwining exponentials are added.

\vskip 0.5cm
\noindent
{\sl Lemma B.8}

\noindent
\it For every function $h(\theta)$  holomorphic and 
uniformly bounded in $ B_{\pi}^{({\rm cut})} $ 
and admitting
continuous boundary values on $\sigma$, the following transform 
\[ [K_{\nu,\mu,r}h] (\theta) =  
{ d\over d\theta}\left[P_{1-\nu}\left({\rm e}^{{\rm i}(r+{\mu})\theta}{\bf P}_{\nu}
\left({\rm e}^{-{\rm i}(\mu -\nu)\theta}h\right)\right)\right](\theta)
\]
is the product of a 
function in 
$\displaystyle \bigcap_{\delta;\ \delta >0}$ $ {\cal O}%
^{0}( B_{\pi- \delta}^{( {\rm cut}) }) $ 
by $\theta^{-\nu}$,   
and also admits  
continuous boundary values on $\sigma$, 
provided one has $0 < \nu <1,\  \mu > \nu -1,\  r\ge 0. $ \rm 

\vskip 0.5cm
\noindent
{\sl Proof:}  

In view of Eqs (B.6) and (B.23), we have:   
\[ [K_{\nu,\mu,r} h](\theta) = {-1 \over{\Gamma(1-\nu) \Gamma(\nu)}}\    
{d \over d\theta} \int_0^{\theta}(\theta-\theta')^{-\nu}
{\rm e}^{{\rm i}(r+\mu)\theta'} d\theta'\ldots
\]

\begin{equation}
\ldots\left[\int_{\pi}^{\theta'}(\cos\theta'-\cos\tau)^{\nu-1} {\rm e}^{-{\rm i}(\mu-\nu)\tau}
h(\tau)\sin\tau d\tau \right] 
\tag{B.28}
\end{equation} 
which is well-defined for $0 < \nu < 1$ and can  
be rewritten as a sum of two terms
\begin{equation}
[K_{\nu,\mu,r} h](\theta)\  =\  h_1(\theta) + h_2(\theta)  
\tag{B.29}
\end{equation} 
corresponding to the following splitting of the integral over $\tau$: 
$\int_{\pi}^{\theta'}=
\int_{0}^{\theta'}+
\int_{\pi}^{0}.$
We shall then study $h_1$ and $h_2$ separately and prove that they both satisfy the 
property to be shown for 
$K_{\nu,\mu,r} h.$   

We first treat the term $h_1$ 
by inverting the order of the integrations over $\theta'$ and $\tau$, which yields: 
\begin{equation}
h_1(\theta)= -{d \over d\theta} \int_0^{\theta} {\cal K}(\theta,\tau)\  
{\rm e}^{{\rm i}(r+\nu)\tau}
h(\tau)\sin\tau d\tau,   
\tag{B.30}
\end{equation} 
where the kernel ${\cal K}$ is defined as follows:
\begin{equation}
{\cal K}(\theta,\tau) =  
{1 \over{\Gamma(1-\nu) \Gamma(\nu)}}    
\int_{\tau}^{\theta}(\theta-\theta')^{-\nu}
(\cos\theta'-\cos\tau)^{\nu-1} 
{\rm e}^{{\rm i}(r+\mu)(\theta'-\tau)} d\theta'.   
\tag{B.31}
\end{equation} 
The validity of Eq (B.30) is submitted to the proof of the regularity
of ${\cal K}$ given below; in particular,
the following alternative to Eq.(B.30) will be justified after checking the 
regularity of 
${\cal K}$ on the diagonal: 
\begin{equation}
h_1(\theta) = -{\cal K}(\theta, \theta)\  
{\rm e}^{{\rm i}(r+\nu)\theta}
h(\theta)\sin\theta 
- \int_0^{\theta} {\partial {\cal K} \over \partial \theta}(\theta, \tau)\  
{\rm e}^{{\rm i}(r+\nu)\tau}
h(\tau)\sin\tau d\tau.    
\tag{B.32}
\end{equation} 
{\sl Study of  
${\cal K}$:}
\ by putting $\theta' =\tau + t(\theta-\tau)$ and passing to the integration 
variable $t\ (0 \le t \le 1)$ in Eq.(B.31), we can rewrite the latter as follows: 
\begin{equation}
{\cal K}(\theta,\tau) =  
{1 \over{\Gamma(1-\nu) \Gamma(\nu)}}    
\int_0^1 \Phi\left(\tau, {t(\theta - \tau) \over 2}\right)
{\rm e}^{{\rm i}(r+\mu)t(\theta-\tau)} 
(1-t)^{-\nu} t^{\nu -1}dt,   
\tag{B.33}
\end{equation} 
where:
\begin{equation}
\Phi(\tau, \zeta) = \left[ -{\sin \zeta \over \zeta} \sin(\tau + \zeta) \right]^{\nu-1}.  
\tag{B.34}
\end{equation} 
One immediately obtains that
${\cal K}(\theta,\theta) = \Phi(\theta,0)= (- \sin \theta)^{\nu -1}$, 
so that the first contribution to $h_1(\theta)$ in Eq.(B.32) is equal to 
$(- \sin \theta)^{\nu} {\rm e}^{{\rm i}(r+\nu)\theta} h(\theta)$. Since $h$ is holomorphic and bounded, 
this function belongs to 
${\cal O}^0 ( B_{\pi}^{({\rm cut})} )$   
( under the assumptions $\nu >0$ and $r \ge 0$);
it also admits continuous boundary values on $\sigma$ like $h$.
One notices that this contribution is the exact analogue of the reproducing expression 
${ d\over d\theta}\left[P_{1-\nu}(P_{\nu}h)\right](\theta) = h(\theta).$  

The study of the second contribution to Eq.(B.32) relies on the following expression 
of $\partial {\cal K} \over \partial \theta $ (deduced from (B.33)): 
\[
{\partial {\cal K} \over \partial \theta}(\theta, \tau) =  
{1 \over{\Gamma(1-\nu) \Gamma(\nu)}}\cdots 
\]
\begin{equation}
\int_0^1 \left\{{\rm i}(r+ \mu) 
\Phi(\tau,\zeta) + {1 \over 2}{\partial \Phi \over \partial \zeta} (\tau, \zeta)\right\}_
{|\zeta = 
{t(\theta - \tau) \over 2}} 
{\rm e}^{{\rm i}(r+\mu)t(\theta-\tau)} 
(1-t)^{-\nu} t^{\nu }dt.    
\tag{B.35}
\end{equation} 
Since $\nu -1 <0,$ Eq.(B.34) implies that the expression inside 
the bracket $\{...\}$ in the latter integral is 
(for each $t \in [0,1]$) a holomorphic function of $\tau$ and $\theta$ 
in the domain $\Delta = \{(\theta,\tau) \in {\bf C}^2;\ \theta \in B_{\pi},\    
\tau \in B_{\pi}, \ 0 < {\rm Im} \tau < 
{\rm Im} \theta\} $  
which is uniformly 
bounded by 
$cst\ {\tau}^{\nu -2} 
{\rm e}^{(t{\rm Im}(\theta-\tau) + {\rm Im}\tau)(\nu-1)}$.  
up to peaks in ${|(\pi \pm \theta)|}^{\nu -2}$ near $\theta - \pm \pi$ (their contributions  
can be factored out in the bounds). 
It directly follows that, under the conditions $r \ge 0,\  \mu > \nu -1,$ 
the complete integrand of (B.35) and thereby 
the kernel  
${\partial {\cal K} \over \partial \theta}(\theta, \tau) $   
are themselves holomorphic 
and uniformly bounded by 
$cst\ {\tau}^{\nu -2}   
{\rm e}^{(\nu -1){\rm Im}\tau}$   
in $\Delta.$ 
One then sees (by using again the condition $r \ge 0$) that 
in the second contribution to the r.h.s. of Eq.(B.32),
the integrand is uniformly bounded by 
$cst\ {\tau}^{\nu -1} ;$  
this contribution is therefore     
holomorphic in $B_{\pi}^{({\rm cut})}$ (except for a branch-point 
with behaviour ${\theta}^{\nu}$ at $\theta = 0$), and uniformly  
bounded there by $cst\ {\theta}^{\nu}$ 
up to the previous peaks in $cst\ {|\pi \pm \theta|}^{\nu -2}$ near $\theta = \pm \pi$. 
It therefore belongs to ${\cal O}^0(B_{\pi-\delta}^{({\rm cut})})$ for all $\delta >0$. Moreover, 
since ${\cal K}$ is analytic 
for $\theta ={\rm i} v,\ \tau = {\rm i}w,\ 0< w\le v$,  
this contribution admits (like  $h$) continuous boundary values on $\sigma$.  
We have thus proved that $h_1(\theta)$ 
satisfies the 
desired properties.

The term $h_2(\theta)$ is treated directly by  
writing (in view of (B.28)):
\begin{equation}
h_2(\theta) = {-1 \over{\Gamma(1-\nu) \Gamma(\nu)}}\    
{d \over d\theta} \int_0^{\theta}(\theta-\theta')^{-\nu}
{\rm e}^{{\rm i}(r+\mu)\theta'} 
\Psi(\theta')
d\theta', 
\tag{B.36}
\end{equation} 
with
\begin{equation}
\Psi(\theta) =
-\int_0^{\pi}(\cos\theta-\cos\tau)^{\nu-1} {\rm e}^{-{\rm i}(\mu-\nu)\tau}
h(\tau)\sin\tau d\tau.   
\tag{B.37}
\end{equation} 
In fact, both functions $\Psi(\theta)$ and 
$\theta {\partial \Psi \over \partial \theta}(\theta)$
are holomorphic in
$B_{\pi}$ (except for branch-points at $0, \pi$ and $ -\pi$)  and  
uniformly bounded by 
$cst\  {\rm e}^{(\nu -1){\rm Im}\theta}$   
in this domain, up to peaks in ${|(\pi \pm \theta)|}^{2\nu -1} $ near
$\theta =\pm \pi$. 
By passing to the integration variable $t= {\theta'\over \theta}, \ 0 \le t \le 1,$ 
in Eq.(B.36), which allows one to derive with respect to $\theta$ under the integral
and to factor out ${\theta}^{-\nu}$, one can 
make use of the previous bounds. In view of the conditions $r \ge 0, \ \mu > \nu-1,$
one checks that the integral is uniformly bounded 
in $B_{\pi -\delta}$ and therefore that  
${\theta}^{\nu} h_2$ belongs to 
${\cal O}^0(B_{\pi-\delta}^{({\rm cut})})$ for all $\delta >0$; 
moreover, $h_2$ is holomorphic on $\sigma$ (like $\Psi$, it has no cut). 

\vskip 1cm
\noindent
{\sl Proof of Proposition B.6:} 

One easily checks that the function ${\underline f}(z)$ defined by
${\underline f}(\cos \theta) = f(\theta)$ is holomorphic in ${\Bbb C}_A,$ with $A= \cosh v_0,\  $
and that it satisfies the assumptions of Lemma B.7. 
This follows from the fact that the sets $B_{\pi}^{(\eta)}$ (see Eq.(B.2)) are equivalent
to the sets ${\Bbb C}_A^{(\phi)}$ of Lemma B.7 by the mapping  $Z: \ \theta \to z=\cos\theta,$
(in the following sense:  
${\Bbb C}_A^{(a\eta)} \subset 
Z( B_{\pi}^{(\eta)})  \subset 
{\Bbb C}_A^{(b\eta)}$ with $\ 0 < a <1 <b$) and that
${\rm a} |{\rm e}^{-{\rm i}\theta}| 
< (1 + |z|) 
<{\rm b}|{\rm e}^{-{\rm i}\theta}|$ ($a,b,$ a and b being fixed numbers). 

It then follows from these facts and from the conclusions of Lemma B.7 (formulas 
(B.25)...(B.27)) that the corresponding functions 
${\hat f}_m^{(\alpha)}(\theta)  = {\rm e}^{{\rm i}(m+\alpha){\theta}} 
\left({\bf P}_{\alpha}{f} \right)(\theta)$ are holomorphic in $B_{\pi}^{(cut)}$ and 
enjoy the following properties:

a) If $\alpha < \beta,\quad    
|{\hat f}_m^{(\alpha)}(\theta)| \le \dfrac{C}{{\eta}^{\beta -\alpha}}\  $     
for 
$\theta \in  B_{\pi}^{(\eta)},\   $ 
which entails that 
${\hat f}_m^{(\alpha)} \in   
{\cal O}^{\beta-\alpha}\ (B_{\pi}^{({\rm cut})})$,   

b) If $\alpha = \beta,\quad      
|{\hat f}_m^{(\alpha)}(\theta)| \le {C}{|\ln \eta|}$      
for 
$\theta \in  B_{\pi}^{(\eta)},\   $ 
which entails that 
${\hat f}_m^{(\alpha)} \in   
{\cal O}^{0*}\ (B_{\pi}^{({\rm cut})})$,   

c) If $\alpha > \beta,\quad      
{\hat f}_m^{(\alpha)}$  is bounded and continuous in the closure of $B_{\pi}^{({\rm cut})}$. 
In order to establish that  
${\hat f}_m^{(\alpha)}$  belongs to the class   
${\cal O}_{\alpha-\beta}\ (B_{\pi}^{({\rm cut})}),$ 
we shall now prove that for all real $\nu$ such that $ 0 < \nu < \alpha-\beta$, the 
function $D_{\nu}{\hat  f}_m^{\alpha}(\theta)$ is the product of $\theta^{-\nu}$ by  
a holomorphic function belonging to  
${\cal O}^0 (B_{\pi- \delta}^{({\rm cut})})$ for all $\delta >0$.
This will be done in three steps: we first give a proof for the case of ordinary derivatives, 
i.e. $\nu = r$ integer; then 
we reduce the proof for 
a general non-integral value of $\nu$ to that  
of a similar property for the corresponding value 
${\nu}_1 = \nu - {\rm E}(\nu)$ and finally we show the latter property for all values
of $\nu_1$ with $0 < \nu_1 <1$.  

1)\ $\nu = r$ integer: 
\ we claim that   
a relation of the following form holds: 
\[
[D_r{\hat f}_m^{(\alpha)}](\theta)  \equiv D_r \left[{\rm e}^{{\rm i}(m+\alpha)\;{\theta}} 
({\bf P}_{\alpha}{f})\right](\theta) 
\]
\[
= \sum_{r'=0}^r X^{(r')}({\rm e}^{{\rm i}\theta})
{\rm e}^{{\rm i}(m+\alpha - r')\ {\theta}} 
[{\bf P}_{\alpha - r'}{f}](\theta)  
\]
\begin{equation}
= \sum_{r'=0}^r X^{(r')}({\rm e}^{{\rm i}\theta})\  
{\hat f}_m^{(\alpha-r')}(\theta),    
\tag{B.38}
\end{equation} 
where each $X^{(r')}$ is a polynomial of degree $2r'$. 
This relation (which is a variant of Eq.(II.43) of [25b)]) is obtained by 
taking the derivative of order $r$ 
with respect to $\theta$ in the integral of (B.21): this is justified 
since $\alpha > \beta + r > r$.  

Eq.(B.38) immediately shows that   
$[D_{r}{\hat f}_m^{(\alpha)}](\theta)$  
is bounded and continuous in the closure of $B_{\pi}^{({\rm cut})}$,     
since each of the factors  
${\hat f}_m^{(\alpha-r')}$ (where $ \alpha-r'> \beta$)  and  
$X^{(r')}({\rm e}^{{\rm i}\theta})$ (with   
$|{\rm e}^{{\rm i}\theta}| = {\rm e}^{-v} < 1 $ in   
$B_{\pi}^{({\rm cut})}$)  
satisfies this property individually. 

2)\ for non-integral $\nu$ , let $\nu = \nu_1 + r$ with 
$r = {\rm E}(\nu) \ge 0,\  0< \nu_1 < 1$.
We apply Eq.(B.9)
(which is legitimate since $\hat f_m^{\alpha}(\theta) $ is holomorphic at 
$\theta = 0$): 
\begin{equation}
[D_{\nu}{\hat  f}_m^{\alpha}](\theta)   
= [D_{r+1} P_{1- {\nu}_1}{\hat  f}_m^{\alpha}](\theta)
= [D P_{1- {\nu}_1} D_r{\hat  f}_m^{\alpha}](\theta)
+ \sum_{p=0}^{r-1} [D_p f](0) \frac{\theta^{p- \nu }}{\Gamma(p- \nu +1)}
\tag{B.39}
\end{equation} 
The sum at the r.h.s. of Eq.(B.39) is the product of ${\theta}^{-\nu}$ by a  
function in ${\cal O}^0 (B_{\pi}^{({\rm cut})})$ (with no cut on $\sigma$). 
In view of Eq.(B.38) we are thus led to show  
the following property (in which we have put $\alpha' = \alpha -r'$,  
with $\alpha' \ge \alpha-r > \beta + \nu_1$):

{\it Let $ 0 < \nu_1 < 1$; then for every $\alpha'$ with $\alpha' > \beta + \nu_1 $ and for every 
$r'\ (r' \ge 0)$, the function $D P_{1-\nu_1}
\left[X^{(r')}({\rm e}^{{\rm i}\theta})\  
{\hat  f}_m^{\alpha'}(\theta)\right]$ 
is the product of  
$\theta^{-\nu_1}$   
by a function in 
$\displaystyle \bigcap_{\delta;\ \delta >0}$ $ {\cal O}%
^{0}( B_{\pi- \delta}^{( {\rm cut}) }) $ 
and it admits continuous boundary values on $\sigma.$}  \  

(When $\nu = \nu_1 < 1$ (i.e. $r=0$), one just uses the latter 
for $D P_{1-\nu}[{\hat f}_m^{\alpha}(\theta)]$). 

3)\ The proof of the previous statement  
relies in a crucial way on Lemma B.8. In fact, it is sufficient 
to replace the polynomial 
$X^{(r')}({\rm e}^{{\rm i}\theta}) $ by a typical term    
${\rm e}^{{\rm i}r\theta},\ r\ge 0, $  and to study the expression:
\begin{equation}
D_{\nu_1}\left[{\rm e}^{{\rm i}r\theta}\  
{\hat  f}_m^{\alpha'}(\theta)\right] 
= {\frac{d}{d\theta}}\left[P_{1-\nu_1} \left({\rm e}^{{\rm i}(r+m+\alpha')\;{\theta}} 
{\bf P}_{\alpha'}(f)\right)\right](\theta).  
\tag{B.40}
\end{equation} 
We can now write 
${\bf P}_{\alpha'}{f} =    
{\underline P}_{\alpha'}{\underline f} =    
{\underline P}_{\nu_1}\ {\underline g}, $ with    
${\underline g} = {\underline P}_{\alpha'-\nu_1}\ {\underline f}$ and notice that, since
$\alpha'-\nu_1 > \beta$, Lemma B.7 c) implies that one can put  
${\underline g}(\cos\theta) =  
{\rm e}^{-{\rm i}(m+\alpha'-\nu_1){\theta}}{h}(\theta),$  {\it with $h$ holomorphic and uniformly bounded
in $B_{\pi}^{({\rm cut})}$, and admitting continuous boundary values on $\sigma$.} 
Eq.(B.40) then becomes:  
\[
D_{\nu_1}\left[{\rm e}^{{\rm i}r\theta}\  
{\hat  f}_m^{\alpha'}(\theta)\right]  
= {\frac{d}{d\theta}}\left[P_{1-\nu_1} \left({\rm e}^{{\rm i}(r+m+\alpha'){\theta}} 
\ {\bf P}_{\nu_1} 
({\rm e}^{-{\rm i}(m+\alpha'-\nu_1) {\theta}} 
\ {h})\right)\right](\theta).  
\]
\begin{equation}
= K_{\nu_1, m+\alpha',r}\  h(\theta)
\tag{B.41}
\end{equation} 
Since $\alpha' > \nu_1$ and $m >-1$, the assumptions of Lemma B.8 are satisfied 
and the announced result follows, which ends the proof
of Proposition B.6.

\vskip 0.5cm
\centerline {\bf ACKNOWLEDGEMENTS}
\vskip 0.5cm
We are very grateful to 
the Italian {\it Istituto Nazionale di Fisica Nucleare  (I.N.F.N.), Sezione di Genova,}
for its financial support to one of us (J.B.) during the elaboration  
of this research project. 

\vskip 0.5cm
\centerline {\bf REFERENCES}

\vskip 0.5cm
\noindent \hangindent=8mm \hangafter=1  
[1] J.R. FORSHAW and D.A. ROSS, ``QCD and the Pomeron'', Cambridge 
University Press, 1997. 

\noindent \hangindent=8mm \hangafter=1  
[2] V.S. FADIN, E.A. KURAEV and L.N. LIPATOV, {\it  Phys. Lett.B} {\bf 60} (1975) p. 50;\  
I.I. BALITSKY and L.N. LIPATOV, {\it  Sov. J. Nucl. Phys.} {\bf 28} (1978) p. 822. 

\noindent \hangindent=8mm \hangafter=1  
[3] M.C. BERG\`ERE and C. GILAIN, ``Regge-pole behaviour in $\phi^3-$field theory'', 
{\it J. Math. Phys.} {\bf 19} (1978) p. 1495-1512;\  
M.C. BERG\`ERE and C. DE CALAN, ``Regge-pole behaviour from perturbative scalar field theories'',
{\it Phys. Rev.} {\bf D 20} (1979) p. 2047-2067. 

\noindent \hangindent=8mm \hangafter=1  
[4] R.K. ELLIS, W.J. STIRLING and B.R. WEBBER ``QCD and Collider 
Physics'', 
Cambridge University Press, 1996.  

\noindent \hangindent=8mm \hangafter=1  
[5] A. DONNACHIE and P.V. LANDSHOFF, {\it Z. f\"ur Phys.C} {\bf 61}, (1994) p.139. 

\noindent \hangindent=8mm \hangafter=1  
[6] H.M. NUSSENZWEIG, ``Diffraction effects in semi-classical Physics'', 
Cambridge University Press, 1992.  

\noindent \hangindent=8mm \hangafter=1  
[7] V, DE ALFARO and T. REGGE, ``Potential Scattering'', North Holland, 
1965. 

\noindent \hangindent=8mm \hangafter=1  
[8] R.G. NEWTON, ``The complex $j-$plane'', W.A. Benjamin, 1964.  

\noindent \hangindent=8mm \hangafter=1  
[9] P.D.B. COLLINS, ``Introduction to Regge theory and High Energy Physics'', 
Cambridge University Press, 1977.  

\noindent \hangindent=8mm \hangafter=1  
[10] J. BROS and G.A. VIANO, ``Complex angular momentum in axiomatic 
quantum field theory'', 
in {\it Rigorous methods in particle physics}, S.Ciulli, 
F. Scheck, W. Thirring Eds.
({\it Springer tracts in Mod. Phys.} {\bf 119} (1990)) p.53-76. 

\noindent \hangindent=8mm \hangafter=1  
[11] M. FROISSART, ``Asymptotic behaviour and subtractions in the Mandelstam  
representation'', 
{\it Phys. Rev.} {\bf 23} (1961) p. 1053-1057.  

\noindent \hangindent=8mm \hangafter=1  
[12] V.N. GRIBOV, ``Partial waves with complex angular momenta and the 
asymptotic behaviour of the
scattering amplitude'', {\it J. Exp. Theor Phys.}
{\bf 14} (1962) p. 1395. 

\noindent \hangindent=8mm \hangafter=1  
[13] H. EPSTEIN, ``Some Analytic Properties of Scattering Amplitudes in Quantum Field Theory'',  
in {\it Axiomatic Field Theory}, M. Chr\'etien and S. Deser Eds, Gordon and Breach, New York, 1966, 
p.3-133.

\noindent \hangindent=8mm \hangafter=1  
[14] J. BROS and G.A. VIANO, in preparation.   

\noindent \hangindent=8mm \hangafter=1  
[15] R. OMNES, ``D\'emonstration des relations de dispersion'' in 
{\it Relations de dispersion et 
particules \'el\'ementaires}, C. De Witt and R. Omnes Eds, Hermann, Paris, 1960, p.317-385. 

\noindent \hangindent=8mm \hangafter=1  
[16] A. MARTIN, {\it Nuovo Cimento} {\bf 42} (1965) p.930 and {\bf 44} (1966) p.1219. 

\noindent \hangindent=8mm \hangafter=1  
[17] O. STEINMANN, {\it Helv. Phys. Acta} {\bf 33} (1960) p.257. 

\noindent \hangindent=8mm \hangafter=1  
[18] D. RUELLE, {\it Nuovo Cimento} {\bf 19} (1961) p.356 and Thesis, Z\"urich, 1959. 

\noindent \hangindent=8mm \hangafter=1  
[19] H. ARAKI, {\it J. Math. Phys.} {\bf 2} (1961) p.163. 

\noindent \hangindent=8mm \hangafter=1  
[20] H. EPSTEIN, V. GLASER, R. STORA, ``General properies of the $n-$point functions in 
local quantum field theory'' 
in {\it Structural Analysis of Collision Amplitudes},  
R. Balian and D. Iagolnitzer Eds, North-Holland, Amsterdam, 1976, p.7-93.  

\noindent \hangindent=8mm \hangafter=1  
[21] O. STEINMANN, {\it Commun. Math. Phys.} {\bf 10} (1968) p.245. 

\noindent \hangindent=8mm \hangafter=1  
[22] A.S. WIGHTMAN, ``Analytic functions of several complex variables'' in 
{\it Relations de dispersion et 
particules \'el\'ementaires}, C. De Witt and R. Omnes Eds, Hermann, Paris, 1960, p.227-315. 

\noindent \hangindent=8mm \hangafter=1  
[23] J. BROS, H. EPSTEIN and V. GLASER, ``Some rigorous analyticity properties 
of the four-point
function in momentum space'',  
{\it Nuovo Cimento}
{\bf 31} (1964) p.1265-1302. 

\noindent \hangindent=8mm \hangafter=1  
[24] J. BROS and G.A. VIANO, ``Connection between the algebra of kernels on the sphere and the 
Volterra algebra on the one-sheeted hyperboloid: holomorphic perikernels'', 
{\it Bull. Soc. Math. France} {\bf 120} (1992) p.169-225. 

\noindent \hangindent=8mm \hangafter=1  
[25] J. BROS and G.A. VIANO, ``Connection between the harmonic analysis 
on the sphere and the harmonic
analysis on the one-sheeted hyperboloid: 
an analytic continuation viewpoint'', 
{\it Forum Math.} a) {\bf 8} (1996) p.621-658, b) {\bf 8} (1996) p.659-722, c) {\bf 9} (1997) 
p.165-191.  

\noindent \hangindent=8mm \hangafter=1  
[26] J. FARAUT, ``Analyse Harmonique et Fonctions Sp\'eciales'', {\it Ecole d'\'et\'e 
d'Analyse Harmonique de Tunis}, 1984.

\noindent \hangindent=8mm \hangafter=1  
[27] N.Ja. VILENKIN, ``Special Functions and the Theory of Group Representations'',  
{\it Amer. Math. Soc. Transl.} {\bf 22}, Providence R.I., 1968. 

\noindent \hangindent=8mm \hangafter=1  
[28] R.F. STREATER and A.S. WIGHTMAN, ``PCT, Spin and Statistics 
and all that'', W.A. Benjamin, 
New York, 1964.  

\noindent \hangindent=8mm \hangafter=1  
[29] R.P. BOAS, ``Entire Functions'', Academic Press, New York, 1954. 

\noindent \hangindent=8mm \hangafter=1  
[30] A. SOMMERFELD, ``Partial Differential Equations in Physics'', Academic 
Press, New York, 1949.  

\noindent \hangindent=8mm \hangafter=1  
[31] G.N. WATSON, ``The diffraction of electric waves by the earth'' {\it Proc. 
Roy. Soc.}, London, 
{\bf 95} (1918) p.83-99. 

\noindent \hangindent=8mm \hangafter=1  
[32] J. FARAUT and G.A. VIANO, ``Volterra algebra and the Bethe-Salpeter 
equation'' 
{\it J. Math. Phys.} {\bf 27} (1986) p.840-848.

\noindent \hangindent=8mm \hangafter=1  
[33] K. OSTERWALDER, R. SCHRADER, {\it Commun. Math. Phys.} {\bf 33} (1973) p.83 
and {\bf 42} (1975) p.281.  

\noindent \hangindent=8mm \hangafter=1  
[34] V. GLASER, {\it Commun. Math. Phys.} {\bf 37} (1974) p.257. 

\noindent \hangindent=8mm \hangafter=1  
[35] J. BROS and V. GLASER, ``L'enveloppe d'holomorphie de l'union 
de deux polycercles'',  
mimeographed report, Saclay 1961. 

\noindent \hangindent=8mm \hangafter=1  
[36] A. ERD\'ELYI, W. MAGNUS, F. OBERHETTINGER, F.G. TRICOMI, 
a) ``Higher Transcendantal Functions'',
Vol.1, McGraw-Hill, New-York, 1953,  
b) ``Tables of Integral Transforms'',
Vol.2, McGraw-Hill, New-York, 1954.

\end{document}